\documentclass[aps,prd,nofootinbib,twocolumn,superscriptaddress,preprintnumbers,letterpaper]{revtex4}

\usepackage{amssymb}
\usepackage{amsmath}
\usepackage{epsfig}
\usepackage{hyperref}
\usepackage{comment}
\usepackage{color}
\usepackage{cleveref}
\usepackage{multirow}
\usepackage{capt-of}
\usepackage{siunitx}
\usepackage{graphicx}
\usepackage{gensymb}
\usepackage{subfigure}
\usepackage[normalem]{ulem}
\usepackage{tabularx}
\usepackage{array}

\makeatletter
\def\simgt{\mathrel{\lower2.5pt\vbox{\lineskip=0pt\baselineskip=0pt
           \hbox{$>$}\hbox{$\sim$}}}}
\def\simlt{\mathrel{\lower2.5pt\vbox{\lineskip=0pt\baselineskip=0pt
           \hbox{$<$}\hbox{$\sim$}}}}
\makeatother

\newcommand{\be}{\begin{equation}}
\newcommand{\ee}{\end{equation}}
\newcommand{\bea}{\begin{eqnarray}}
\newcommand{\eea}{\end{eqnarray}}

\preprint{MIT-CTP/4797}

\begin{document}

\title{The Darkest Hour Before Dawn: Contributions to Cosmic Reionization from Dark Matter Annihilation and Decay}

\author{Hongwan Liu}
\email{hongwan@mit.edu}
\affiliation{Center for Theoretical Physics, Massachusetts Institute of Technology, Cambridge, MA}

\author{Tracy R. Slatyer}
\email{tslatyer@mit.edu}
\affiliation{Center for Theoretical Physics, Massachusetts Institute of Technology, Cambridge, MA}

\author{Jes\'{u}s Zavala\footnote{Marie Curie Fellow}}
\email{jzavala@dark-cosmology.dk}
\affiliation{Dark Cosmology Centre, Niels Bohr Institute, University of Copenhagen, Juliane Maries Vej 30, 2100 Copenhagen, Denmark}

\begin{abstract} 
Dark matter annihilation or decay could have a significant impact on the ionization and thermal history of the universe. In this paper, we study the potential contribution of dark matter annihilation ($s$-wave- or $p$-wave-dominated) or decay to cosmic reionization, via the production of electrons, positrons and photons. We map out the possible perturbations to the ionization and thermal histories of the universe due to dark matter processes, over a broad range of velocity-averaged annihilation cross-sections/decay lifetimes and dark matter masses. We have employed recent numerical studies of the efficiency with which annihilation/decay products induce heating and ionization in the intergalactic medium, and in this work extended them down to a redshift of $1+z = 4$ for two different reionization scenarios. We also improve on earlier studies by using the results of detailed structure formation models of dark matter haloes and subhaloes that are consistent with up-to-date $N$-body simulations, with estimates on the uncertainties that originate from the smallest scales. We find that for dark matter models that are consistent with experimental constraints, a contribution of more than 10\% to the ionization fraction at reionization is disallowed for all annihilation scenarios. Such a contribution is possible only for decays into electron/positron pairs, for light dark matter with mass $m_\chi \lesssim \SI{100}{MeV}$, and a decay lifetime $\tau_\chi \sim 10^{24} - 10^{25}\SI{}{s}$. 
\end{abstract}


\maketitle

\section{Introduction}
\label{sec:Introduction}

The epoch of reionization and the emergence of the universe from the cosmic dark ages is a subject of intense study in modern cosmology. As baryonic matter began to collapse around initial fluctuations in the dark matter (DM) density seeded by inflation, the earliest galaxies in our universe began to form. These structures, perhaps accompanied by other sources, eventually began to emit ionizing radiation, creating local patches of fully ionized hydrogen gas around them. These patches ultimately grew to encompass the entire universe, leading to the fully ionized intergalactic medium (IGM) that we observe today. 

While the process of reionization is broadly understood, the exact details of how and when reionization occurred are still somewhat unclear. Quasars and the earliest stars certainly played a part in reionization, but their relative energy contributions to the process are still a matter of ongoing research. Some studies have found \cite{Fan2001} that a significant population of dim and unobserved quasars must be present in order for them to completely reionize the universe. Similar conclusions have  been drawn for star-forming galaxies \cite{Robertson2013}. This uncertainty has resulted in some interest in other sources of energy that might contribute to reionization. 

DM provides a particularly compelling candidate, and has been considered several times in the literature. Many models allow DM to annihilate or decay into Standard Model particles, which in turn can deposit energy into the IGM through ionization, heating or other processes. The annihilation rate, which scales as the square of the density, rises substantially with the onset of structure formation and the collapse of DM into dense haloes, potentially yielding a large energy injection in the reionization epoch.

Our current knowledge of reionization can already place interesting constraints on DM properties. Constraints from optical depth and the temperature of the IGM placed strong constraints on DM models \cite{Cirelli:2009bb} that could generate the cosmic ray excesses observed by PAMELA \cite{Adriani:2008zr} and Fermi+HESS \cite{Abdo:2009zk,Collaboration:2008aaa,Aharonian:2009ah}. IGM temperature data as well as CMB power spectrum measurements can also be used to constrain the properties of $p$-wave annihilating and decaying DM \cite{Diamanti2014}. More recently, it has been shown that with improved measurements of the optical depth to the surface of last scattering and near-future probes of the cosmic ionization history, it should be possible to set new and significant constraints on the properties of annihilating or decaying DM \cite{Kaurov2015}. 

Turning the question around, the potential role that DM may have played in reionization has also been broadly explored. Earlier papers in the literature were able to find possible scenarios in which annihilating DM could contribute significantly to reionization, once structure formation was taken into account \cite{Chuzhoy2008,Natarajan:2008pk}. Subsequently, \cite{Belikov:2009qx} included the important effect of inverse Compton scattering off the cosmic microwave background (CMB) photons, and showed that weakly interacting massive particle (WIMP) DM candidates could play a dominant role in reionization. More recently, studies of $s$-wave annihilation of dark matter using an analytic description for the boost to the DM density during structure formation found that an unrealistic structure formation boost to the annihilation rates or an overly large cross-section was required for a DM-dominated reionization scenario consistent with existing experimental results from the CMB \cite{Poulin2015,Lopez-Honorez:2013lcm}. Multiple authors \cite{Mapelli:2006ej,Hansen:2003yj,Kasuya2004} have also shown that a significant contribution from decaying DM to reionization in a manner consistent with WMAP results is possible using specific DM decay rates and products.

In this paper, we examine the potential contribution of dark matter toward reionizing the universe, but improve on previous results in four crucial ways:

\begin{enumerate}
\item We consider an extremely wide range of DM masses, from 10 keV to TeV scales, and rather than selecting specific annihilation/decay channels, we consider the impact of electrons, positrons and photons injected at arbitrary energies. This allows us to place general, model-independent constraints on DM annihilation or decay, beyond the WIMP paradigm;

\item In addition to $s$-wave annihilation, we consider energy injection into the IGM through $p$-wave annihilation and decay. Energy injections in these scenarios have a different dependence on redshift and on the details of structure formation compared to the case of $s$-wave annihilation: consequently, different constraints dominate. We improve on these earlier results by performing a more accurate calculation of the energy injection/deposition rates and by taking into account the relevant constraints in each energy injection channel; 

\item The details of structure formation and its uncertainties are critical in determining the $s$-wave and $p$-wave annihilation rates \cite{Mack2014}. We use a detailed and up-to-date prescription of structure formation for our calculations, including the contribution of substructure in haloes (previous studies on substructure include \cite{Bartels:2015uba,Moline:2016pbm}). By calculating the boost factor to DM annihilation assuming two different halo profiles (consistently applied to both haloes and subhaloes) as well as the difference to the boost factor that results from including substructure effects, these results also allow us to estimate the uncertainties associated with structure formation, including uncertainties related to the subhalo boost factor;

\item We use the latest results presented in \cite{Slatyer2015} to determine how energy injection from annihilations or decays is eventually deposited into the IGM via ionization and heating. We have extended the code to be applicable even when the universe is completely ionized, allowing us to determine how energy is deposited into the IGM at redshifts below $1+z=10$ (the previous lower limit for the code) assuming different reionization scenarios. This improvement allows us to use astrophysical constraints from $z \lesssim 6$ with confidence, and to estimate the sensitivity of our constraints to the details of the (re)ionization history.
\end{enumerate} 

Our paper is structured as follows: in Section \ref{sec:ExptConstraints}, we will review the main existing results that will be used to set constraints on the DM contribution to reionization. Section \ref{sec:EnergyInjection} gives a brief overview of energy injection from $s$-wave annihilation, $p$-wave annihilation and decays, for an unclustered/homogeneous distribution of DM. Our structure formation prescription is detailed in Section \ref{sec:StructureFormation}, while Section \ref{sec:fz} explains how we determine the heating and ionization deposited to the IGM, given an energy injection history and a structure formation model. Section \ref{sec:FreeEleFrac} outlines the three-level atom model for hydrogen used to determine the ionization and IGM temperature history from the energy deposition history. Finally, Section \ref{sec:Constraints} shows our derived constraints for each of the DM processes considered here, with our conclusions following in Section \ref{sec:Conclusion}. 

Throughout this paper, we make use of the central values for the cosmological parameters derived from the TT,TE,EE+lowP likelihood of the Planck 2015 results \cite{PlanckCollaboration2015}. This is obtained from a combination of the measured TT, TE and EE CMB spectra for $l \geq 30$ and a temperature and polarization pixel-based likelihood for $l<30$. Specifically, our choice of parameters are $H_0 = \SI{67.27}{km \s^{-1} Mpc^{-1}}$, $\Omega_m = 0.3156$, $\Omega_b h^2 = 0.02225$ and $\Omega_c h^2 = 0.1198$. These values give a present day atomic number density of $n_A = 0.82 \rho_c \Omega_b/m_p = \SI{2.05E-7}{\centi\meter^{-3}}$.

\section{Constraints from Experimental Results}
\label{sec:ExptConstraints}

To understand how significant a role DM can play in the process of reionization, we must first examine the current experimental constraints on both reionization and DM. 

Extensive astrophysical observations of early quasars and the IGM around them have enhanced our understanding of the process of reionization. By studying quasars at redshift $z \sim 6$ and hydrogen Ly$\alpha$ absorption in their spectra due to the Gunn-Peterson effect, multiple groups have shown that reionization of hydrogen was mostly complete by $z \sim$ 6 \cite{Becker2001,Fan2006,Ota2008}. Observations from even larger redshifts $z\sim 7-8$ indicate that hydrogen reionization occurred relatively quickly, with the neutral hydrogen fraction rising to 0.34 at $z\sim 7$ and exceeding $0.65$ at $z \sim 8$ \cite{Schenker2014}. Neutral helium became reionized at a similar time compared to hydrogen due to their relatively similar ionization energies, but a harder spectrum of ionizing radiation is required to doubly-ionize neutral helium atoms \cite{Loeb2013,Choudhury2006}. Work done on the helium Ly$\alpha$ spectra for quasars at lower redshifts has shown that helium was completely reionized by $z\sim 3$ \cite{Zheng2004}, when quasars could produce the required ultraviolet spectrum. 

Another quantity important to understanding reionization is the IGM temperature, $T_{\text{IGM}}$. Energy deposited into the IGM can both ionize and heat the gas, and the rate of ionization and heating are both highly dependent on $T_{\text{IGM}}$. Measurements of $T_{\text{IGM}}$ place interesting constraints on processes that inject energy into the IGM at redshifts $z \lesssim 6$, since a large injection of energy at these redshifts would result in excessive heating of the IGM. For example, in the case of potential DM contributions, \cite{Diamanti2014} made use of $T_{\text{IGM}}$ measurements to constrain the velocity-averaged cross-section of MeV-TeV DM undergoing $p$-wave annihilation into lepton pairs, as well as the decay lifetimes for MeV-TeV DM decaying into lepton pairs. They found that bounds from $T_{\text{IGM}}$ considerably improved the constraints set by measurements from the CMB and from baryon acoustic oscillations, strengthening the constraints set for the $p$-wave annihilation cross-section by more than an order of magnitude over the full range of DM masses considered. 

Several measurements of $T_{\text{IGM}}$ as a function of redshift have been performed in the last two decades. Earlier studies \cite{Schaye2000} measured the distribution of widths in Ly$\alpha$ absorption spectra from quasars in the redshift range $z = 2.0 - 4.5$ to determine the history of $T_{\text{IGM}}$ in this range, and determined that $\SI{5100}{\kelvin} \leq T_{\text{IGM}}(z=4.3) \leq \SI{20000}{\kelvin}$. More recent studies \cite{Becker2011,Bolton2011} of the IGM temperature from the Lyman-$\alpha$ forest \cite{Becker2011} and from quasars \cite{Bolton2010,Bolton2011} have pushed these measurements back to $z \sim 6$, with the two measurements of $T_{\text{IGM}}$ at the largest redshifts given by (errors reflect 95\% confidence):

\begin{alignat}{1}
	\log_{10} \left( \frac{T_{\text{IGM}}(z=6.08)}{\text{K}} \right) &= 4.21^{+0.06}_{-0.07}, \nonumber \\
	\log_{10} \left( \frac{T_{\text{IGM}}(z=4.8)}{\text{K}} \right) &= 3.9 \pm 0.1.
	\label{eqn:TIGMConstraints}
\end{alignat}

The first measurement, discussed in \cite{Bolton2011}, is almost certainly an overestimate of the true IGM temperature at that redshift: this result does not account for photo-heating of HeII around the quasar being measured, which would result in the measured temperature being significantly higher than the temperature of the IGM away from these quasars. Nonetheless, it serves as a conservative upper bound on $T_{\text{IGM}}$. 

Aside from direct astrophysical measurements, the CMB can also reveal much about reionization. One important aspect of this epoch that can be measured from the CMB is the total optical depth $\tau$ since recombination, given by
\begin{align}
	\tau = -\int_0^{z_\text{CMB}} dz \, n_e(z) \sigma_T \frac{dt}{dz},
\label{eqn:OpticalDepth}
\end{align}
where $n_e$ is the number density of free electrons, $\sigma_T$ is the Thomson scattering cross-section and $z_{\text{CMB}}$ is the redshift of recombination. Scattering of CMB photons off free electrons present after reionization suppresses the small-scale acoustic peaks in the power spectrum by a factor of $e^{-2\tau}$. The Planck collaboration reports the measured optical depth to be \cite{PlanckCollaboration2016}
\begin{align}
	\tau = 0.058 \pm 0.012.
\label{eqn:measuredOpticalDepth}
\end{align}
Planck has also been able to determine a reionization redshift $z_{\text{reion}}$, assuming a step-like reionization transition modeled by a $\tanh$ function and characterized by some width parameter $\delta z = 0.5$ (referred to as the ``redshift-symmetric'' parameterization in \cite{PlanckCollaboration2016}). $z_{\text{reion}}$ is the redshift at which the free electron fraction $x_e \equiv n_e/n_{\text{H}} = 0.54$. Here $n_{\text{H}}$ is the number density of hydrogen (both neutral and ionized) and $n_e$ is the number density of free electrons. $x_e=1.08$ upon complete reionization after taking into account the complete (single) ionization of helium as well. Based on the measured optical depth, the derived $z_{\text{reion}}$ assuming a redshift-symmetric parameterization of the reionization is
\begin{alignat}{1}
  z_{\text{reion}} = 8.8 \pm 0.9.
\end{alignat}
We can factor out the uncertainty associated with reionization after $z = 6$ and its contribution to the optical depth by writing:
\begin{multline}
	\tau = -\int_0^3 dz \left[n_{\text{H}}(z) + 2n_{\text{He}}(z) \right] \sigma_T \frac{dt}{dz} \\
	- \int_3^6 dz\, [n_{\text{H}} (z) + n_{\text{He}}(z)]  \sigma_T \frac{dt}{dz} \\
	- \int_6^{z_{\text{CMB}}} dz\, n_e(z) \sigma_T \frac{dt}{dz},
\end{multline}
where $n_{\text{He}}$ is the redshift-dependent number density of helium (both neutral and ionized). The first two terms are the contribution to the optical depth from reionized hydrogen and helium, while the last term is the contribution from the unknown ionization history of the universe above $z = 6$. The first two terms can be directly evaluated given the baryon number density today, and give a total contribution of $\delta \tau_0 = 0.038$. The remaining measured optical depth must therefore have come from contributions prior to $z=6$, i.e. 
\begin{align}
	\delta \tau = -\int_6^{z_{\text{CMB}}} dz\, n_e(z) \sigma_T \frac{dt}{dz} \leq 0.044,
\label{eqn:ExcessOpticalDepth}
\end{align}
in order for $\tau$ to be within the experimental uncertainty of equation (\ref{eqn:measuredOpticalDepth}) at the 95\% confidence level.

For the case of $s$-wave annihilation, the CMB power spectrum also provides a robust constraint on the velocity-averaged annihilation cross-section $\langle \sigma v \rangle$, since additional ionization of the IGM at high redshifts induces a multipole-dependent modification to the temperature and polarization anisotropies \cite{Padmanabhan:2005es}. The Planck collaboration \cite{PlanckCollaboration2015} has placed an upper bound on $p_{\text{ann}}$, defined as
\begin{alignat}{1}
	p_{\text{ann}} (z) = f_{\text{eff}} \frac{\langle \sigma v \rangle}{m_\chi},
\end{alignat}
where $f_{\text{eff}}$ is a constant proxy for $f(z)$, the efficiency parameter that describes the ratio of total energy deposited to  total energy injected at a particular redshift $z$, and $m_\chi$ is the mass of the DM particle. The CMB power spectra are most sensitive to redshifts $z \sim 600$ (for $s$-wave annihilation), and so the constraint on $\langle \sigma v \rangle$ can be estimated from that redshift \cite{Finkbeiner2012}. Using the TT,TE,EE+lowP Planck likelihood, the 95\% upper limit on this parameter at $z=600$ was found to be:
\begin{alignat}{1}
	p_{\text{ann}}(z = 600) < \SI{4.1E-28}{cm^3 s^{-1} GeV^{-1}}.
	\label{eqn:pann}
\end{alignat}

Given $f_{\text{eff}}$ for $s$-wave annihilation, which in turn is obtained from $f(z)$, this leads immediately to a constraint on $\langle \sigma v \rangle$ as a function of $m_\chi$. $f(z)$ has been calculated for arbitrary injections of electrons, positrons  and photons in the 10 keV-TeV range; in this paper we will thus refer to injections of electron/positron pairs ($e^+e^-$) and photon pairs ($\gamma \gamma$), while keeping in mind that more general DM annihilation/decay channels can be represented as linear combinations of photons/electrons/positrons at different energies.\footnote{See \cite{Slatyer2012,Slatyer2015} and the publicly available results and examples found at \texttt{http://nebel.rc.fas.harvard.edu/epsilon} for further information on how this is done.} This approach neglects the contribution of protons and antiprotons, which is generally quite small \cite{Weniger2013}.

\begin{figure*}
	\subfigure{
		\includegraphics[scale=0.58]{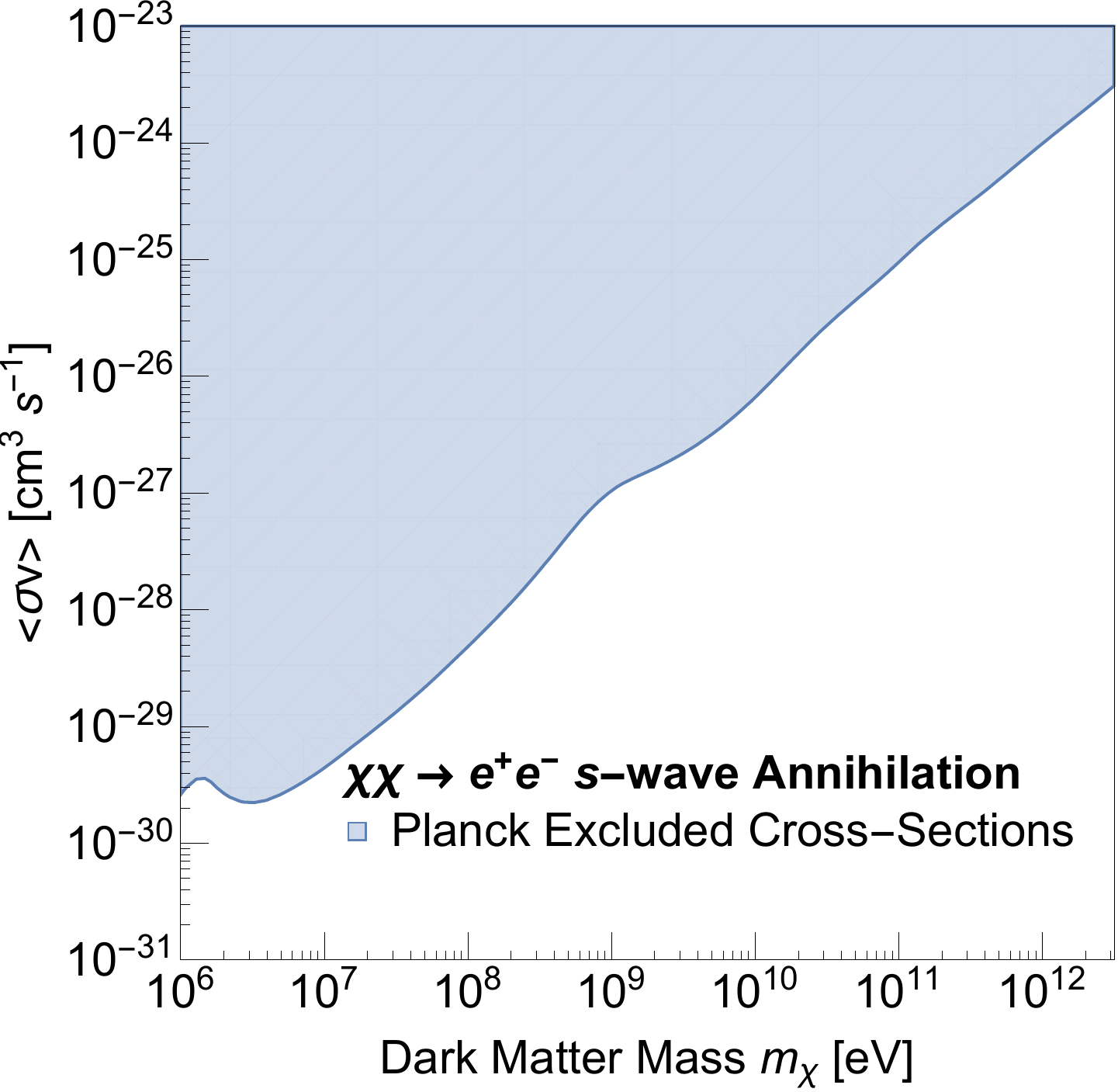}
	}
	\subfigure{
		\includegraphics[scale=0.58]{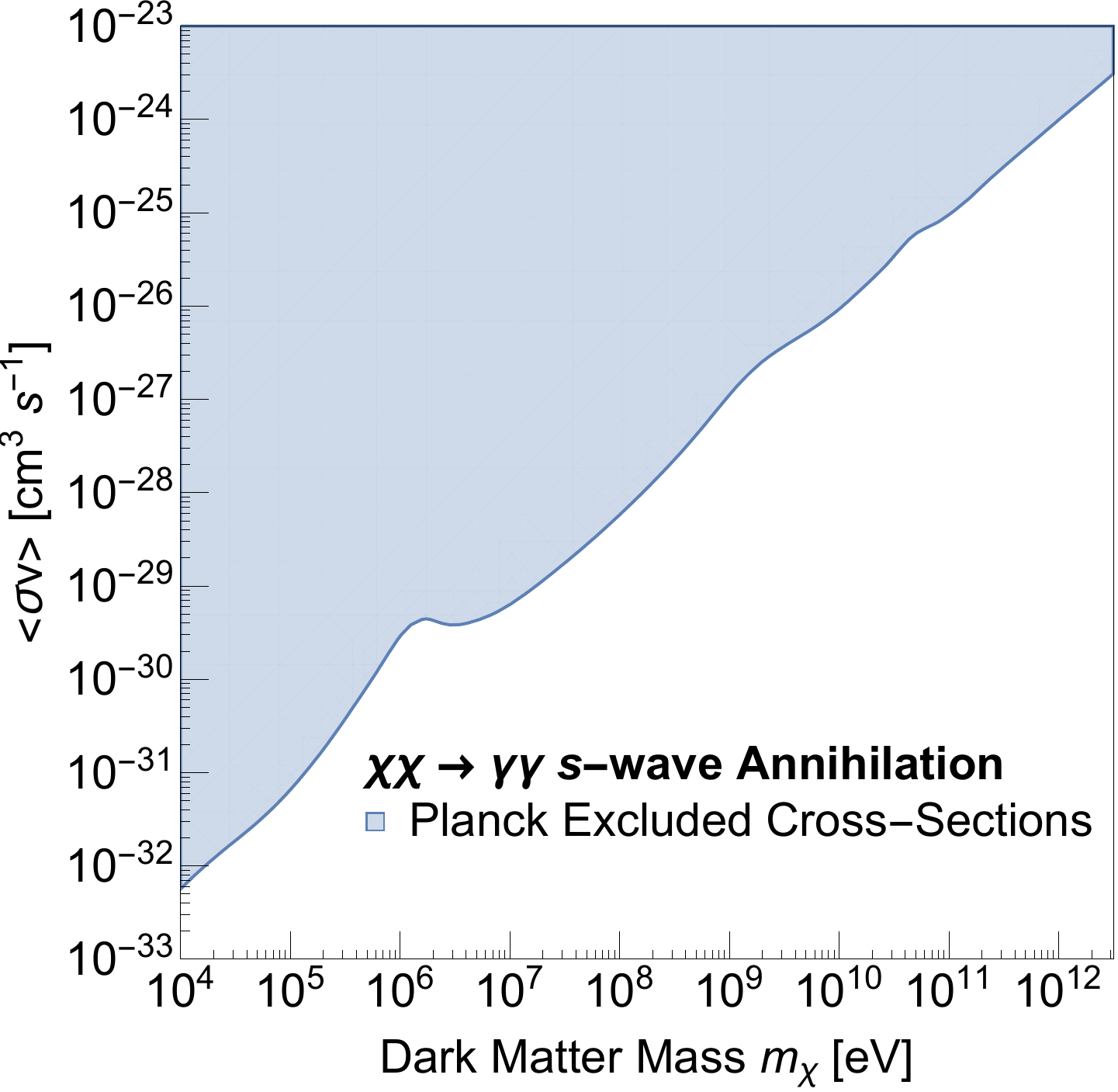}
	}
	\caption{\footnotesize{The $95$\% excluded cross-section based on Planck's upper limit given by equation (\ref{eqn:pann}) for (left) $ \chi \chi \to e^+e^-$ and (right) $\chi \chi \to \gamma \gamma$ $s$-wave annihilation.}}
	\label{fig:excludedXSec}
\end{figure*}

In Section \ref{sec:fz}, we will give a brief summary of our calculation of $f(z)$, which is based on the work detailed in \cite{Slatyer2012,Slatyer2015}. The full details of obtaining an actual value for $f_{\text{eff}}$ from our calculation of $f(z)$ across a large range of DM masses can be found in \cite{Slatyer2015a}. Figure \ref{fig:excludedXSec} shows the constraints on $s$-wave annihilation into $e^+e^-$ (left panel) and $\gamma \gamma$ (right panel), based on the CMB power spectrum data from Planck. 

\section{Unclustered Dark Matter Energy Injection Scenarios}
\label{sec:EnergyInjection}

In this paper, three scenarios by which DM can inject energy into the IGM are considered: $s$-wave annihilation, $p$-wave annihilation and decay. The total energy injected by both $s$- and $p$-wave annihilation of uniformly distributed DM is given by
\begin{alignat}{1}
	\left( \frac{dE}{dV dt} \right)_{\text{ inj}} = \rho^2_{\chi,0} (1+z)^6 \frac{\langle \sigma v \rangle}{m_{\chi}},
	\label{eqn:injRateSmooth}
\end{alignat}
 where $m_{\chi}$ is the DM particle mass and $\rho_{\chi,0} = \rho_{c} \Omega_c$ is the overall smooth density of DM today, with $\rho_c$ being the critical density of the universe today. In $s$-wave annihilation, $\langle \sigma v \rangle$ is constant, while in $p$-wave annihilation, $\sigma v \propto v^2$. This velocity dependence can be factored out by assuming a Maxwellian velocity distribution, which
simplifies the calculation since we can take the 1D velocity dispersion ($\sigma_{1\text{D}}$) as a proxy for the velocity enhancement/suppression in the thermal average:
\begin{alignat}{1}\label{proxy_p}
	\langle \sigma v \rangle_p \propto\int_0^1v^2f_{\rm MB}(v)dv=\sigma_{1D}^2.
\end{alignat}
We can then write, by picking a reference dispersion velocity $\sigma_{1\text{D,ref}}$:
\begin{alignat}{1}
	\langle \sigma v \rangle_{p,B} = \left(\frac{\sigma_{1\text{D,B}}}{\sigma_{1\text{D,ref}}}\right)^2  (\sigma v)_{\text{ref}},
\end{alignat}
where $\sigma_{1\text{D,B}}$ is the one-dimensional characteristic dispersion velocity of unclustered DM. This quantity is redshift dependent, but assuming thermal equilibrium of the DM distribution, $ \sigma_{1\text{D,B}}^2 \propto T$, which for non-relativistic DM scales as $T \propto (1+z)^2$. Thus the energy injection rate for $p$-wave annihilation for uniformly distributed DM can be written as
\begin{alignat}{1}
	\left(\frac{dE}{dV dt} \right)_{p\text{ inj}} = \rho^2_{\chi,0} (1+z)^8 \frac{(\sigma v)_{\text{ref}}}{m_\chi} \left(\frac{\sigma_{1\text{D,B}} (z=0)}{\sigma_{1\text{D,ref}}}\right)^2,
	\label{eqn:smoothpwave}
\end{alignat}
where $\sigma_{1\text{D,B}}(z=0)$ is the present-day value of $\sigma_{1\text{D,B}}$. Throughout this paper, we choose $\sigma_{1\text{D,ref}} = 100 \mathrm{km/s}$ (a value consistent with \cite{Diamanti2014}), which is roughly the present-day DM dispersion velocity in haloes with a mass comparable to the Milky Way ($\lesssim10^{12}$M$_\odot$) today.

Finally, the energy injected from the decay of DM is given by
\begin{alignat}{1}
	\left(\frac{dE}{dV dt} \right)_{d \text{ inj}} = \rho_{\chi,0}(1+z)^3 \frac{1}{\tau_{\chi}},
\end{alignat}
where $\tau_\chi$ is the decay lifetime, which is taken to be much longer than the age of the universe so that the change in DM density due to decay is negligible. This assumption is valid given known limits on the decay lifetime deduced from Planck and WMAP \cite{Diamanti2014} as well as gamma-ray experiments \cite{Dugger:2010ys,Essig2013} for a large range of decay channels.

We have thus far only considered unclustered DM distributions, where the comoving DM density is constant, but structure formation causes the local density and velocity dispersion of DM to deviate strongly from the expected value for a homogeneous distribution. The onset of structure formation thus significantly changes the energy injection history due to $s$- and $p$-wave annihilations. However, the previous notation is still useful: once we have obtained a structure formation history, we can characterize the energy injection from a realistic DM distribution by replacing equations (\ref{eqn:injRateSmooth}) and (\ref{eqn:smoothpwave}) with effective multipliers to the unclustered DM density. A realistic structure formation history is thus crucial in calculating the energy injection rate from DM.

\section{Structure Formation}
\label{sec:StructureFormation}

In the Cold Dark Matter (CDM) scenario, DM clusters into gravitationally self-bound haloes across a very large range of scales, from the (model-dependent) minimum limit set by DM kinetic decoupling ($10^{-11}-10^{-3}$M$_\odot$ for WIMPs \citep[e.g.][]{Bringmann2009}) to $10^{15}$M$_\odot$ cluster-size haloes. $N$-body simulations can accurately follow DM structure formation but only in a limited mass range: it is not yet possible to cover the full dynamical range corresponding to CDM particles. In order to explore the unresolved regime, hybrid approaches which have a core analytical model calibrated against numerical simulations must be used, e.g., the well-known halo model \citep[e.g.][]{Seljak_00}, or the recently introduced $P^2SAD$ (clustering in phase space) \citep{Zavala2015}. We will follow these two
approaches in this paper, describing their most relevant elements.
 
We assume that after recombination, structure formation is described by linear perturbation theory followed by the immediate formation (collapse) of haloes. In this scenario, haloes collapse (form) at a redshift $z_{\rm col}$ with an average overdensity $\bar{\rho}_h=\Delta\rho_{c}(z_{\rm col})$, where $\rho_{c}$ is the critical density of the universe. The choice of the overdensity $\Delta$ varies in the literature, but for simplicity we will use the redshift independent, widely used value of $\Delta=200$. 
The formation redshift is given by the spherical collapse model, which connects the linear power spectrum with the epoch of collapse, resulting in a hierarchical picture of structure formation. In particular, the halo collapses when the 
rms linear overdensity $\sigma(M,z)$ (mass variance) crosses the linear overdensity threshold $\delta_c\sim1.686$:
\begin{equation}\label{sigma_rms}
	\sigma^2(M,z)=\int d^3{\bf k}\,P(k,z)W^2(k,M),
\end{equation}
where $W(k,M)$ is a filter function in Fourier space, and $P(k,z)$ is the linear CDM power spectrum. For the spherical collapse model, the window function is a top-hat filter in real space. 
We compute the primordial matter power spectrum with the code CAMB \citep{2000ApJ...538..473L} with a cosmology consistent with Planck data.

\subsection{Halo Model}

{\bf (i) Flux multiplier.} For the purposes of this work, we are interested in computing the excess DM annihilation over the contribution from the smooth background due to the collapse of DM into haloes. Following the notation of \cite{Taylor2003},\footnote{To avoid conflicting with notation used in later sections, we use the letter $\mathcal{B}$ to refer to the flux multiplier instead of the letter $f$ as in \cite{Taylor2003}.} we write this excess (flux multiplier) for a particular redshift as:
\begin{eqnarray}\label{flux_cosmic}
	\mathcal{B}(z)&=&\frac{1}{\rho_B^2V_B}\int_{m_{\rm min}}^{\infty}\left(V_B\frac{dn}{dM}dM\right)\bar{\rho}^2_hV_h(M)B_{h}(M)\nonumber\\
	&=&\frac{\Delta}{\Omega_m^2\rho_{\rm crit}}\int_{m_{\rm min}}^{\infty}MB_h(M)\frac{dn}{dM}dM,
\end{eqnarray}
where $\left(V_B\frac{dn}{dM}dM\right)$ is the number of haloes in the cosmic volume $V_B$, with a background matter density $\rho_B=\Omega_m\rho_c$. Each halo is assumed to be spherical with a radial density profile $\rho(r)$ truncated at a virial radius $r_{200}$. The annihilation rate in the halo is enhanced over the rate based on the average DM density by an amount
\begin{equation}\label{flux_halo}
	B_h(M)=\frac{4\pi}{\bar{\rho}^2_hV_h(M)}\int_0^{r_{200}}\rho^2(r)r^2\,dr.
\end{equation}

{\bf (ii) Density profile.} In most of the resolved mass regime of current simulations, haloes are well-fitted by a {\it universal} two-parameter NFW density profile \citep{Navarro:1996gj}. An even better fit is that of a three-parameter Einasto profile \citep{Einasto}. The simplicity of the NFW profile and, more importantly, its reduction to an almost one-parameter profile makes it an appealing choice in analytic studies. We will consider these two profiles for this study except at very low halo masses near the filtering mass scale, where recent simulations of the formation of the first haloes (microhaloes) indicate that their inner density profiles might be cuspier than the NFW profile \citep[e.g.][]{Anderhalden2013,2014ApJ...788...27I}. Although these simulations can follow the evolution
of microhaloes only until $z\sim30$ (due to limited resolution, since long wavelength perturbations comparable to the box size cannot be neglected at lower redshifts), we assume that the density profile of these microhaloes can be described by these results all the way down to $z=0$. 

{\it NFW profile and microhaloes.} We use the density profile given by
\begin{equation}\label{rho_smooth}
\rho(x)=\frac{\rho_s}{x^\alpha(1+x)^{3-\alpha}},
\end{equation}
 where $x\equiv r/r_s$, and $r_s$ and $\rho_s$ are the scale radius and density, respectively. Setting $\alpha=1$ gives the NFW profile, which adopt for haloes and subhaloes. For haloes near the filtering mass scale, we follow \cite{2014ApJ...788...27I}, which states that $\alpha$ scales as a power law of the halo mass:
 \begin{equation}\label{alpha_micro}
 	\alpha=-0.123~{\rm log}\left(\frac{M}{10^{-6}M_\odot}\right)+1.461
 \end{equation}
for $M<10^{-3}$$M_\odot$. Above this scale, we set $\alpha=1$. Substituting equation (\ref{rho_smooth}) into equation (\ref{flux_halo}), we have:
\begin{equation}\label{flux_halo_power}
B_h(M)=\frac{c^3}{3m^2(c)}\int_0^c\frac{x^2 dx}{x^{2\alpha}(1+x)^{6-2\alpha}},
\end{equation}
where $c\equiv r_{200}/r_s$ is the concentration parameter, which is a function of halo mass (see below), and:
\begin{equation}\label{flux_halo_power_2}
	m(c)=\int_0^c\frac{x^2 dx}{x^\alpha(1+x)^{3-\alpha}}.
\end{equation}
Equations (\ref{flux_halo_power}) and (\ref{flux_halo_power_2}) both have analytic solutions.

{\it Einasto profile.} The density profile is given by:
\begin{equation}\label{einasto_eq}
\rho(r)=\rho_{-2}\,{\rm exp}\left(\frac{-2}{\alpha_e}\left[\left(\frac{r}{r_{-2}}\right)^{\alpha_e}-1\right]\right),
\end{equation}
where $\rho_{-2}$ and $r_{-2}$ are the density and radius at the point where the logarithmic density slope is -2, and $\alpha_e$ is the Einasto
shape parameter. This three-parameter profile is reduced to only two parameters once the total mass $M\equiv M_{200}$ of a halo is fixed. In particular
we can write:
\begin{multline}
	M_{200} = \frac{4\pi r_{-2}^3\rho_{-2}}{\alpha_e}{\rm exp}\left(\frac{3{\rm ln\alpha_e}+2-{\rm ln} 8}{\alpha_e}\right) \\
	\times \gamma\left[\frac{3}{\alpha_e},\frac{2}{\alpha_e}\left(\frac{r_{200}}{r_{-2}}\right)^{\alpha_e}\right].
\end{multline} 
The parameter $\alpha_e$ and the ``concentration'' $c_e=r_{200}/r_{-2}$ are connected to $M_{200}$ through $\sigma(M,z)$ as we describe below. 
Once these parameters are known, we can compute the boost to the annihilation rate over the average in a halo by solving equation (\ref{flux_halo}) numerically.

The cosmic annihilation flux multiplier given by equation (\ref{flux_cosmic}) due to the population of haloes above a minimum mass $M_{\rm min}$ is fully determined once we specify the halo mass function $dn/dM$ and the properties of the density profiles. In the Extended Press-Schechter (EPS) formalism, both of these are fully determined for a given halo mass. More specifically, they can be written as formulae that depend on $\sigma(M,z)$.

{\bf (iii) Mass function.} The mass function in the case of ellipsoidal collapse is given by \citep{ST1999}:
\begin{align}\label{eq_mf}
	\frac{dn}{d{\rm ln}M}&=\frac{1}{2}f(\nu)\frac{\rho_B}{M}\frac{d{\rm ln} (\nu)}{d{\rm ln}M},\\
	f(\nu)&=A\sqrt{\frac{2q\nu}{\pi}}\left[1+\left(q\nu\right)^{-p}\right]{\rm exp}^{-q\nu^2},
\end{align}
 with $A=0.3222$, $p=0.3$, and $q=1$, and:
 \begin{equation}
 	\nu\equiv\frac{\delta_c(z)^2}{\sigma(M,z)^2},
 \end{equation}
 where $\delta_c(z)=1.686/D(z)$ is the linearly extrapolated threshold for spherical collapse, with $D(z)$ being the growth factor normalized to unity at $z=0$.
 
Free-streaming of DM particles prevents the formation of haloes below a (filtering) scale, which depends on the mass of the DM particle. This results in a cutoff to the primordial power spectrum at the filtering scale. The difference between a CDM power spectrum with a filtering scale and without (i.e. setting the mass of the DM particles effectively to zero) is typically given in terms of the transfer 
function $T^2_\chi=P_{\rm m_\chi}/P_{\rm m_\chi\rightarrow0}$, which for neutralino DM has the form \citep{Green2005}:
\begin{equation}\label{transfer_func}
	T_\chi(k)=\left[1-\frac{2}{3}\left(\frac{k}{k_A}\right)^2\right]{\rm exp}\left[-\left(\frac{k}{k_A}\right)^2-\left(\frac{k}{k_B}\right)^2\right],
\end{equation}
where
\begin{multline}\label{transfer_func_2}
	k_A= 2.4\times10^6\left(\frac{m_\chi}{100~{\rm GeV}}\right)^{1/2}\nonumber\\
		\times\frac{(T_{\rm kd}/30~{\rm MeV})^{1/2}}{1+{\rm ln}(T_{\rm kd}/30~{\rm MeV})/19.2}~{\rm Mpc}/h,
\end{multline}
\vspace{-0.6cm}
\begin{alignat}{1}
	k_B&=5.4\times10^7\left(\frac{m_\chi}{100~{\rm GeV}}\right)^{1/2}\left(\frac{T_{\rm kd}}{30~{\rm MeV}}\right)^{1/2}~{\rm Mpc}/h, 
\end{alignat}
and $T_{\rm kd}$ is the (model-dependent) kinetic decoupling temperature.

To include the effect of free-streaming into the mass function, we use the code provided by \citep{2013MNRAS.433.1573S}, which computes the mass function following equation (\ref{eq_mf}) using a {\it sharp-k} window function for the mass variance calibrated to match the results of simulations that include a cutoff in the power spectrum as given by the transfer function in equation (\ref{transfer_func}). We note that $T_{\rm kd}$ and $m_\chi$ together determine the minimum self-bound halo mass $M_{\rm min}$. Choosing a different $M_{\min}$ changes the global contribution of (sub)haloes by some overall factor in a redshift-independent manner.
We take $m_\chi=100$~GeV and $T_{\rm kd}=28$~MeV to compute the cutoff to the primordial power spectrum given by equations~(\ref{transfer_func}-\ref{transfer_func_2}).\footnote{ For neutralino dark matter, the kinetic decoupling temperature generally increases with particle mass, although a broad range of values for a fixed mass is allowed. Based on Fig. 2 of \cite{Bringmann2009} we have chosen a typical value within that range for $m_\chi=100$~GeV.} This results in a damping scale due to free streaming with a characteristic mass of $M_{\rm min}=10^{-6}$M$_\odot$ \citep[see equation (13) and Fig. 3 in Ref.][]{Bringmann2009}, which is the canonical value for WIMPs. The impact of choosing different values of $M_{\rm min}$ will be studied later in this section.

{\bf (iv) Parameters of the density profiles.} The median density profile of haloes with a given mass is fully specified by one parameter, typically the halo mass. Since CDM haloes form hierarchically, low mass haloes are more concentrated than more massive ones. This specifies the second parameter (concentration) of the profile. Ultimately, this parameter is connected to the density of the Universe at the (mass-dependent) time of collapse for a given halo. 

{\it NFW profile and microhaloes}. The concentration of an NFW halo is a  strong function of halo mass that has been explored in great detail in the literature using analytical and numerical methods. We use the model by \cite{2012MNRAS.423.3018P} to compute the concentration-mass 
relation. The model is calibrated to recent simulations down to their resolution limit ($M\sim10^{10}$~M$_\odot$), but more importantly, it is physically motivated since it uses $\sigma(M,z)$ as the main quantity connected to the concentration. In this way, it takes into account the flattening of the linear power spectrum towards smaller halo masses. We refer the reader to Section 5 of 
\cite{2012MNRAS.423.3018P} for the formulae that lead
to the computation of $c(M,z)$. We only consider haloes with a ``peak-height'' $\nu\equiv\delta_c/\sigma$ up to $3\sigma$. The larger $\nu$ is, the rarer and the more massive the halo is relative to the characteristic clustering mass defined by $\nu=1$. 
 
 For microhaloes, we make a correction to the NFW concentrations given by the Ref.~\cite{2012MNRAS.423.3018P} model to take into account the steeper profiles of microhaloes. To do so, we follow
 the results from \cite{2014ApJ...788...27I} (see their Figure 9). In particular, for $\alpha=1.5,1.4,1.3,1.0$ in equation~(\ref{alpha_micro}), they find $c_{\rm NFW}=2.0c_{\rm micro},1.67c_{\rm micro}, 1.43c_{\rm micro}, 1.0c_{\rm micro}$; we use these values to interpolate for a given microhalo mass.
 
 {\it Einasto profile.} In this case we follow the work by \cite{Klypin2014} to connect the parameters $\alpha_e$ and $c_e$ (concentration) with $\sigma(M)$. These authors use a similar
 analysis as that of \cite{2012MNRAS.423.3018P}, and find the following empirical relations:
 \begin{eqnarray}
 	\alpha_e&=&0.015+0.0165\nu^2,\nonumber \\
	r_{200}/r_{-2}&=&6.5\nu^{-1.6}(1+0.21\nu^2).
 \end{eqnarray}
 Note that $\alpha_e$ approaches a constant value asymptotically for low $\nu$ (i.e. low halo masses), which implies that low mass haloes of a given mass only differ in one parameter, their concentration (as in the NFW case).
 
 {\bf (v) Substructure.} Each DM halo is composed of a smooth DM distribution and a hierarchy of subclumps that merged into the main halo at some point in the past and have been subjected to 
 tidal disruption. The modeling of the abundance of main haloes and their inner smooth structure have been described previously, and we now consider the impact of substructure on the annihilation rate.
 
 To account for the self-annihilation of DM in substructures, we define a {\it subhalo boost} over the flux multiplier of a main halo (i.e. over $B_h(M)$ in equation~(\ref{flux_halo})):
 \begin{multline}\label{sub_boost}
 \mathcal{B}(m_{\rm sub})=\frac{1}{B_h(M)} \int_{m_{\rm min}}^{m_{\rm max}}\frac{\bar{\rho}_{\rm sub}(m_{\rm sub})}{\bar{\rho}_h} \\
 \times B_{\rm sub}(m_{\rm sub})m_{\rm sub} \frac{dN}{dm_{\rm sub}}dm_{\rm sub},
 \end{multline}
 where $dN/dm_{\rm sub}$ is the subhalo mass function and $\bar{\rho}_{\rm sub}$ and $B_{\rm sub}$ are the average density within a subhalo and its flux multiplier of mass $m_{\rm sub}$, respectively. Because of tidal disruption, these quantities depend in principle on the distance 
 of the subhalo relative to the halo center, but since we are interested in the total subhalo boost to the annihilation rate, we can assume that most of the boost comes from subhaloes near the virial radius of the host. This is a good approximation since tidal disruption considerably reduces the abundance of subhaloes near the halo center. For instance, looking at Figure 3 of 
Ref.~\cite{Springel:2008cc}, we
 see that only $\sim30\%$ of the annihilation rate in subhaloes comes from within 100 kpc ($\sim0.4r_{200}$) of a Milky Way-sized halo. On the other hand, near the virial radius of a host with an assumed NFW profile, the tidal radius for a
 subhalo of mass $m_{\rm sub}$ is approximately given by \citep[e.g. equation (12) of][]{Springel:2008cc}
 \begin{eqnarray}
 	r_t &=& \left(\frac{m_{\rm sub}}{\left[2-\frac{d{\rm ln}M}{d{\rm ln} r}\right]M(<r)}\right)^{1/3}r \qquad \qquad \qquad \qquad  \nonumber
 \end{eqnarray}
 \vspace{-0.6cm}
 \begin{multline}
 	\quad \sim \left(\frac{m_{\rm sub}}{M}\right)^{1/3} r_{200} \\ \qquad \times \left(2-\frac{c^2}{(1+c)^2}\frac{1}{{\rm ln}(1+c)-c/(1+c)}\right)^{-1/3},
 \end{multline}
where $c\equiv c(M,z)$ is the concentration of the host. We can then substitute $\frac{\bar{\rho}_{\rm sub}}{\bar{\rho}_h}$ for the following in equation~(\ref{sub_boost}):
\begin{eqnarray}
	\left.\frac{\bar{\rho}_{\rm sub}(<r_t)}{\bar{\rho}_h}\right\vert_{r_{200}}=
	2-\frac{c^2}{(1+c)^2}\frac{1}{{\rm ln}(1+c)-c/(1+c)}.\nonumber\\
\end{eqnarray}
This density ratio has only small variations around 2 with low mass haloes being more overdense on average than more massive subhaloes. 

The {\it subhalo mass function} is in principle also a function of halocentric distance, but it becomes the global subhalo mass function under the approximation that subhaloes near the virial radius dominate the annihilation rate. The subhalo mass function
has a similar functional form as the halo mass function. In particular, it is approximately a power law (except at very large masses) with a similar slope to the halo mass function, 
$dN/dm_{\rm sub}\propto m_{\rm sub}^{-1.9}$ \citep{Springel:2008cc}; the normalization however is different. This functional form is nearly universal if $m_{\rm sub}$ is scaled to the host mass.\footnote{This universality is even clearer if the ratio of maximum circular velocities is used instead of the masses to define the subhalo mass function \citep[e.g.][]{Cautun2014}.} We use the fitting formulae for the subhalo mass function given by \cite{Gao2011}, which is based on a suite of high resolution simulations covering a large dynamical range of masses and is valid for $z\leq2$; for higher redshift we assume that the formulae at $z=2$ holds (our results are actually not very sensitive to this assumption). We assume also that these formulae are preserved in the unresolved regime, down to the filtering mass scale, and apply the same cutoff at low masses due to free streaming (or kinetic decoupling) as that for the halo mass function.
 
 To calculate the subhalo flux multiplier $B_{\rm sub}$, we assume the same density profiles as in the case of main haloes, i.e. we use equations (\ref{flux_halo_power}) and (\ref{flux_halo_power_2}) in the case of the NFW profile and the microhaloes, and find the result numerically in the case of the Einasto profile. 
 This is a good approximation since, as we mentioned before, the subhaloes that contribute most to the signal are those near the virial radius of the host. Thus, tidal disruption would not have transformed their
 inner structure significantly, particularly their inner regions, which strongly dominate the annihilation rate. However, in the case of the NFW profile, we do account for a slight modification to the concentration-mass relation in the form of an upscaling of a factor of 2.6 to the characteristic density $\rho_s$ (which is roughly a $30\%$ increase in concentration, see Figure 28 of Ref.~\cite{Springel:2008cc}). This modification is because for a given mass, subhaloes (even near the virial radius) are slightly more concentrated than isolated haloes. For the case of the Einasto profile, we do not
 make this correction since there is no systematic study about this. We note however that this correction to the overall flux multiplier $\mathcal{B}(z)$ is relatively small. 
 
 \subsection{The Particle Average Phase Space Density ($P^2SAD$) Approach}

Instead of modeling the clustering of DM indirectly as a collection of haloes (and subhaloes) with a certain internal DM distribution, one can model it directly by looking at the DM two point correlation function $\xi(\Delta x)$ (or its Fourier transform, the power spectrum). It has been shown that the flux multiplier, defined in equation (\ref{flux_cosmic}), is equal to the limit of $\xi$ when the separation between particles $\Delta x$ goes to zero \cite{Serpico2012}:
\begin{equation} \label{eq_p2sad}
	\mathcal{B}={\rm lim}_{\Delta x\rightarrow 0} \xi(\Delta x).
\end{equation}
Thus, if one can directly obtain a prediction of the DM power spectrum in the deeply non-linear regime, then it is possible to directly compute the flux multiplier without the many steps and approximations involved in the halo model.

This approach has been developed recently by analyzing the coarse-grained phase space distribution directly from DM simulations. In particular, by measuring the two dimensional particle phase space average density ($P^2SAD\equiv\Xi(\Delta x, \Delta v)$, where $\Delta x$ and $\Delta v$ are the distance and relative speed between particles) in high resolution simulations, it has been possible to physically model this new statistic of DM clustering and predict the right hand side of equation~(\ref{eq_p2sad}) \cite{Zavala2014a,Zavala2014b,Zavala2015}. In particular one can write:
\begin{equation}\label{real_2pcf_std}
	\xi(\Delta x)_{{\cal V}_6} = \frac{\langle\rho\rangle_{{\cal V}_6}}{\rho_B^2}\int d^3{\bf \Delta v}~\Xi(\Delta x, \Delta v)_{{\cal V}_6} - 1,
\end{equation}
where $\langle\rho\rangle_{{\cal V}_6}$ is the average DM density within the phase space volume (${\cal V}_6$) over which $P^2SAD$ is averaged. In a cosmic volume $V_B$ we can write:
\begin{equation}\label{normalization}
	\frac{\langle\rho\rangle_{{\cal V}_6}}{\rho_B^2}=\frac{1}{\rho_B}\frac{M_{V_B}}{\rho_B V_B}=\frac{\mathcal{F}_{\rm subs}(V_B)}{\rho_B},
\end{equation}
where $\mathcal{F}_{\rm subs}(V_B)$ is the mass fraction contained in substructures within the cosmic volume $V_B$ that is calculated using the subhalo and halo mass functions, described above
in the halo model section:
\begin{equation}\label{norm_p2sad}
	\mathcal{F}_{\rm subs}(V_B)=\frac{1}{\rho_B}\int_{M_{\min}}^{\infty}M\frac{dn}{dM}\mathcal{F}_{\rm s,h}(M)dM,
\end{equation} 
where $\mathcal{F}_{\rm s,h}(M)$ is the mass fraction within subhaloes in a halo of mass $M$ (computed from the subhalo mass function). 

$P^2SAD$ can be described with a physically motivated model that combines the stable clustering hypothesis in phase space, the spherical collapse model and tidal 
disruption of subhaloes \cite{Zavala2014b,Zavala2015}. This model has 7 free parameters, which have been calibrated in \cite{Zavala2015}  for DM particles inside
subhaloes exclusively. Since the clustering of DM at very small scales is dominated precisely by these particles, we can use this model to predict the global flux
multiplier in a cosmic volume. We note that although $P^2SAD$ has remarkably universal structural properties (this is the reason why it is a powerful 
statistic to predict the nonlinear power spectrum at unresolved scales), the parameters of its modeling have only been calibrated at relatively low redshifts. We therefore
warn that above $z=1$, its predictions remain uncertain at this point. Since we are particularly interested in DM annihilation at higher redshift in this paper, we assume
that the parameters of the physical model of $P^2SAD$ calibrated at $z=0$ remain unchanged. 

Overall, because of its direct connection with the annihilation signal, there is significantly less uncertainty associated with $P^2SAD$ compared to the more traditional halo models used to calculate the boost factor described earlier. With proper calibration at higher redshifts, $P^2SAD$ could have been used as the main method in this paper, but owing to the current limitations, we use it only as a sanity check on the results obtained from the halo model approach, and as a brief introduction to a powerful new method of obtaining boost factors that may become useful in future work. 

\subsection{The Effective Density for Dark Matter Annihilation due to Structure Formation}

\begin{figure}
\center{
\includegraphics[height=8.0cm,width=8.0cm]{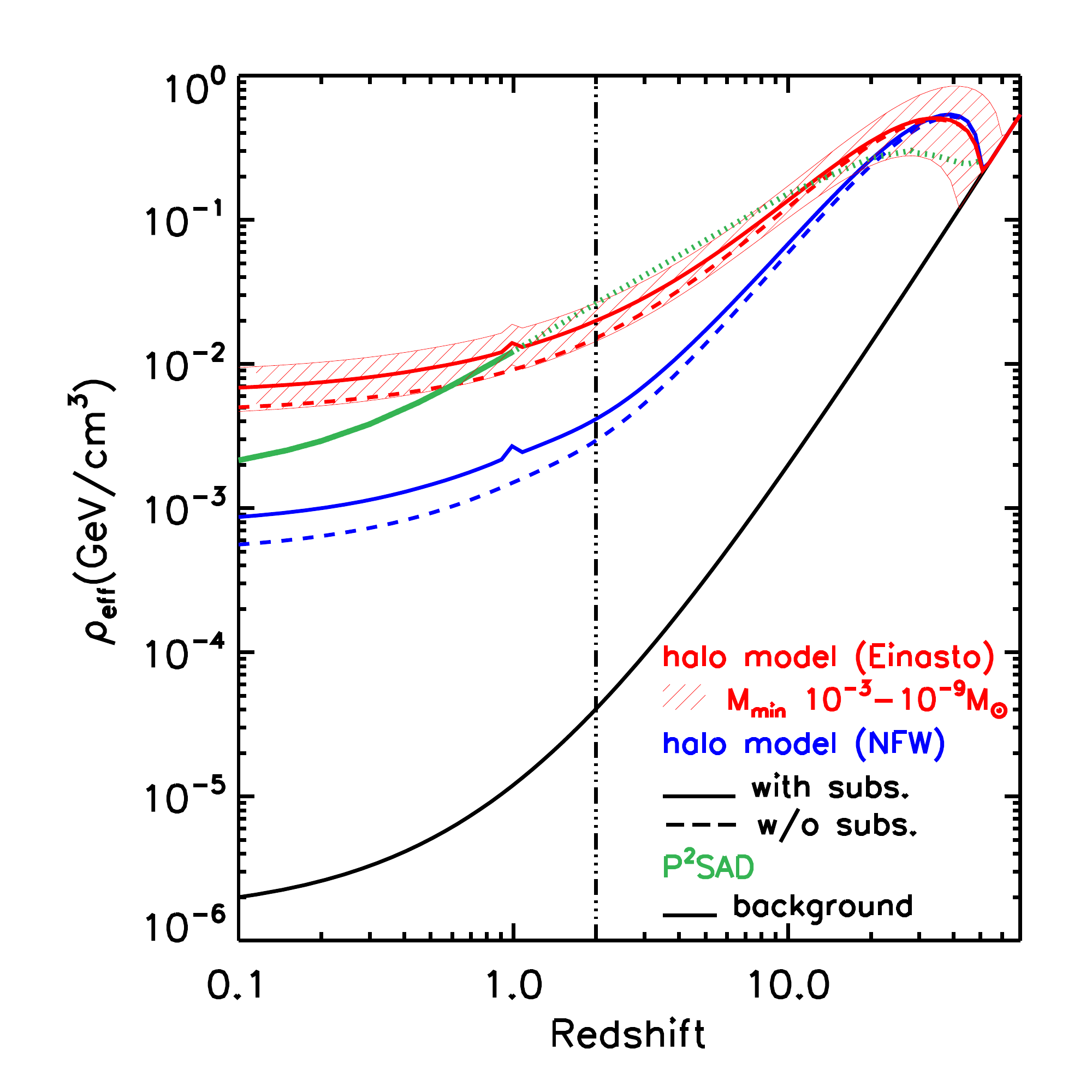} 
}
\caption{The effective DM density as a function of redshift (relevant for $s$-wave annihilation). The blue and red lines show the predictions from the {\it halo model} of structure formation with (solid) and
without (dashed) substructures. The blue (red) line uses an NFW (Einasto) profile for the haloes with parameters given by Ref.~\cite{2012MNRAS.423.3018P} (\cite{Klypin2014}). The green line shows the prediction by a new approach based on the clustering of phase space ($P^2SAD$, Ref.~\cite{Zavala2015}). This approach has only been calibrated at
low redshifts, and thus is uncertain for $z>1$ (green dotted line). The vertical dot-dashed line marks the maximum redshift where the subhalo mass function we have used has been calibrated.  In the case of the halo model with the Einasto profile, we also show with a hatched area the impact of varying $M_{\rm min}$ by 6 orders of magnitude, from $10^{-3}$M$_\odot$ (lower contour) to $10^{-9}$M$_\odot$ (upper contour). For all the other cases, we have used $M_{\rm min}=10^{-6}$M$_\odot$. The solid black line shows the average smooth background density.}
\label{fig_rho_eff} 
\end{figure}

Having described our modeling of the flux multiplier, we can finally write the effective DM density $\rho_{\text{eff}}$ as a boost over the background due to structure formation, which we will then use to compute
the DM annihilation rate as a function of redshift:
\begin{equation}
	\rho_{\rm eff}(z)=\rho_B(z)\left(1+\mathcal{B}_s(z)\right)^{1/2},
	\label{eqn:rhoeff}
\end{equation}
where $\rho_B(z)= \rho_{\chi,0}(1+z)^3$ and $\mathcal{B}_s=\mathcal{B}$ (defined in equation~(\ref{flux_cosmic})).

The predictions for $\rho_{\text{eff}}$ for the two structure formation models are shown in Figure \ref{fig_rho_eff}. The predictions of the  {\it halo model} are in blue (``conservative'', or low-boost) and red (``stringent'', or high-boost), corresponding to the cases where (sub)haloes are modeled with an NFW profile with a concentration mass relation as given by the model in \cite{2012MNRAS.423.3018P}  and with an Einasto profile with parameters given in \cite{Klypin2014} respectively. In the plot we show these cases with (solid) and without (dashed) substructure. Beyond $z=2$ (vertical dot-dashed line), the 
parameters of the fitting formulae for the subhalo mass function have not been calibrated and the predictions are thus more uncertain, but at higher redshifts the impact of substructure on the global annihilation rate is minimal. The large difference between the red and blue curves is actually not caused directly by the use of different density profiles (Einasto vs NFW), but by the relatively different concentrations of low mass haloes predicted by the formulae in Refs.~\cite{2012MNRAS.423.3018P} and \cite{Klypin2014}. We have also explored variations over the minimum self-bound halo mass, varying $M_{\rm min}$ by 6 orders of magnitude. The impact of this on $\rho_{\text{eff}}$ is shown by the hatched area for the Einasto halo model with substructures (the other cases show a similar variation). Although $M_{\rm min}$ plays a role in setting the value of $\rho_{\rm eff}$, varying $M_{\rm min}$ between $10^{-9}$ to $10^{-3} M_\odot$ changed $\rho_{\rm eff}$ by only a factor of approximately 2.15, with the effect being larger at larger redshifts, since a larger value of $M_{\text{min}}$ leads to a delay in the onset of structure formation. This effect is relatively minor compared to the uncertainties in the halo model, at least at $z<10$. We have also found that for both $s$-wave and $p$-wave annihilation, the level of variation in $M_{\rm min}$ explored here produced only percent-level variations in the ionization and thermal histories, and consequently none of our subsequent results are sensitive to our choice of $M_{\min}$. We therefore adopt the canonical value of $M_{\rm min}=10^{-6}$M$_\odot$ for the rest of this paper.

The approach based on the DM clustering in phase space, $P^2SAD$, is shown with a solid green line, and with a dotted green line beyond the reach where it has been calibrated. It predicts a behavior for $\rho_{\rm eff}$
that lies in between the {\it halo model} predictions. It does seem to favor a larger annihilation rate (i.e. ultimately larger halo concentrations) than the model with the smallest structure formation boost (blue), given that it lies closer to the model with the largest structure formation boost (red). This approach is however only certain close to $z=0$, where the green line is lower than the red one by a significant amount. We will take the difference between the red and the blue line as our degree of uncertainty in the predictions of the structure formation prescriptions.

Equation~(\ref{eqn:rhoeff}) is the quantity of relevance for the case of $s$-wave annihilation, where the astrophysical part of the signal scales as $\rho_{\rm eff}^2$. In the case of $p$-wave annihilation, given the velocity dependence of the
astrophysical signal, we can write instead
\begin{equation}
	(\rho v/c)_{\rm eff}(z)=\rho_B(z)(\sigma_{\rm 1D, B}(z)/c)\left(1+\mathcal{B}_p(z)\right)^{1/2},
	\label{eqn:rhoeff_p}
\end{equation}
where we assume that the velocity distribution of the DM particles is Maxwellian, as in equation (\ref{proxy_p}). In particular, $\sigma_{\rm 1D, B}(z)=\sigma_{\rm 1D, B}(z=0)(1+z)=10^{-11}c({\rm GeV}/m_\chi)^{1/2}(1+z)$ is the velocity dispersion of unclustered DM, and $\mathcal{B}_p$ is given by multiplying the halo and subhalo flux multipliers by $(\sigma_{1D, h}/c)^2$. We have approximated the average 1D velocity dispersion of the (sub)halo by $\sigma_{1D, h}\sim V_{\rm max,h}/\sqrt{3}$, with $V_{\rm max, h}$ being the maximum circular velocity of the (sub)halo computed from its density profile.

Notice that while we have characterized the structure formation contribution as a boost factor multiplying the smooth background contribution, in reality this is an additive contribution: $(\rho v/c)_{\text{eff}}$ within the haloes does not depend on $\sigma_{\text{1D,B}}(z)$, since once structure formation sets in, the characteristic velocity of dark matter particles is set by gravity and not by the primordial thermal motion of unclustered dark matter. Thus the exact value of $\sigma_{\text{1D,B}}(z)$ is important only before the onset of structure formation at $z \gtrsim 50$. Throughout this paper, we have used the value of $\sigma_{\text{1D,B}}(z = 0)$ computed with $m_\chi = \SI{100}{GeV}$ and $T_\mathrm{kd} = 28$ MeV. This choice results in a highly suppressed annihilation rate prior to structure formation, and results in ionization histories that are indistinguishable from an ionization history with no dark matter at redshifts $z \gtrsim 50$. We have also investigated the effects of adopting larger values of $\sigma_{\text{1D,B}}(z=0)$ corresponding to smaller $m_\chi$ or $T_\mathrm{kd}$, but have found that our present choice is optimistic for producing significant ionization just prior to reionization in a manner that is consistent with the optical depth constraints. Further discussion of this matter can be found in Section \ref{sec:Constraints}.

\begin{figure}
\center{
	\includegraphics[height=8.0cm,width=8.0cm]{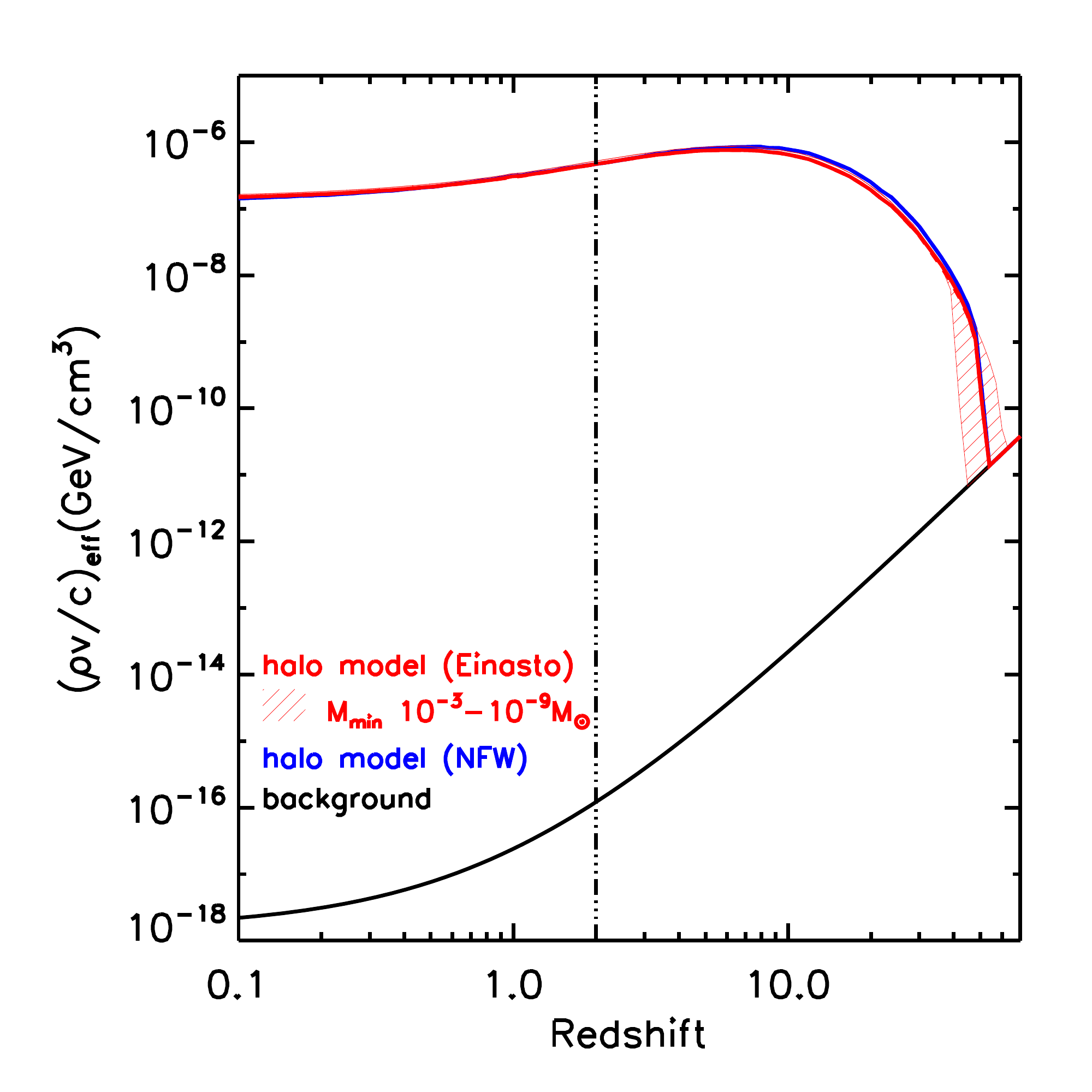} 
}
\caption{The effective DM density $\times$ velocity as a function of redshift (equivalent to Fig.~\ref{fig_rho_eff} but for the case of $p$-wave annihilation given by equation~(\ref{eqn:rhoeff_p})). All the line styles and colors are as in Fig.~\ref{fig_rho_eff}.  In the case of the halo model with the Einasto profile, we also show with a hatched area the impact of varying $M_{\rm min}$ by 6 orders of magnitude, from $10^{-3}$M$_\odot$ (lower contour) to $10^{-9}$M$_\odot$ (upper contour). For all the other cases, we have used $M_{\rm min}=10^{-6}$M$_\odot$. The background is normalized to the thermal velocity dispersion of DM particles with $m_\chi=100$~GeV.}
\label{fig_rho_eff_pwave} 
\end{figure}

We show the effective DM density $\times$ velocity in Figure \ref{fig_rho_eff_pwave}, defined in equation (\ref{eqn:rhoeff_p}). The uncertainties in the structure formation scenario in this case are minimal since annihilation in massive, resolved haloes dominates the overall flux. The uncertain contribution for haloes below the resolution limit of current simulations is minimal. This is why the predictions from the halo model for the two cases we have considered nearly overlap each other, and is the reason why there is a negligible impact of substructures (the lines showing the effect overlap completely with those without substructures in Figure \ref{fig_rho_eff_pwave}). A different value of $M_{\rm min}$ is only important at the redshifts closest to the onset of structure formation. Still, within the 6 orders of magnitude of variation of $M_{\rm min}$, we have found no important changes in our main results.

\section{Effective Deposition Efficiency}
\label{sec:fz}

\subsection{$f_c(z)$ for Smooth Dark Matter Distributions}

Energy injected by DM annihilation or decay at any given redshift is not immediately deposited into the IGM. At certain redshifts and input energies, the characteristic time for a photon to completely deposit its energy can be comparable to or greater than the Hubble time, making the `on-the-spot' approximation for the deposition of energy problematic \cite{Slatyer2009}. Moreover, the efficiency at which injected energy is deposited into various channels (e.g. ionization of the IGM vs. heating of the IGM) is generically a complicated function of redshift, the energy of the injected particles, and the background level of ionization.

The details of the deposition process can be distilled into a single quantity $f_c(z)$, the ratio between energy deposited in channel $c$ and the injected energy at a given redshift $z$, i.e. 
\begin{alignat}{1}
	\left(\frac{dE}{dtdV} \right)_{c,\text{dep}} = f_c(z) \left(\frac{dE}{dtdV} \right)_{\text{inj}}
	\label{eqn:fcz}
\end{alignat}
where the channels considered are ionization of H (H ion), ionization of He (He ion), Lyman-$\alpha$ excitation of H atoms (Ly$\alpha$), heating of the IGM (heat), and energy converted into continuum photons that we observe as distortions to the CMB energy spectrum (cont). 

To calculate $f_c(z)$, we first need to calculate $T_c(z_{\text{inj}},z_{\text{dep}},E) \, d \log(1+z_{\text{dep}})$, the fraction of energy injected at redshift $z_{\text{inj}}$ that is deposited at redshift $z_{\text{dep}}$ into channel $c$ due to an injection of particles with individual energy $E$, discretized into redshift bins of size $d \log(1+z_{\text{dep}})$. This is done using the code developed in \cite{Slatyer2012,Slatyer2015}, and only a brief summary of the code is given here. Starting with some injection of an $e^+ e^-$ or $\gamma \gamma$ pair at $z_{\text{inj}}$, the code tracks the cooling of particles and all of the secondary particles produced in these cooling processes in steps of $d \log(1 + z_{\text{dep}}) = 10^{-3}$. Photons that can efficiently photoionize HI, HeI and HeII in the IGM are removed from the main code and are considered to be ``deposited'', together with all electrons (including secondary electrons from photoionization) below \SI{3}{keV}. The proportion of energy deposited into each channel $c$ from the deposited photons and electrons is then determined by a separate low-energy code, which is described in full detail in \cite{Slatyer2015}. The code assumes only small modifications to the ionization history of the universe from DM, since large modifications are ruled out by observational constraints. With this assumption, any arbitrary injection history with an arbitrary energy spectrum of particles can then be treated as a linear combination of individual injections of fixed energy at particular redshifts.

In the original code, $T_c (z_{\text{inj}},z_{\text{dep}},E) \, d \log(1+z_{\text{dep}})$ was computed from $1+z = 3000$ to $1+z=10$ for both injection and deposition redshift, over a large range of particle kinetic energies ($E \sim 10$ keV to $\SI{}{TeV}$).  Below $1+z_{\text{dep}} = 10$, the ionization history becomes much less certain due to the process of reionization. The exact details of the ionization history can have a significant impact on our calculation of $f_c(z)$: $f_{\text{H ion}}$, for example, should decrease significantly when $x_e \equiv n_e/n_{\text{H}}$ is close to 1. However, in order to make use of constraints on $T_{\text{IGM}}$ and $\delta \tau$, the code has to be extended down to lower redshifts. Given this uncertainty, we defer a discussion of how these results are extended down to $1+z_{\text{dep}} = 4$ to the following sub-section.

At the end of the calculation, we would have determined the fraction of energy injected at $z_{\text{inj}}$ that is deposited at some deposition redshift $z_{\text{dep}}$, broken down by deposition channel. Determining the total deposited energy at some redshift $z_{\text{dep}}$ therefore requires knowledge of the full injection history. To relate the deposited energy to the current injected energy and obtain $f_c(z)$ as defined in equation (\ref{eqn:fcz}), we have to integrate $T_c(z_{\text{inj}},z_{\text{dep}},E) d\log(1+z_{\text{dep}})$ over all injection redshifts prior to $z_{\text{dep}}$. For any arbitrary DM energy injection process, the spectrum of particles injected has a typical redshift dependence $dN/(dE\, dV\, dt) \propto (1+z)^\alpha$, where $\alpha = 6$ for $s$-wave annihilation, $\alpha = 8$ for $p$-wave annihilation and $\alpha = 3$ for decay. In each case, we can factor the spectrum into a redshift-dependent factor multiplied by an energy spectrum $d\bar{N}/dE$ that is independent of redshift. Doing this, one can show \cite{Slatyer2012} that
\begin{multline}
  f_c(z) = \frac{H(z)}{(1+z)^{\alpha-3} \sum\limits_{\text{species}}\int E \frac{d\bar{N}}{dE} dE} \,  \\
  \times \sum_{\text{species}} \int \frac{(1+z')^{\alpha-4}}{H(z')} dz' \int T_c(z',z,E) E \frac{d\bar{N}}{dE} dE,
\end{multline}
where the sum over species indicates that we are combining effects from all species produced in the annihilation process. For this paper, we only consider the case where DM annihilates or decays into $e^+e^-$ or $\gamma \gamma$, with each particle having fixed, identical total energy $E=m_\chi$ for annihilations or $E = m_\chi/2$ for decays. In this case, $f_c(z)$ further simplifies to
\begin{alignat}{1}
	f_{c}(z,E) = \frac{H(z)}{(1+z)^{\alpha-3}} \int \frac{(1+z')^{\alpha-4}}{H(z')} T_{c}(z',z,E)\, dz'
\end{alignat}
for each of the injection species being considered. The quantity $f_c(z,E)$ for the injection species $e^+e^-$ and $\gamma\gamma$ will be denoted by a subscript $e$ and $\gamma$, respectively. While the spectrum of particles associated with any DM injection process may be significantly more complicated, ultimately any such process deposits energy into the IGM via $e^+e^-$ pairs or photon pairs. Understanding the energy deposition efficiency through $e^+e^-$ or $\gamma\gamma$ is thus sufficient to understand the effect of DM annihilation/decay on the IGM, since the energy deposition efficiency of any annihilation/decay process is simply an appropriate sum over $f_{c,e/\gamma}(z,E)$ over injection species and all relevant energies.

\subsection{$f_c(z)$ at Low Redshifts}

We defer a full treatment of calculating $f_c(z)$ to low redshifts to an upcoming paper, and instead give a brief summary of the method here. We have computed $f(z)$ down to a redshift of $1+z = 4$ in three different scenarios: (i) instantaneous and complete reionization at $z = 6$, which is close to the expected redshift of reionization from astrophysical measurements of $T_{\text{IGM}}$; (ii) instantaneous and complete reionization at $z = 10$, which is close to the expected redshift of reionization from measurements of the CMB power spectrum; and (iii) no reionization. These different reionization conditions were used not just for the deposition of energy by low-energy photons and electrons, but also for the high-energy code which tracks high-energy electrons and photons as they cool over time, since the photoionization rate of high-energy photons depend strongly on the ionization history. Previous studies typically assume that $f_c(z)$ can be written as a redshift- and model-dependent efficiency function $f(z)$, which describes the efficiency with which high-energy particles are degraded to low energies and is independent of the deposition channel. This function multiplies a channel-dependent factor $\chi_c(x_e(z))$ that depends only on the free electron fraction and describes the absorption of low-energy particles into each of the deposition channel.\footnote{One popular choice is the scheme called the ``SSCK approximation'' in \cite{Slatyer2015a}, where a fraction $(1-x_e)/3$ is deposited into ionization and excitation each, with the remaining $(1+2x_e)/3$ going into heating.} However, our calculation of $\chi_c(z)$ depends on the low-energy photon spectrum at each redshift, and so depends on both $x_e$ and the injection history in a non-trivial way. The $f_c(z)$ results found in \cite{Slatyer2015} took these effects into account assuming the standard \texttt{RECFAST} ionization history, and can be used for small perturbations about that scenario. However, when considering reionization and markedly different reionization scenarios, $f_c(z)$ must be re-computed in each case by re-calculating the cooling in both the high-energy and low-energy regimes.

In order to perform these calculations, we also assume simultaneous reionization of neutral helium (HeI) at the same redshift as HI reionization. After HI and HeI reionization, low-energy photons can deposit their energy through (i) the ionization of singly-ionized helium (HeII); (ii) excitations to HeII; or (iii) distortions of the CMB energy spectrum. 

After reionization, the high energy code tags photons as deposited only when they can efficiently photoionize HeII. Thus any ``deposited'' photon with energy $E > \SI{54.4}{eV}$ corresponds to a HeII ionization and consequently gives rise to a secondary low-energy electron spectrum. Photons below this threshold cannot ionize anything else, and are assigned to the excitation or distortion channels. Low-energy electrons, including the secondary spectrum produced by photoionizing photons, deposit energy according to the same model used in \cite{Slatyer2015}, which is in turn based on \cite{Valdes:2007cu,Valdes:2009cq,MNR:MNR20624}. In accordance with these results, once full reionization occurs, the electrons deposit their energy into the IGM solely through heating, since there are no longer any neutral hydrogen atoms to ionize or excite.

We note here that prior to the instantaneous reionization, the code assumes a standard ionization history computed by the recombination code \texttt{RECFAST}. Furthermore, we have assumed the instantaneous reionization of HeII at $1+z = 4$, which is not a fully realistic model. Once the contribution to $x_e$ from DM annihilations become significant enough, our calculation for $f_c(z)$ based on the \texttt{RECFAST} result will not reflect the true $f_c(z)$ for the new ionization history that includes the DM contribution, and likewise for a HeII reionization scenario that differs significantly from instantaneous reionization at $1+z = 4$. 

In principle, this means that $f_c(z)$ should be calculated iteratively: after calculating $x_e(z)$ for a certain DM model using the $f_c(z)$ obtained from the \texttt{RECFAST} ionization history, $f_c(z)$ should be recalculated with the new $x_e(z)$, with this process repeated until convergence of $x_e(z)$ is achieved. However, we stress that such a computationally intensive process is unnecessary, since calculating $f_c(z)$ assuming a \texttt{RECFAST} ionization history results in an $x_e$ ($T_{\text{IGM}}$) prior to reionization that is always larger (smaller) than what we would get with an iterative calculation. This ensures that we have not unintentionally ruled out any DM model with a significant contribution to reionization consistent with the $T_{\text{IGM}}$ constraints, even without performing an iterative calculation of $f_c(z)$. This behavior can be seen in Figure \ref{fig:freeEleFracDecayAllowedRegion}, which shows a comparison of the ionization and thermal history computed with $f_c(z)$ after one iteration with the default $f_c(z)$ used in the rest of the paper. This point will be discussed further in Section \ref{sec:Constraints}.

\subsection{$f_c(z)$ Including Structure Formation}

The formation of structures at late times gives rise to local densities that greatly exceed the cosmological DM density $\rho_{\chi,0}$, accompanied by an increase in the velocity dispersion of DM particles within haloes. This has no effect on the rate of energy injection from DM decay, since the average rate of decays per unit volume across the universe remains the same. 
In the case of DM $s$-wave annihilation, however, the increased density increases the rate of interaction, while for $p$-wave annihilation both the increased density and increased velocity dispersion dramatically enhance the annihilation rate. These effects cause a significant deviation from the expected energy injection due to a smooth/homogeneous DM distribution. 

The increase in the density can be parameterized by an effective density $\rho_{\text{eff}}(z)$ for $s$-wave annihilation (equation (\ref{eqn:rhoeff}) and Figure \ref{fig_rho_eff}), and an effective density times velocity dispersion $(\rho v/c)_{\rm eff}(z)$ for $p$-wave annihilation (equation (\ref{eqn:rhoeff_p}) and Figure \ref{fig_rho_eff_pwave}). 

With these effective quantities, the energy injection rate can be written as a boost factor multiplied by the unclustered distribution injection rate:
\begin{alignat}{1}
	\left(\frac{dE}{dV dt}\right)_{\text{inj}} &= \left(\frac{dE_s}{dV dt}\right)_{\text{inj}}[1 + \mathcal{B}_{s,p}(z)],
\end{alignat}
where the subscript $s$ in $E_s$ indicates the energy injection due to a smooth distribution of DM given by equations (\ref{eqn:injRateSmooth}) and
(\ref{eqn:smoothpwave}) for the $s$- and $p$-wave cases, respectively. The effective deposition efficiency can now be re-defined as
\small
\begin{alignat}{1}
	f_c(z) &= \frac{H(z)}{(1+z)^{\alpha-3}}  \int \frac{(1+z')^{\alpha-4}}{H(z')} T_c(z',z,E) [1 + \mathcal{B}_{s,p}(z')] \, dz',
	\label{eqn:fz}
\end{alignat}
\normalsize
so that
\begin{alignat}{1}
	\left(\frac{dE}{dV dt} \right)_{c,\text{dep}} = f_c(z) \left(\frac{dE_s}{dV dt} \right)_{\text{inj}}.
\end{alignat}
$f_c(z)$ is now the ratio of the energy deposited in channel $c$ including structure formation effects to the injected energy due only to the smooth DM distribution, which has a simple analytic form. For $s$-wave annihilation, the boost factor is 
\begin{alignat}{1}
	1 + \mathcal{B}_s(z) = \frac{\rho_{\text{eff}}^2(z)}{(1+z)^6 \rho_{\chi,0}^2},
\end{alignat}
where $\rho_{\text{eff}}$ is shown in Figure \ref{fig_rho_eff}. For $p$-wave annihilation, the effect of structure formation is parametrized not only by an effective density $\rho_{\text{eff}}$, but also by the characteristic one-dimensional velocity of the DM particles. The boost factor is:
\begin{alignat}{1}
	1 + \mathcal{B}_p(z) = \frac{(\rho v/c)_{\text{eff}}^2 (z)} {(1+z)^8 \rho_{\chi,0}^2(\sigma_{1D,B}(z=0)/c)^2}.
	\label{eqn:pwaveInj}
\end{alignat}
where $(\rho v/c)_{\text{eff}}$ is shown in Figure \ref{fig_rho_eff_pwave}.

Contour plots of $f_c(z)$ for all of the DM energy injection processes producing $e^+e^-$ or $\gamma \gamma$, including the effects of structure formation where relevant, are shown in Appendix \ref{app:fz}. 

\section{Free Electron Fraction and IGM Temperature History}
\label{sec:FreeEleFrac}

\subsection{The Three-Level Atom}

In order to compute the contribution of DM annihilation to the optical depth and IGM temperature, the hydrogen atoms in the IGM are modeled using the effective 3-level atom model for hydrogen, first described in \cite{Peebles1968,Zeldovich1969}. Equations describing the rate of change of $x_e$ and $T_{\text{IGM}}$ as a function of redshift can be derived from this model, and are given in many studies that calculate the ionization history of the universe. These equations form the basis of the \texttt{RECFAST} \cite{Seager:2000} code: they are relatively easy to integrate, and show good agreement with the full \texttt{RECFAST} code in computing $x_e(z)$. We have checked that our integrated ionization history of the universe with neither DM nor ionization is in good agreement with the result produced by \texttt{RECFAST}. These equations can also be easily modified to include energy injection from DM with the full $f_c(z)$ dependence of equation (\ref{eqn:fz}). We have verified that after including DM injection, our results are in good agreement with the ionization history obtained by \texttt{RECFAST} with the inclusion of DM.  

A full description of the three-level atom is given in \cite{AliHaimoud:2010dx}. All hydrogen atoms are described by a ground state ($n=1$) and a first excited state $(n=2)$, with all excited states being in thermal equilibrium with the continuum. Direct recombination from the continuum to the ground state is assumed to have no net effect on $x_e$, as each photon produced quickly ionizes another hydrogen atom. Without DM, the net rate of ionization in this model is given by
\begin{alignat}{1}
	\frac{dx_e}{dz} \frac{dz}{dt} = I_3(z) = C \left[\beta_e(1-x_e) e^{-h\nu_\alpha/k_BT} - \alpha_e x_e^2 n_{\text{H}} \right].
	\label{eqn:I3}
\end{alignat}
where $\nu_\alpha$ is the Lyman-$\alpha$ frequency. The rate of ionization is described by just a single recombination coefficient $\alpha_e$ and a single ionization coefficient $\beta_e$. As pointed out in \cite{Chluba:2015lpa}, $\beta_e$ should be evaluated at the CMB temperature and not at the electron temperature as in the \texttt{RECFAST} code; this is consistent with the implementation of the \texttt{RECFAST} calculation in the \texttt{HyREC} code. $C$ is a factor dependent on redshift and $x_e$, given by
\begin{alignat}{1}
	C = \frac{\Lambda n_{\text{H}}(1-x_e) + 8\pi \nu_\alpha^3 H}{\Lambda n_{\text{H}}(1-x_e) + 8\pi \nu_\alpha^3 H + \beta_e n_{\text{H}} (1-x_e)}.
\end{alignat}
where $\Lambda = \SI{8.23}{s^{-1}}$ is the decay rate of the metastable $2s$-state in hydrogen to the ground state. The $C$ factor is the ratio of the recombination rates (from $n=2$ to $n=1$) to all possible transition rates from $n=2$, and characterizes the probability of achieving recombination from $n=2$. 

Our analysis should in principle include ionized helium, but assuming that helium remains neutral prior to reionization is justified for several reasons. First, the helium ionization fraction has been shown to have little influence on the total free electron fraction, assuming a standard recombination history obtained from the more sophisticated \texttt{RECFAST} calculation. Even after including unclustered DM annihilation with a large annihilation parameter of $p_{\text{ann}} = \SI{1.8E-27}{cm^3 s^{-1} GeV^{-1}}$, setting the helium ionization fraction to be a constant anywhere in the range $10^{-10}$ to $10^{-3}$ resulted in a difference of at most 0.2\% in the calculated free electron fraction at all redshifts \cite{Galli2013}. Moreover, $f_{\text{He ion}}(z)$ is small compared to the other channels; this, together with the significantly smaller number density compared to hydrogen, means that helium ionization is a relatively unimportant process even with large energy injections from DM. This allows us to safely assume that helium remains neutral prior to reionization in the three-level atom equations, although our calculation of $f_c(z)$, which features in the DM injection rate, does not make this assumption.

Below $1+z=10$, in the three scenarios we consider, the expression for $I_3(z)$ with only neutral helium continues to be valid until instantaneous reionization occurs. After reionization, $x_e$ is instantaneously set to 1.08, and $I_3(z)$, together with any other terms that contribute to changing $x_e$, are set to zero, since we assume the universe remains ionized from then on. Only $T_{\text{IGM}}$ will continue to evolve after reionization. 

\subsection{Heating of the IGM}

The evolution of $x_e$ depends on $T_{\text{IGM}}$, and so $T_{\text{IGM}}$ also needs to be determined as a function of redshift in order to obtain the ionization history. The rate of change of $T_{\text{IGM}}$ without energy injection from DM can be written as the sum of two separate processes affecting the temperature: 
\begin{alignat}{1}
	\frac{dT_{\text{IGM}}}{dz} \frac{dz}{dt} = Q_{\text{adia}}(z) + Q_{\text{CMB}}(z). 
	\label{eqn:TIGMEvolutionNoDM}
\end{alignat}
$Q_{\text{adia}}(z)$ represents the cooling of the IGM due to the expansion of the universe, and is simply given by
\begin{alignat*}{1}
	Q_{\text{adia}}(z) = \frac{2T_{\text{IGM}}}{1+z} \frac{dz}{dt},
\end{alignat*}
so that without any contribution from other sources, $T_{\text{IGM}} \propto (1+z)^2$, as is expected from adiabatic cooling of the baryons in the IGM. The second term, $Q_{\text{CMB}}(z)$, is the rate of change of temperature as a result of energy transfer to or from the CMB via Compton scattering processes. The rate of energy transfer from these processes is \cite{Weymann1965}:
\begin{alignat}{1}
	\frac{dE}{dV dt} = 4\sigma_T a T_{\text{CMB}}^4 x_e n_{\text{H}} (1+z)^3 \left(\frac{T_{\text{CMB}} - T_{\text{IGM}}}{m_e} \right),
\end{alignat}
where $\sigma_T$ is the Thomson scattering cross-section and $a$ is the radiation constant. This energy transfer leads to the following increase in temperature of the IGM: 
\begin{alignat*}{1}
	\frac{dE}{dV} = \frac{3}{2} n_{\text{tot}} (1+z)^3 dT_{\text{IGM}}.
\end{alignat*}
Here, $n_{\text{tot}}$ is the total number density $n_{\text{tot}} = n_e + n_{\text{HII}} + n_{\text{HI}} + n_{\text{He}} = (x_e + 1 + 0.079)n_{\text{H}}$. This gives
\begin{alignat}{1}
	Q_{\text{CMB}}(z) = \left(\frac{8\sigma_T a T^4_{\text{CMB}}}{3m_e} \right) \frac{n_{\text{H}}}{n_{\text{tot}}}(T_{\text{CMB}} - T_{\text{IGM}})x_e.
\end{alignat}

\subsection{Energy Deposition from Dark Matter}

We will now make use of $f_c(z)$ to translate the energy injection into terms that alter the rate of change of $x_e$ and $T_{\text{IGM}}$. The total amount of energy deposited into HI ionization leads straightforwardly to an increase in $x_e$: 

\begin{alignat}{1}
	I_{\chi,\text{ion}}(z) = \left(\frac{dE}{dV dt} \right)_{\text{inj}} \frac{f_{\text{H ion}} (z)}{V_{\text{H}} n_{\text{H}} (1+z)^3}\, ,
\end{alignat}
where $V_{\text{H}} = \SI{13.6}{eV}$ is the ionization potential of hydrogen. The factor of $1/n_H(1+z)^3$ normalizes the total energy to the density of hydrogen at that redshift. This term adds straightforwardly to the ionization rate of the IGM given by equation (\ref{eqn:I3}). 

Energy going into Lyman-$\alpha$ excitations also changes the rate of ionization, since hydrogen becomes easier to ionize. The total contribution to $x_e$ is given by
\begin{alignat}{1}
	I_{\chi,\text{Ly}\alpha}(z) = \left(\frac{dE}{dV dt}\right)_{\text{inj}} \frac{(1-C) f_{\text{Ly}\alpha}(z) }{h\nu_\alpha n_{\text{H}} (1+z)^3} \,,
\end{alignat}
where the $1-C$ factor is the probability of ionization from the excited hydrogen atom at energy level $n=2$ and hence the contribution to $x_e$. 

Finally, DM annihilation can deposit energy directly into heating at a rate
\begin{alignat}{1}
	Q_\chi(z) = f_{\text{Heat}}(z) \left(\frac{dE}{dV dt}\right)_{\text{inj}}  \frac{2}{3 n_{\text{tot}} (1+z)^3}.
\end{alignat}

To summarize, the coupled differential equations that need to be integrated simultaneously to obtain $x_e$ and $T_{\text{IGM}}$ are
\begin{alignat}{1}
	\frac{dx_e}{dz} \frac{dz}{dt} &= I_3 (z) + I_{\chi,\text{ion}} (z) + I_{\chi,\text{Ly}\alpha}(z) \,, \\
	\frac{dT_{\text{IGM}}}{dz} \frac{dz}{dt} &= Q_{\text{adia}}(z) + Q_{\text{CMB}}(z) + Q_\chi(z) \,.
\end{alignat}

Aside from DM and the instantaneous reionization scenarios considered, no further sources of heating or reionization (e.g. star-forming galaxies and other stellar phenomena) are included in these equations.\footnote{See \cite{Poulin2015} for an example of how heating from astrophysical sources can be included in a similar analysis.} This simplification is consistent with our computation of $f_c(z)$ using the standard ionization history, which overestimates the true contribution of $x_e(z)$ from DM, while underestimating the corresponding $T_{\text{IGM}}(z)$ contribution. A full treatment including astrophysical sources of heating and ionization would require a better understanding of $f_c(z)$ in situations where reionization is gradual, and we defer such a study to future work.

The initial conditions used for the integration are $x_e(z=1700)=1$ and $T_{\text{IGM}} = T_{\text{CMB}}(z=1700)$, corresponding to the state of baryonic matter prior to recombination. The contribution to the optical depth by DM annihilation/decay $\delta \tau$, at a given $\langle \sigma v \rangle$ or $\tau_\chi$ and mass $m_\chi$ is then determined by integrating equation (\ref{eqn:OpticalDepth}) up to $z=1700$ and subtracting the residual integrated optical depth that is already present when there is no DM. Note that when we consider reionization at $z = 10$, we do not include the contribution to $\delta \tau$ from $x_e$ between $z = 6$ and 10.\footnote{Note that the optical depth contribution from instantaneous reionization at $z = 10$ exceeds the Planck optical depth measurement, and thus would leave no room for any contribution from DM at all. However, we do not use the optical depth constraint in this manner.} We will discuss the calculation of $\delta \tau$ and the use of the optical depth constraints given by equation \ref{eqn:ExcessOpticalDepth} further in Section \ref{sec:Constraints}.


\section{Results}
\label{sec:Constraints}

We now calculate the integrated free electron fraction $x_e$ and IGM temperature $T_{\text{IGM}}$ as a function of redshift in each of the three DM energy injection scenarios considered ($s$-wave annihilation, $p$-wave annihilation and decay), for a wide range of $\langle \sigma v \rangle$ and decay lifetimes $\tau_\chi$, and $m_\chi$ between $\sim 10$ keV and $\sim 1$ TeV. As we discussed in Section \ref{sec:fz}, we have neglected any additional $x_e$ contribution from DM processes in our computation of $f_c(z)$, even though DM energy injection can produce significant deviations from the standard ionization history prior to reionization. Moreover, even after reionization occurs, the prescription for HeII reionization could affect the energy deposition. Thus the $f_c(z)$ curves we compute may not be completely accurate for an ionization history that is significantly different from the \texttt{RECFAST} result, or where HeII reionization cannot be approximated as occurring instantaneously at $1+z=4$.

Fortunately, our $f_c(z)$ calculations underestimate the contribution of DM to reionization, as more realistic ionization histories would generally have \emph{higher} ionization fractions, which in turn would suppress the additional ionization from DM. With a higher ionization fraction for HI (HeII), the energy deposited into ionization of HI (HeII) decreases, since there are fewer HI (HeII) atoms to ionize or excite prior to reionization (after reionization), while energy going into heating increases in both cases. This intuitive explanation of the behavior of $f_c(z)$ is consistent with the results used in our low-energy code to assign deposited energy from low-energy electrons into the various channels, where the MC results show that all of the energy from low-energy electrons go into collisional heating processes as $x_e$ tends to 1. Thus the $f_c(z)$ curves calculated under our assumptions consistently overestimate the rate of energy deposition into ionization, while underestimating the rate of energy deposited as heat. 

This means that if the contribution to reionization is small with the $f_c(z)$ values used here for a given cross-section/lifetime and mass, then a more accurately computed $f_c(z)$ assuming an elevated $x_e$ will have an even smaller contribution to $x_e$ and a larger contribution to $T_{\text{IGM}}$, making the result more constrained by the $T_{\text{IGM}}$ limits. Similarly, including other conventional sources of ionization would only decrease the contribution that DM can make to reionization: the presence of other sources would produce a larger $x_e$ than we have assumed, which again suppresses the energy deposition fraction into ionization while enhancing the fraction into heating. 

To check the robustness of our constraints, we have also repeated our calculations considering: 

\begin{enumerate}
	\item Different reionization conditions, namely (i) instantaneous and complete reionization at $z=6$; (ii) instantaneous and complete reionization at $z=10$; and (iii) no reionization, to see how sensitive our results are to the uncertainty in the specifics of reionization and in particular in the redshift at which reionization occurs. For each reionization condition, $\delta \tau$ is integrated appropriately over $x_e(z)$, after which the optical depth from $x_e(z)$ without DM is subtracted. This includes the optical depth contribution from redshifts after reionization, where $x_e = 1.08$. Each reionization scenario results in a different $T_{\text{IGM}}(z)$ evolution after reionization occurs, and also has a different redshift at which we assess the contribution of DM to reionization (more details below); 

	\item A range of structure formation scenarios that bracket the uncertainties on the properties of low-mass (sub)haloes, below the resolution of current cosmological simulations; and

	\item Two different IGM temperature constraints as shown in equation (\ref{eqn:TIGMConstraints}), namely (i) $T_{\text{IGM}}(z=6.08) = \SI{18621}{K}$; (ii) $T_{\text{IGM}}(z = 4.8) = \SI{10000}{K}$, where we have taken the upper bound at 95\% confidence. We do not make use of the lower bound, since $f_{\text{Heat}} (z)$ is likely to be an underestimate for reasons outlined above. The second temperature measurement is more constraining and will be used as the main temperature constraint, but constraints obtained from both temperature limits will be shown for the main $p$-wave result.   
\end{enumerate}

The three main quantities of interest are: (i) $x_e$ at a redshift just prior to the assumed instantaneous reionization at $z=6$ or $z=10$, or at $z=6$ for the case of no reionization, since hydrogen reionization is known to be complete by then; (ii) $T_{\text{IGM}}$ at $z=6.08$ and $z=4.8$ for comparison with the results shown in equation (\ref{eqn:TIGMConstraints}); and (iii) the total integrated optical depth $\delta \tau$. If DM with a given $\langle \sigma v \rangle$ or $\tau_\chi$ and $m_\chi$ can produce $x_e > 0.1$ just before reionization (or at $z=6$ for the case of no reionization) we consider this a possible scenario in which DM can contribute significantly to reionization. The 10\% level used in this paper is arbitrary, and we will also present results for contributions ranging from 0.025\% to 90\% in the form of color density plots for all injection species and all DM processes.

A few remarks should be made about the calculation of optical depth and the use of the optical depth constraints in this paper. To compute $\delta \tau$, we integrate the optical depth due to DM annihilation/decay from $z_{\text{reion}}$ to recombination.\footnote{When there is no reionization, we start integrating from $z = 6$, making $\delta \tau$ identical to the case with $z_{\text{reion}} = 6$.} We then compare $\delta \tau$ to the bound on excess optical depth from redshifts $z > 6$, assuming full ionization for $z \leq 6$; that is, for the purposes of computing the maximum allowed exotic contribution to optical depth, we essentially treat $z_{\text{reion}} = 6$ for all scenarios, even when $\delta \tau$ includes only DM contributions from $z > 10$. This allows us to understand how our limits could weaken if the reionization history were different: including gradual reionization from astrophysical sources between $z = 6$ and $z = 10$, for example, would likely suppress the contribution to reionization and hence optical depth from DM annihilation during this period, resulting in a smaller contribution from DM to reionization than would have been determined with instantaneous reionization at $z_{\text{reion}} = 6$. By taking $z_{\text{reion}} = 10$ and not considering the contribution to optical depth for $z < 10$, we obtain the weakest constraints from the $\delta \tau$ bound given in equation (\ref{eqn:ExcessOpticalDepth}). In this way, these two reionization scenarios bracket the possible contribution of DM to reionization. Thus, although including the optical depth due to complete, instantaneous reionization at $z = 10$ would exceed the Planck optical depth measurement, we still consider this scenario in order to study the DM contribution to reionization in a model-independent way. Assuming two different instantaneous reionization scenarios also allows us to probe the possible effects of earlier reionization on the DM contribution to the temperature evolution. 

We will choose as our benchmark the scenarios where the largest $x_e$ just prior to reionization can be obtained from the {\it smallest} $\langle \sigma v \rangle$ or {\it longest} decay lifetimes, since various experimental constraints set upper bounds on the cross-sections and lower bounds on the decay lifetimes. In all cases, reionization at $z = 6$ is more realistic than no reionization and is also more easily achieved than at $z = 10$, making it the main reionization scenario to consider. The structure formation scenario with the largest boost factor allows for reionization with a smaller cross-section, and thus we choose this as our benchmark (for $s$-wave annihilation this is the ``stringent'' case shown with a solid red line in Figure \ref{fig_rho_eff}, while for $p$-wave annihilation all scenarios give the same boost).

\subsection{$s$-wave Annihilation}

\begin{figure*}[t!]
	\subfigure{
		\includegraphics[scale=0.59]{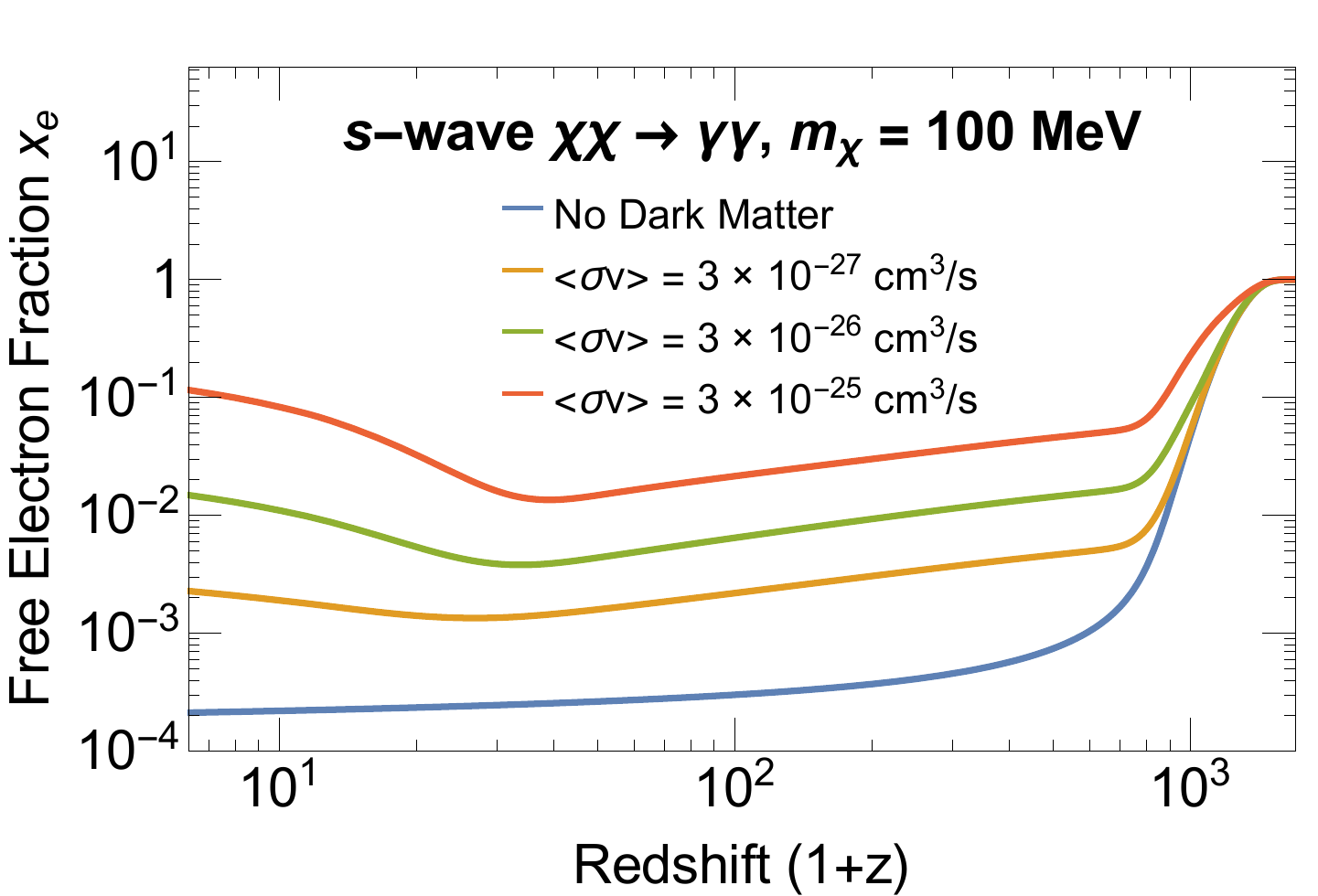}
	}
	\subfigure{
		\includegraphics[scale=0.58]{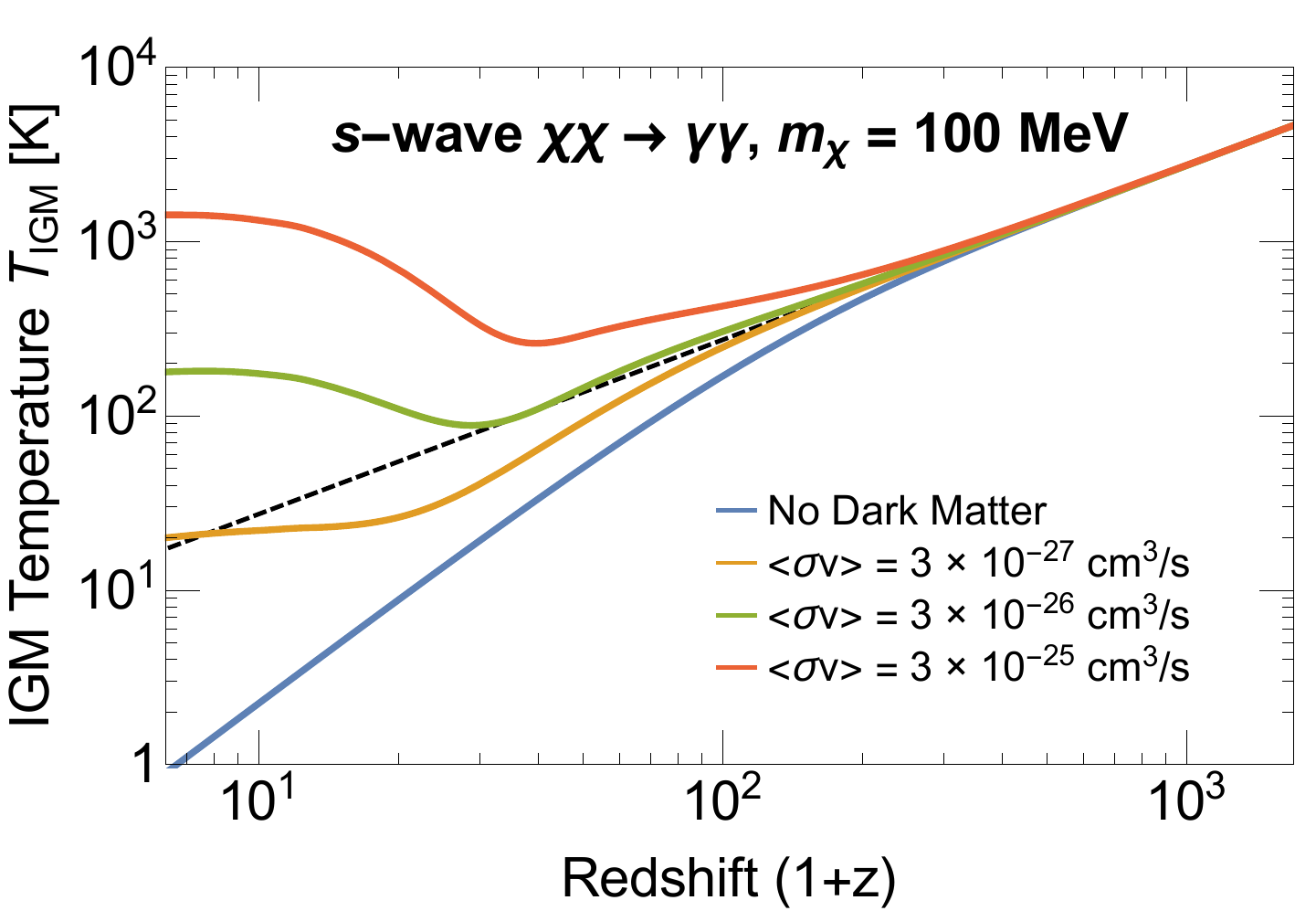}
	}
	\caption{\footnotesize{Integrated free electron fraction $x_e$ and IGM temperature $T_{\text{IGM}}$ for $\chi \chi \to \gamma \gamma$ $s$-wave annihilation for $m_\chi = \SI{100}{\mega \eV}$ with (from bottom to top): no DM; $\left<\sigma v\right> = $ \SI{3E-27}{cm^3 s^{-1}}; \SI{3E-26}{cm^3 s^{-1}} and \SI{3E-25}{\cm^3 \s^{-1}} respectively. The CMB temperature is shown as a dashed line for reference. No reionization is assumed.}}
	\label{fig:freeEleFracsWave}
\end{figure*}

Figure \ref{fig:freeEleFracsWave} shows the integrated free-electron fraction $x_e$ for the particular case of DM with $m_\chi = \SI{100}{MeV}$ undergoing $s$-wave annihilation into a pair of $\SI{100}{MeV}$ photons with a cross-section ranging from $\SI{3E-27}{}$ to $\SI{3E-25}{cm^3 s^{-1}}$, as well as the case with no DM for comparison. These curves show the result with no reionization: different reionization conditions are identical up to the redshift of reionization $z_{\text{reion}}$, whereupon $x_e$ instantaneously becomes 1 until the present day. These curves are representative of the $x_e$ histories across all DM masses and cross-sections for $s$-wave annihilation. At $z \sim 20$, structure formation becomes important, which greatly increases $f_c(z)$ in all channels, leading to an increase in $x_e$. $s$-wave annihilation of the smooth distribution of DM results in a larger baseline $x_e$ after recombination, which is higher for larger $\langle \sigma v \rangle$ at the same $m_\chi$. 

Along with $x_e$, the IGM temperature history $T_{\text{IGM}}(z)$ is also simultaneously integrated. The IGM temperature curves for DM undergoing $s$-wave annihilation into \SI{100}{MeV} photons for cross-sections ranging from $\SI{3E-27}{}$ to $\SI{3E-25}{cm^3 s^{-1}}$ are shown in the same figure and are also representative of IGM temperature histories across a broad range of $\langle \sigma v \rangle$ and $m_\chi$. The CMB temperature is included for reference. The IGM is initially coupled to the CMB, but once recombination occurs, the temperature starts to fall more rapidly than the CMB temperature. DM $s$-wave annihilations decrease the fall-off in temperature at relatively large redshifts. At $z\sim 20$, the impact of structure formation once again increases the IGM temperature significantly relative to the case with no DM.

The contribution of DM to reionization through $s$-wave annihilation is significantly constrained by the CMB power spectrum measurements derived by Planck 2015 \cite{PlanckCollaboration2015}, as well as by the measured total integrated optical depth. The cross-section for annihilation must be large enough for significant ionization to occur at redshifts near reionization; however, increasing the cross-section also increases the residual free electron fraction during the cosmic dark ages. This residual $x_e$ is constrained severely by the CMB anisotropy spectrum, which is sensitive to any additional ionization near redshifts $z \sim 600$. A large $x_e$ during the cosmic dark ages also contributes significantly to the optical depth. Since $n_e(z) \propto x_e(z)(1+z)^3$ and $dt/dz \propto (1+z)^{-5/2}$, the integrand in equation (\ref{eqn:OpticalDepth}) is proportional to $x_e(z)(1+z)^{1/2}$. The significantly elevated $x_e$ baseline means that the dominant contribution to $\delta \tau$ comes from early times when $z$ is large: since structure formation is relevant at later times, it does not add significantly to $\delta \tau$.  

\begin{figure*}[t!]
	\subfigure{
		\includegraphics[scale=0.59]{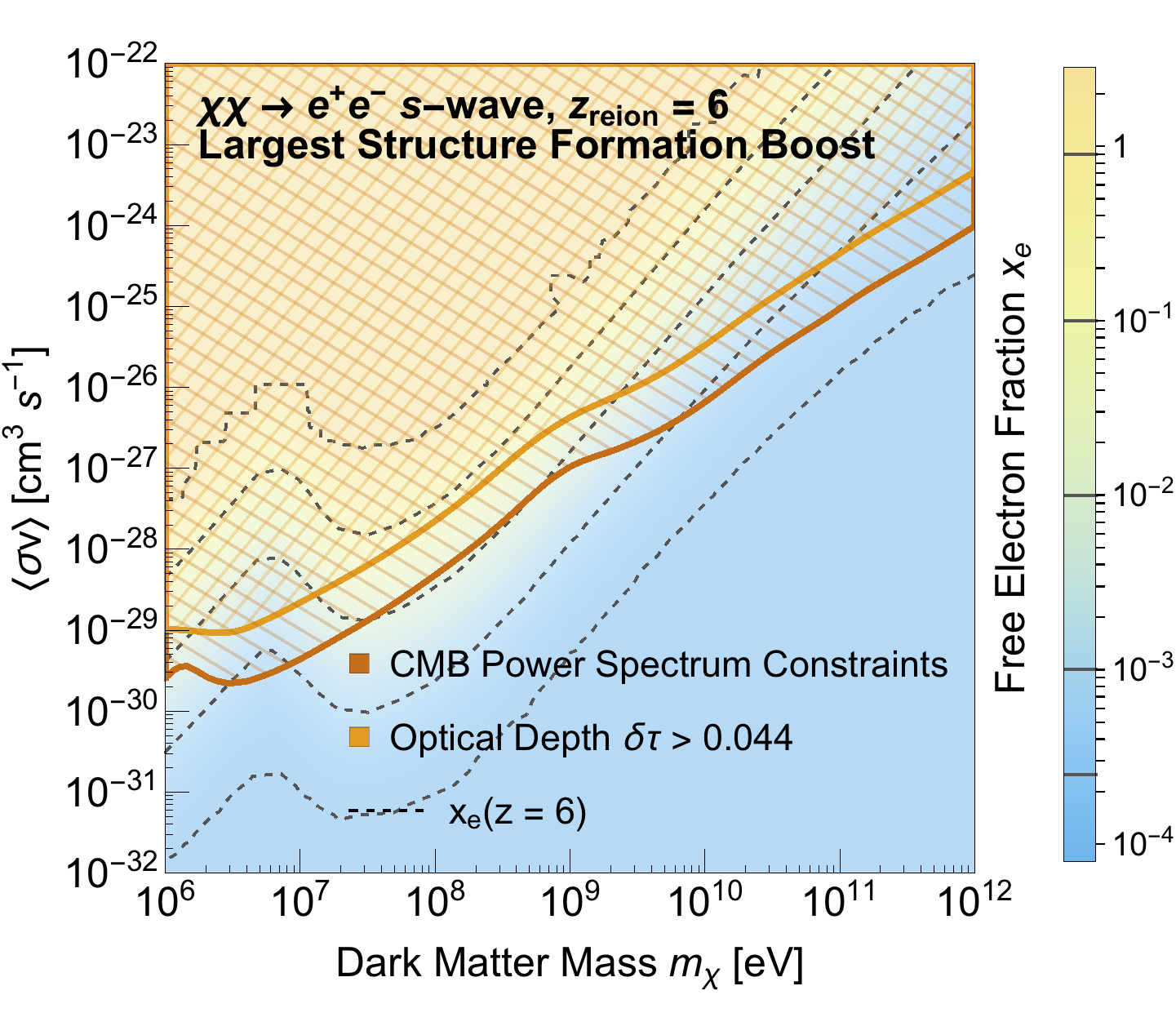}
	}
	\subfigure{
		\includegraphics[scale=0.59]{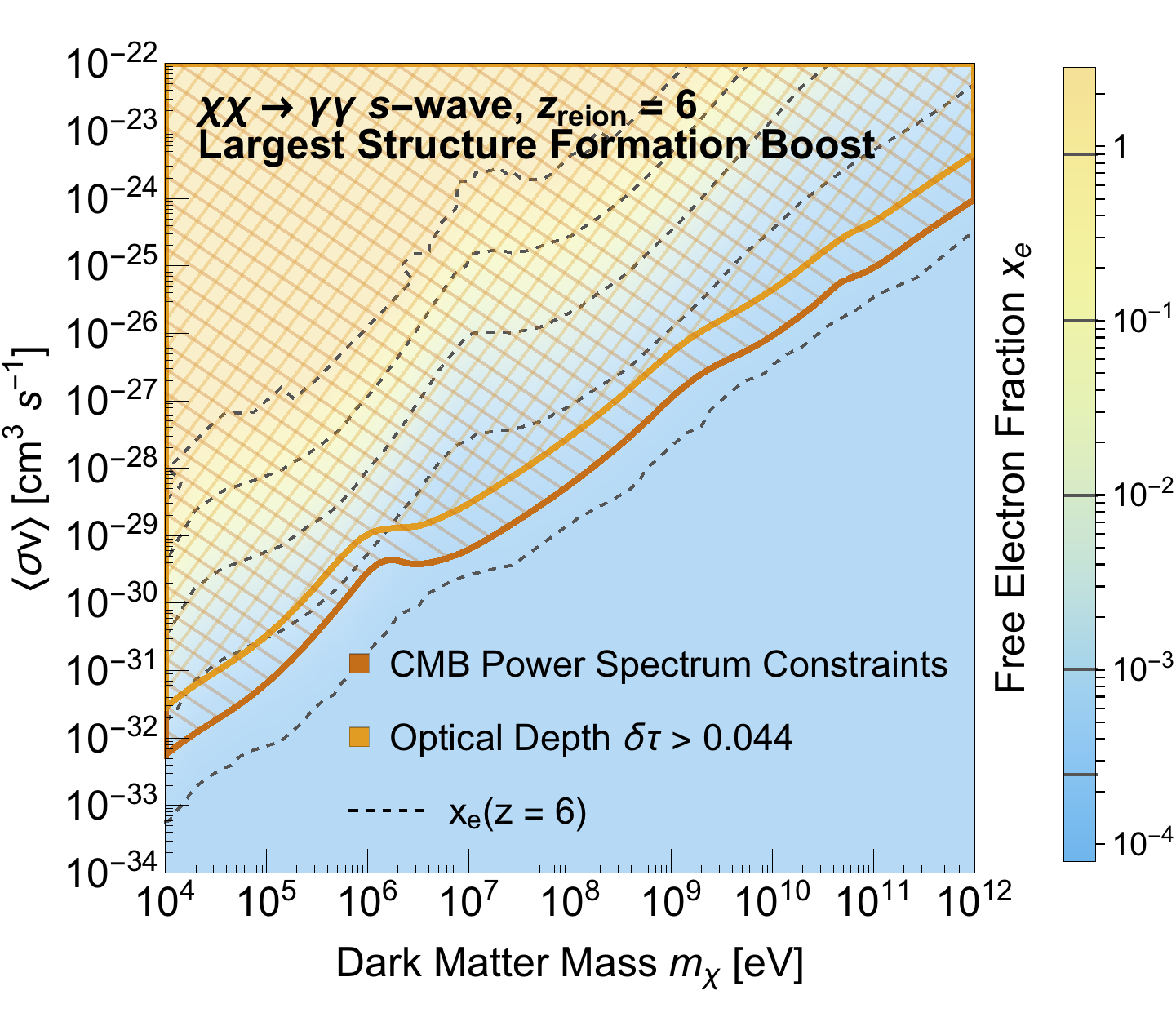}
	}
	\caption{\footnotesize{DM contribution to reionization for $\chi \chi \to e^+e^-$ (left) and $\chi \chi \to \gamma \gamma$ (right) $s$-wave annihilation, benchmark scenario. The hatched regions correspond to parameter space ruled out by the CMB power spectrum constraints as measured by Planck (red) and optical depth constraints (orange) respectively. The color density plot shows the DM contribution to $x_e$ just prior to reionization at $z = 6$, with contours (black, dashed) shown for a contribution to $x_e(z = 6) = $ 0.025\%, 0.1\%, 1\%, 10\% and 90\% respectively.}}
	\label{fig:xeConstraintsPlot_sWave}
\end{figure*}

We performed the integration of $x_e(z)$ and $T_{\text{IGM}}(z)$ over a broad range of masses and cross-sections, and computed the optical depth from $x_e(z)$ using equation (\ref{eqn:OpticalDepth}). Figure \ref{fig:xeConstraintsPlot_sWave} shows the free electron fraction just prior to reionization $x_e(z=6)$ for the benchmark scenario of both $\chi \chi \to e^+e^-$ and $\chi \chi \to \gamma \gamma$, as well as the excluded cross-sections due to constraints from the CMB power spectrum as measured by Planck and from the integrated optical depth. Constraints from $T_{\text{IGM}}$ are presented in Appendix \ref{app:additionalConstraints}. These bounds are less constraining, but unlike the CMB and optical depth constraints, they are sensitive to the low redshift behavior of $s$-wave annihilations: increasing the boost from structure formation beyond the value used here may relax the CMB and optical depth bounds, but this would strengthen the $T_{\text{IGM}}$ constraints. 

Although we have shown the results for these two processes ($\chi \chi \to e^+e^-$ and $\chi \chi \to \gamma \gamma$) as a function of $\langle \sigma v \rangle$ and $m_\chi$, we stress that these constraints go beyond these two annihilation channels. We discuss this point and present bounds on $\langle \sigma v \rangle/m_\chi$ as a function of the injection energy of the final products (which may in general be very different from $m_\chi$) in Appendix \ref{app:additionalConstraints} of this paper.

In both annihilation channels, there is no parameter space where a significant contribution to reionization occurs while being consistent with either the CMB power spectrum or optical depth bounds, with the CMB power spectrum bounds being approximately one order of magnitude stronger than the optical depth bounds. We stress that the optical depth constraints are similar regardless of reionization conditions, since $\delta \tau$ is the additional contribution from DM only, and is therefore not affected by the period where $x_e = 1$ after reionization. As a result, the true optical depth limits for reionization at $z = 10$ are likely stronger than what is shown here, since we do not include the additional contribution to optical depth from the fully ionized universe between $z = 6$ and $z = 10$. Furthermore, $\delta \tau$ is dominated by contributions from larger redshifts ($z \gtrsim 100$) and is relatively insensitive to the exact details of reionization and structure formation at $z \lesssim 20$. At the maximum $\langle \sigma v \rangle$ allowed by the CMB power spectrum bound, the DM contribution to $x_e$ just prior to reionization is below 2\% for $\chi \chi \to e^+e^-$ and below 0.1\% for $\chi \chi \to \gamma \gamma$ across all $m_\chi$ considered. These results are shown in Figure \ref{fig:xeMaxConstraints} in the conclusion.

\begin{figure*}[t!]
	\subfigure{
		\includegraphics[scale=0.59]{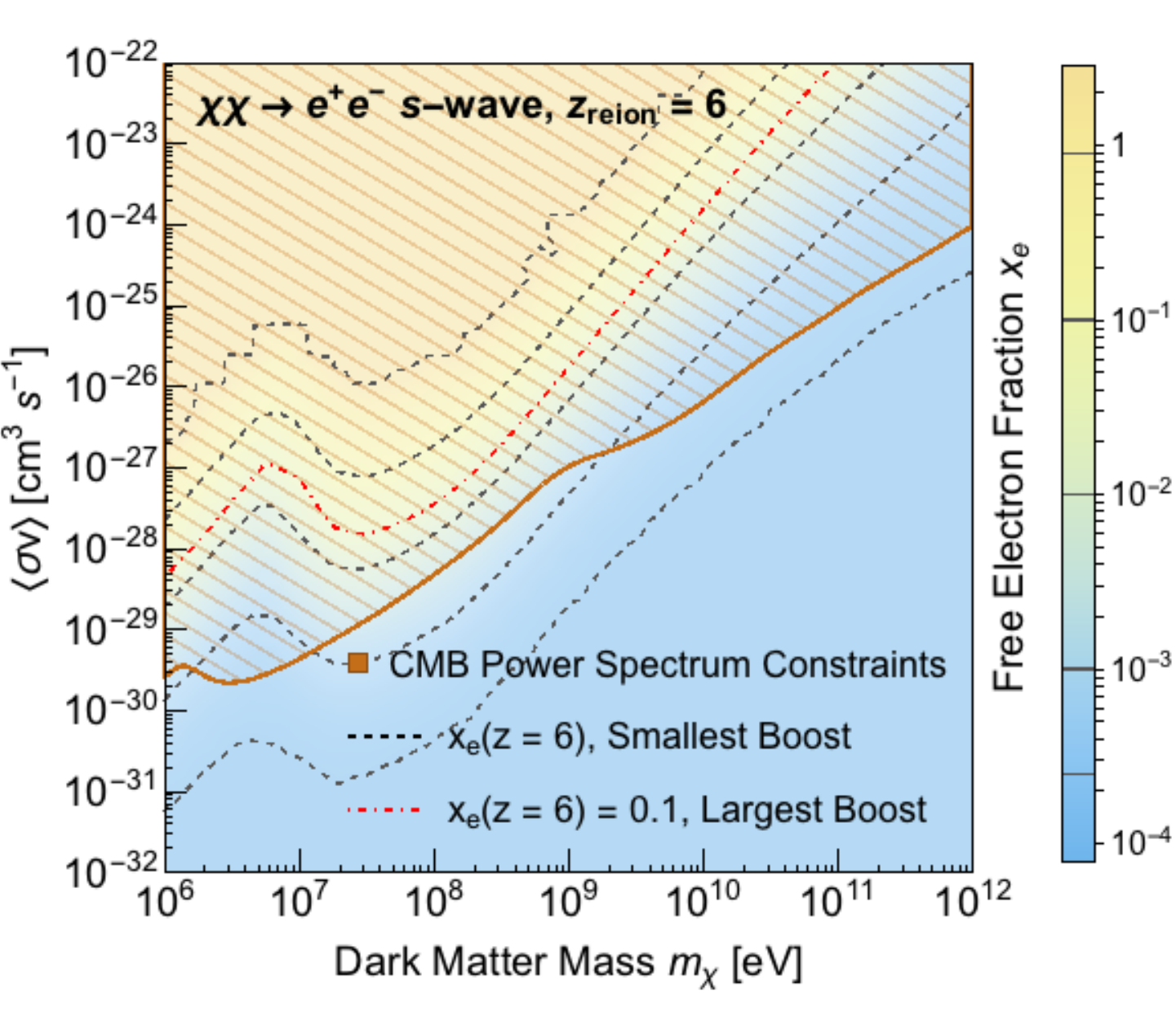}
	}
	\subfigure{
		\includegraphics[scale=0.59]{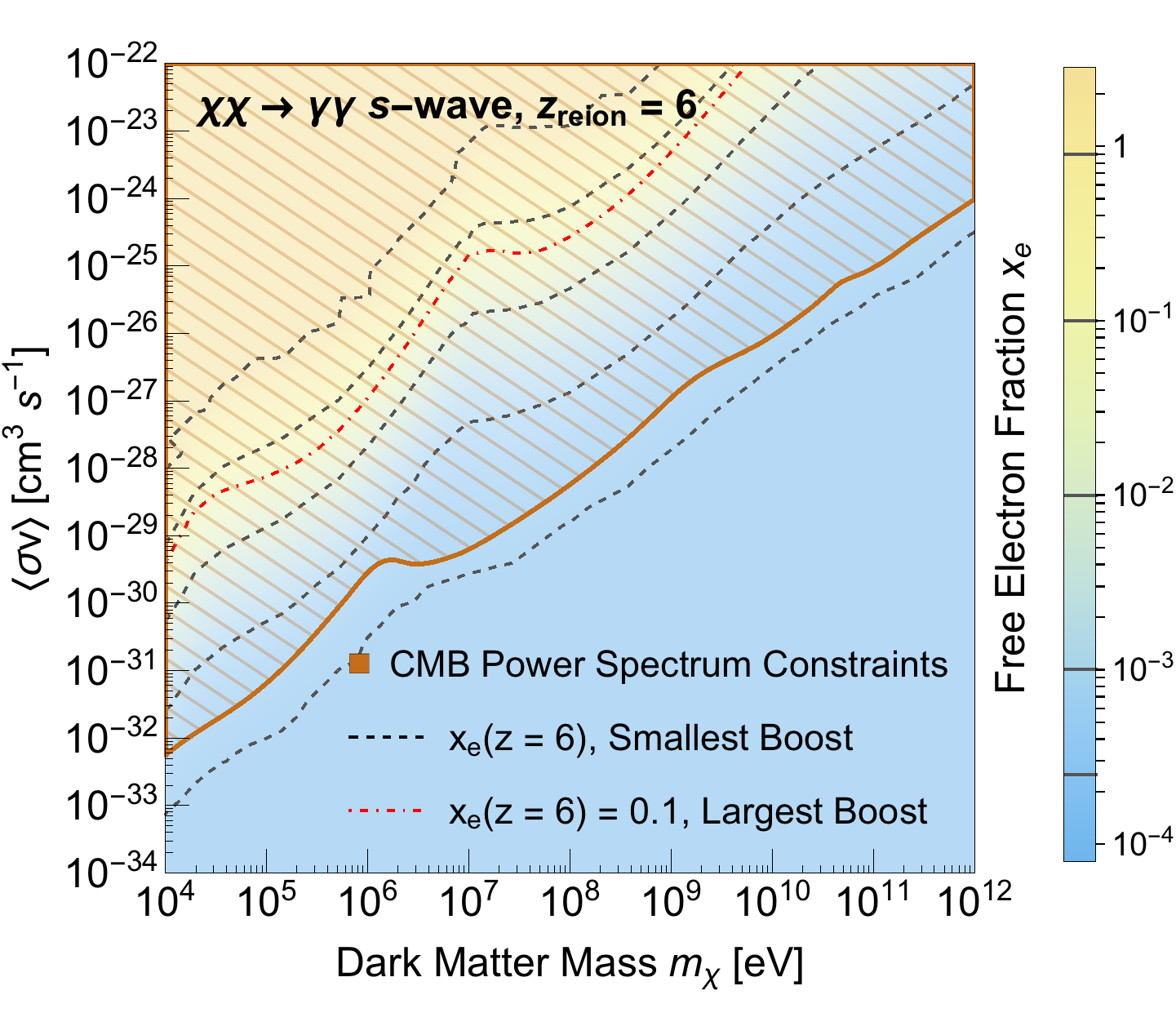}
	}
	\caption{\footnotesize{DM contribution to reionization for $\chi \chi \to e^+e^-$ (left) and $\chi \chi \to \gamma \gamma$ (right) $s$-wave annihilation assuming a different structure formation prescription. The color density plot shows the DM contribution to $x_e$ just prior to reionization at $z = 6$ assuming an NFW profile without subhaloes, with contours (black, dashed) shown for a contribution to $x_e(z = 6) = $ 0.025\%, 0.1\%, 1\%, 10\% and 90\% respectively. The red, dot-dashed contour for $x_e(z = 6) = 0.1$ assuming the benchmark Einasto profile with subhaloes, which has the largest boost factor at all redshifts, is also shown for comparison. The CMB power spectrum constraints obtained by Planck are shown by the hatched red region. }}
	\label{fig:xeConstraintsStructSysPlot_sWave}
\end{figure*}

Figure \ref{fig:xeConstraintsStructSysPlot_sWave} shows the reionization constraints on $s$-wave annihilation for the structure formation prescriptions with the smallest and largest boost factor (used as the benchmark). As expected, significant ionization prior to reionization can be achieved at lower cross-sections in the benchmark model, making it the most likely structure formation prescription for evading the constraints. Differences in structure formation can increase the value of $\langle \sigma v \rangle$ at which ionization becomes significant by less than an order of magnitude, and all of the regions with a significant contribution to reionization in either structure formation scenario are firmly ruled out by the Planck constraints.

\begin{figure*}[t!]
	\subfigure{
		\includegraphics[scale=0.59]{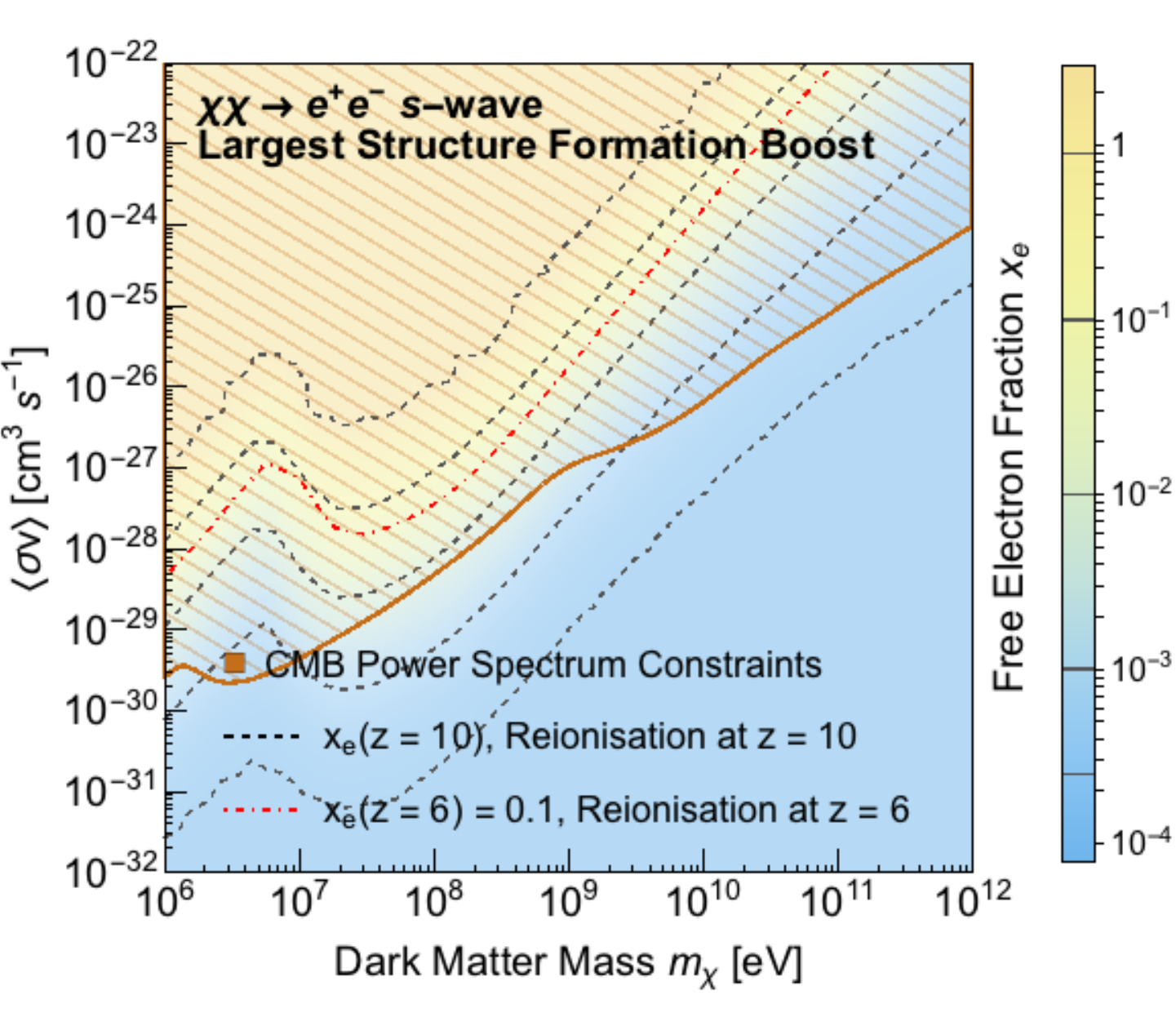}
	}
	\subfigure{
		\includegraphics[scale=0.59]{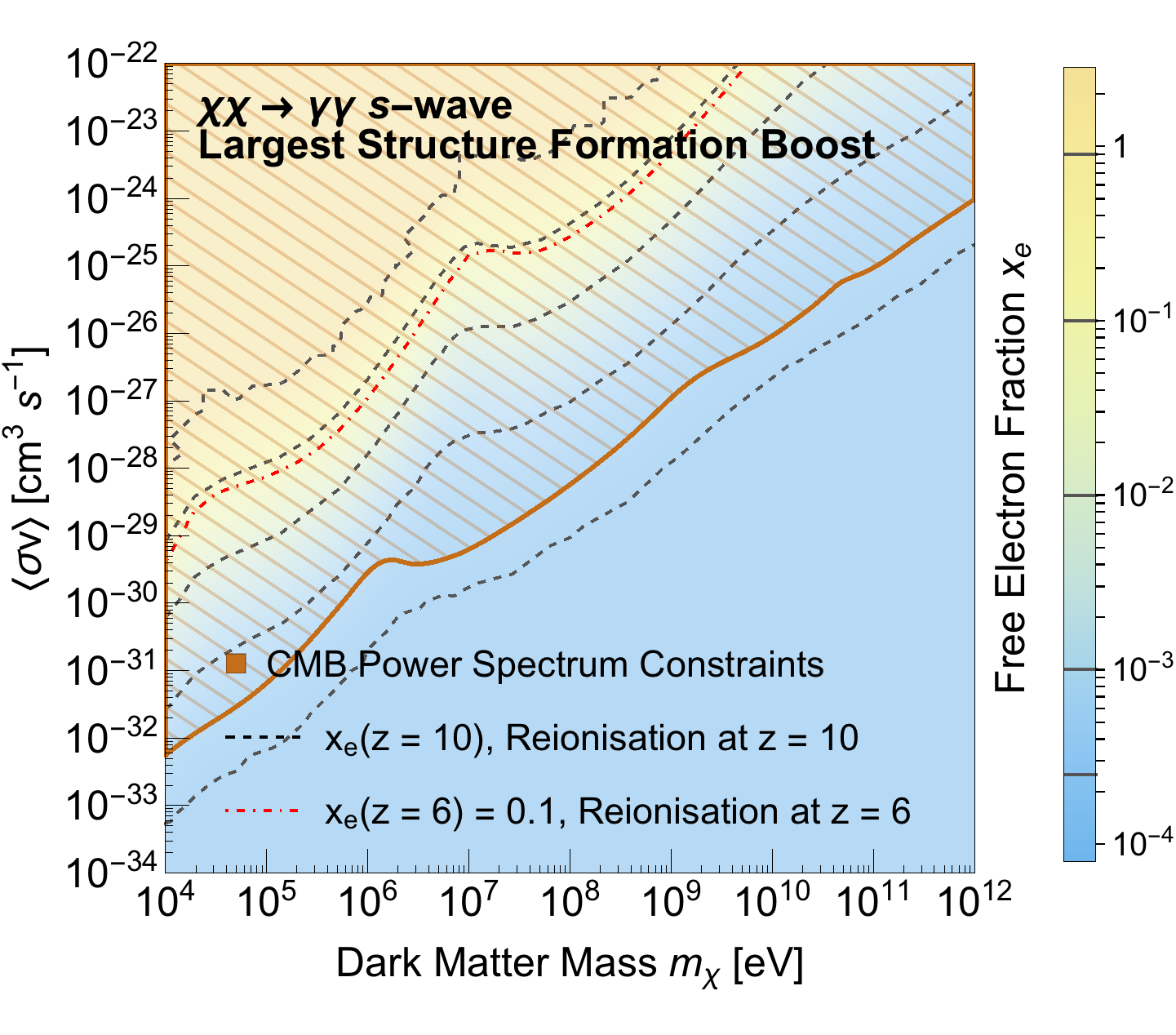}
	}
	\caption{\footnotesize{DM contribution to reionization for $\chi \chi \to e^+e^-$ (left) and $\chi \chi \to \gamma \gamma$ (right) $s$-wave annihilation, assuming a different reionization scenario.The color density plot shows the DM contribution to $x_e$ just prior to reionization at $z = 10$, with contours (black, dashed) shown for a contribution to $x_e(z = 10) = $ 0.025\%, 0.1\%, 1\%, 10\% and 90\% respectively. The red, dot-dashed contour shows $x_e(z = 6) = 10\%$ with reionization at $z = 6$ for comparison. The CMB power spectrum constraints obtained by Planck are shown by the hatched red region. }}
	\label{fig:xeConstraintsReionSysPlot_sWave}
\end{figure*}

Similarly, differences in reionization redshifts do little to change the result. Since $x_e(z)$ is identical in all three reionization scenarios until the point of reionization, there is no difference between $x_e(z=6)$ with reionization at $z=6$ and no reionization. With reionization at $z=10$, $x_e(z=10)$ is always less than $x_e(z=6)$ as $x_e$ increases rapidly between $z = 6$ and $10$, and so the region in parameter space where significant contribution to reionization occurs decreases when choosing an earlier redshift of reionization. Figure \ref{fig:xeConstraintsReionSysPlot_sWave} summarizes these results. 

To conclude, any significant contribution to reionization through $s$-wave DM annihilation is severely constrained by the cross-section bounds from the Planck CMB power spectrum measurement as well as the expected integrated optical depth to the surface of last scattering. For values of $\langle \sigma v \rangle$ that are consistent with the Planck CMB power spectrum constraints, we can only expect a contribution of no more than 2\% of the total ionization just prior to reionization (see Figure \ref{fig:xeMaxConstraints}). Our results are consistent with the conclusion reached in \cite{Poulin2015}. We have also shown that these results are robust to our assumptions on the structure formation scenario and on the redshift of reionization.

\subsection{$p$-wave Annihilation}\label{sec_pwave}

\begin{figure*}[t!]
	\subfigure{
		\includegraphics[scale=0.59]{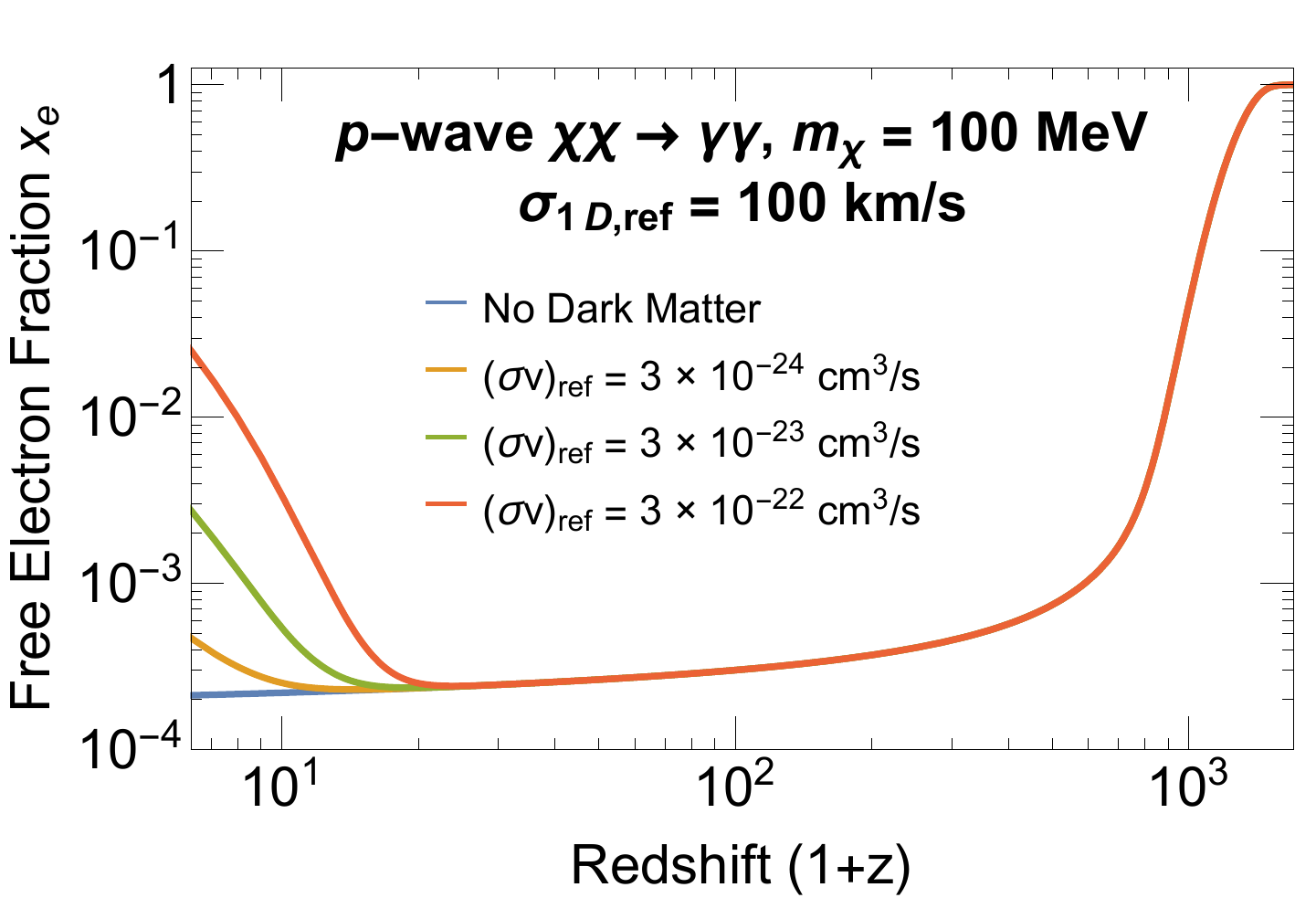}
	}
	\subfigure{
		\includegraphics[scale=0.58]{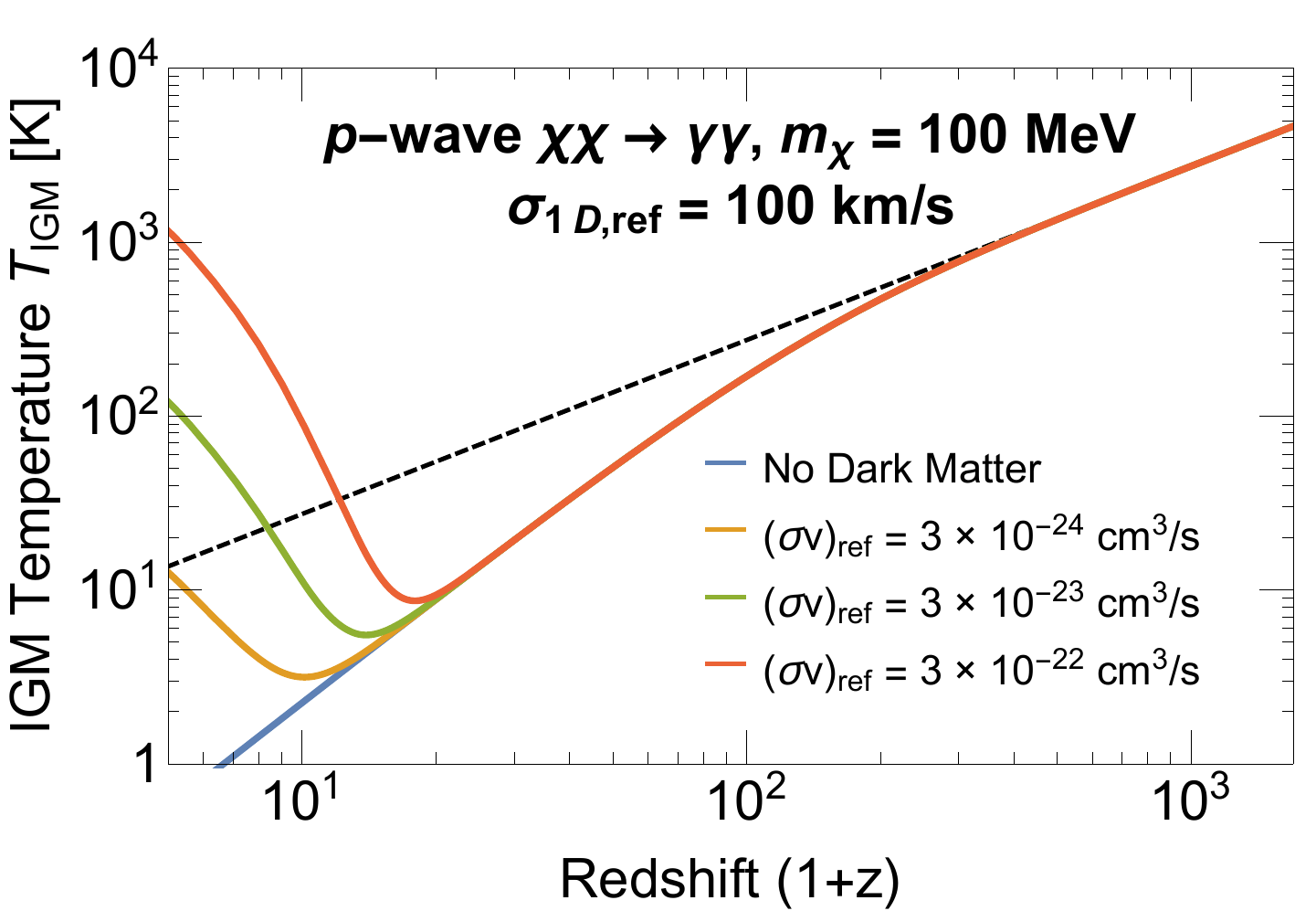}
	}
	\caption{\footnotesize{Integrated free electron fraction $x_e$ and IGM temperature $T_{\text{IGM}}$ for $\chi \chi \to \gamma \gamma$ $p$-wave annihilation for $m_\chi = \SI{100}{\mega \eV}$ with (from bottom to top): (blue) no DM; $(\sigma v)_{\text{ref}} = \SI{3e-24}{cm^3 s^{-1}}$, $(\sigma v)_{\text{ref}} = \SI{3e-23}{cm^3 s^{-1}}$ and $(\sigma v)_{\text{ref}} = \SI{3e-22}{cm^3 s^{-1}}$ respectively. The CMB temperature is shown as a dashed line. No reionization is assumed. }}
	\label{fig:freeEleFracpWave}
\end{figure*}

In $p$-wave annihilation, the $v^2$ dependence of the cross-section results in a $v^2/v_{\text{ref}}^2$ suppression of the energy injection rate, given in equation (\ref{eqn:pwaveInj}). Figure \ref{fig:freeEleFracpWave} shows the integrated $x_e$ for the case of $\chi \chi \to \gamma \gamma$ $p$-wave annihilation with $(\sigma v)_{\text{ref}}$ between \SI{3E-24}{cm^3 s^{-1}} and \SI{3E-22}{cm^3 s^{-1}}. Prior to the relevance of structure formation, the velocity suppression is a large effect, resulting in no additional contribution to $x_e$ unless the cross-section is exceptionally large. Once structure formation occurs, however, the velocity dispersion of DM particles within haloes increases significantly, increasing in turn the energy injection rate from $p$-wave annihilation. This results in a sudden and large increase in both $x_e$ and $T_{\text{IGM}}$ at $z \sim 20$. 

As we discussed earlier in section \ref{sec:StructureFormation}, the annihilation rate prior to structure formation is dependent on our choice of $\sigma_{\text{1D,B}}$, which we have taken to be the velocity dispersion for unclustered DM with $m_\chi = \SI{100}{GeV}$ and $T_{\text{kd}} = \SI{28}{MeV}$. Choosing a significantly smaller value of $m_\chi$ or $T_{\text{kd}}$ increases $\sigma_{\text{1D,B}}$, which in turn increases the annihilation rate prior to structure formation. With a sufficiently small value of $m_\chi$ and/or $T_{\text{kd}}$, $x_e$ will stay at a value significantly above the expected $x_e$ with no dark matter, similar to the ionization histories typical of $s$-wave dark matter shown in Figure \ref{fig:freeEleFracsWave}. While this leads to an increase in $x_e$ just prior to reionization, the optical depth bounds that we considered for $s$-wave annihilations become very constraining, particularly with the sharp increase in $x_e$ after structure formation that is not present in the $s$-wave case. Decreasing $m_\chi$ and/or $T_{\text{kd}}$ therefore makes it harder for a significant contribution to be made to reionization in a way that is consistent with the optical depth limits, making our unclustered velocity dispersion choice an optimistic one.

Unlike $s$-wave annihilation, constraints from the CMB power spectrum on the contribution of DM to reionization for $p$-wave annihilation are velocity-dependent, and depend strongly on the ``coldness'' of DM particles, i.e. on their unclustered velocity dispersion. Significant $x_e$ at low redshifts can be achieved without any significant increase in the free electron fraction at redshift $z \sim 600$ by choosing a  small enough $m_\chi$ so that the velocity dispersion prior to structure formation is small. Optical depth constraints are also weaker since there is no increase in the baseline ionization during the cosmic dark ages, unlike in $s$-wave annihilation. 
Instead, the IGM temperature after reionization has been shown to be a significantly more important constraint on the $p$-wave annihilation cross-section than bounds obtained from the CMB power spectrum \cite{Diamanti2014}. Once the effect of structure formation becomes relevant, the late-time energy injection results in significant heating of the IGM. Figure \ref{fig:freeEleFracpWave} shows this behavior for the case of $\chi \chi \to \gamma \gamma$ $p$-wave annihilation with $\sigma v_{\text{ref}}$ between \SI{3E-24}{cm^3 s^{-1}} and \SI{3E-22}{cm^3 s^{-1}}. At large enough cross-sections, $T_{\text{IGM}}$ after reionization exceeds the limits set by equation (\ref{eqn:TIGMConstraints}). 

\begin{figure*}[t!]
	\subfigure{
		\includegraphics[scale=0.59]{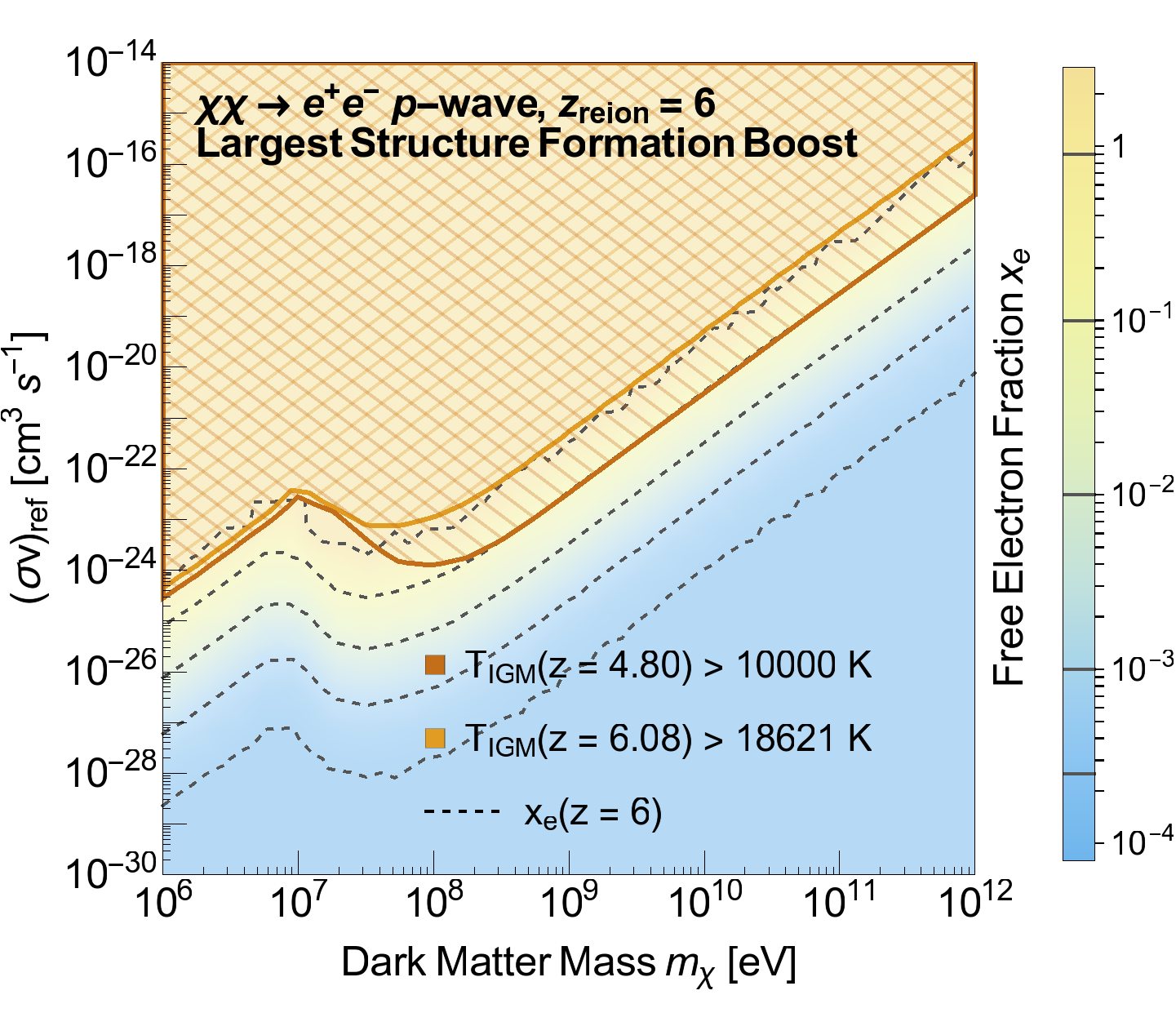}
	}
	\subfigure{
		\includegraphics[scale=0.59]{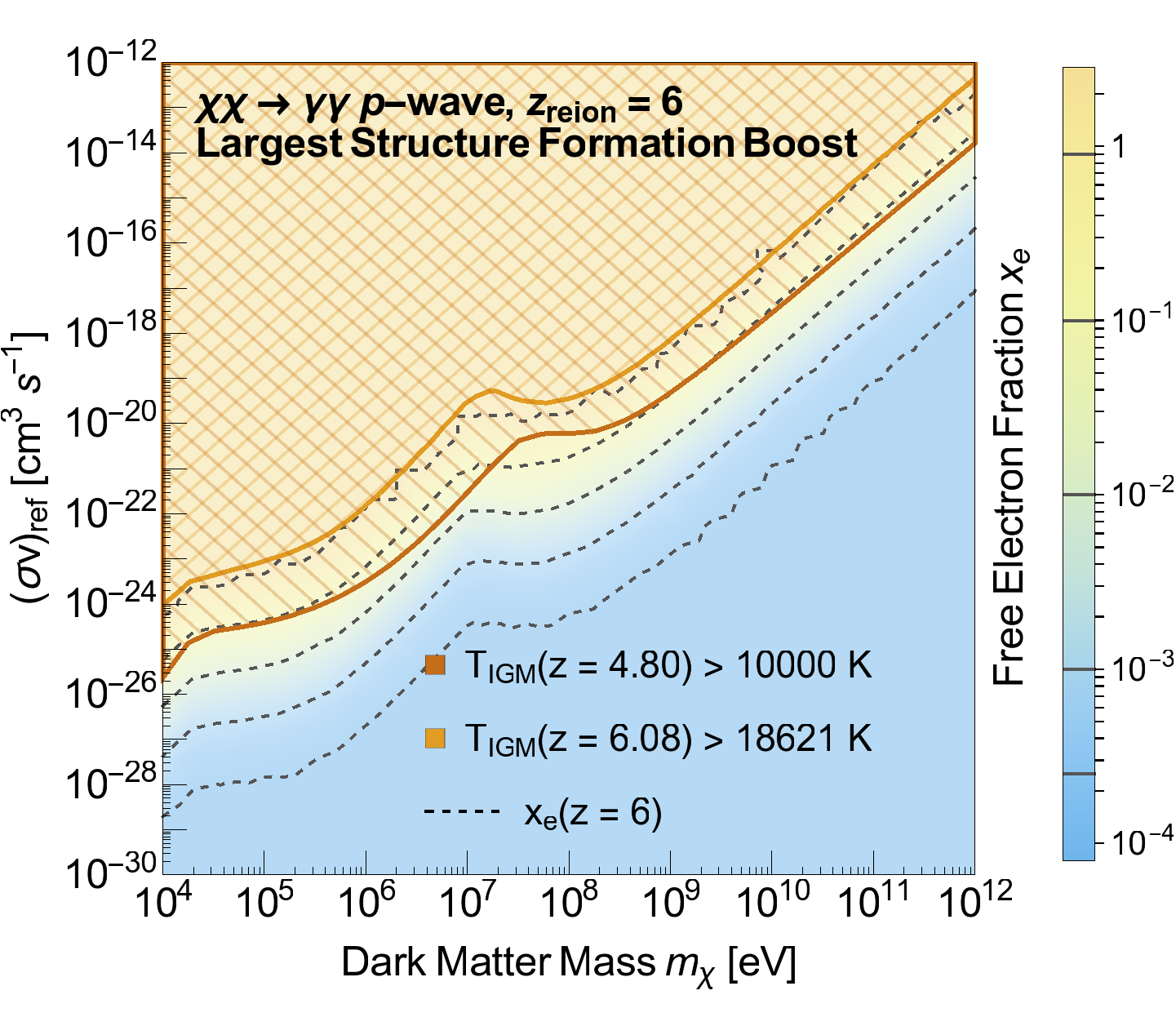}
	}
	\caption{\footnotesize{DM contribution to reionization for $\chi \chi \to e^+e^-$ (left) and $\chi \chi \to \gamma \gamma$ (right) $p$-wave annihilation, benchmark scenario. The hatched regions correspond to parameter space ruled out by $T_{\text{IGM}}(z = 4.80) < \SI{10000}{K}$ (red) and $T_{\text{IGM}}(z = 6.08) < \SI{18621}{K}$ (orange) respectively. The color density plot shows the DM contribution to $x_e$ just prior to reionization at $z = 6$, with contours (black, dashed) shown for a contribution to $x_e(z = 6) = $ 0.025\%, 0.1\%, 1\%, 10\% and 90\% respectively.}}
	\label{fig:xeConstraintsPlot_pWave}
\end{figure*}

Figure \ref{fig:xeConstraintsPlot_pWave} shows $x_e(z=6)$ just prior to reionization for our benchmark scenario in the $(\sigma v)_{\text{ref}}$ - $m_\chi$ parameter space, as well as the excluded parameter space due to constraints from $T_{\text{IGM}}(z=6.08)$ and $T_{\text{IGM}}(z=4.8)$. The same results on the parameter space of $(\sigma v)_{\text{ref}}/m_\chi$ and injection energy of the annihilation products are shown in Appendix \ref{app:additionalConstraints}. Masses above \SI{100}{MeV} for $\chi \chi \to e^+ e^-$ and almost all $m_\chi$ for $\chi \chi \to \gamma \gamma$ are excluded by the benchmark IGM temperature constraint, $\log_{10} T_{\text{IGM}}(z=4.8) < 4.0$. The most likely region in parameter space that can still result in reionization is in the $\chi \chi \to e^+e^-$ channel with $m_\chi < \SI{100}{MeV}$ and $(\sigma v)_{\text{ref}}$ between $10^{-25}$ and $10^{-23} \SI{}{cm^3 s^{-1}}$, and in the $\chi \chi \to \gamma \gamma$ channel with $m_\chi \sim \SI{100}{MeV}$ and $(\sigma v)_{\text{ref}} \sim 10^{-21} \SI{}{cm^3 s^{-1}}$. These cross-sections are much larger than a thermal relic cross-section, but can be accommodated in a large variety of DM models, including any non-thermally produced DM or forbidden DM \cite{DAgnolo2015}. 

The sudden relaxation of the $T_{\text{IGM}}$ constraints below $m_\chi \sim \SI{100}{MeV}$ and the corresponding decrease in $x_e(z=6)$ for $\chi \chi \to e^+e^-$ deserve a special mention here. DM particles with $m_\chi < \SI{100}{MeV}$ annihilating into electrons lose their energy principally through inverse Compton scattering off CMB photons, which by $z \sim 10$ mainly produces photons close to or below the ionizing threshold for hydrogen. After reionization, photoionization by these secondary photons is suppressed further, as the only remaining neutral species is HeII, which has a larger ionization energy. Thus, only a small fraction of the energy goes into collisional heating (due to secondary electrons) of the IGM, with most of the energy from the DM annihilation being deposited as continuum photons. This results in a decrease in IGM temperature after the reionization redshift. At higher DM masses, in contrast, the lower-redshift IGM temperature bound is significantly more constraining, as the IGM temperature invariably continues to increase even after reionization: the $e^+e^-$ pair produced by the annihilation can now upscatter photons to energies above the ionization threshold of HeII. These photoionization events produce low-energy secondary electrons even after reionization, which in turn can collisionally heat the IGM.

\begin{figure*}[t!]
	\subfigure{
		\includegraphics[scale=0.59]{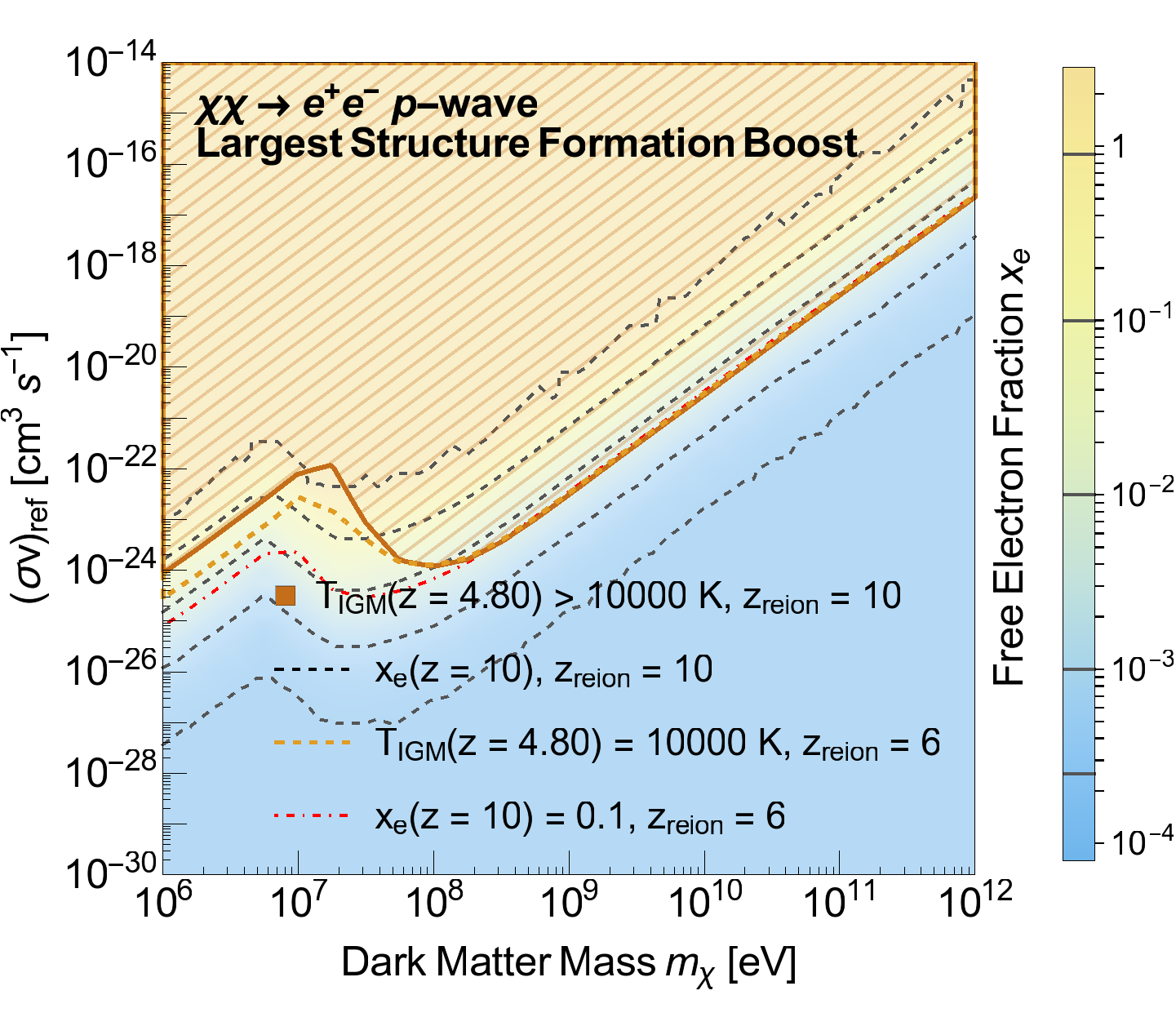}
	}
	\subfigure{
		\includegraphics[scale=0.59]{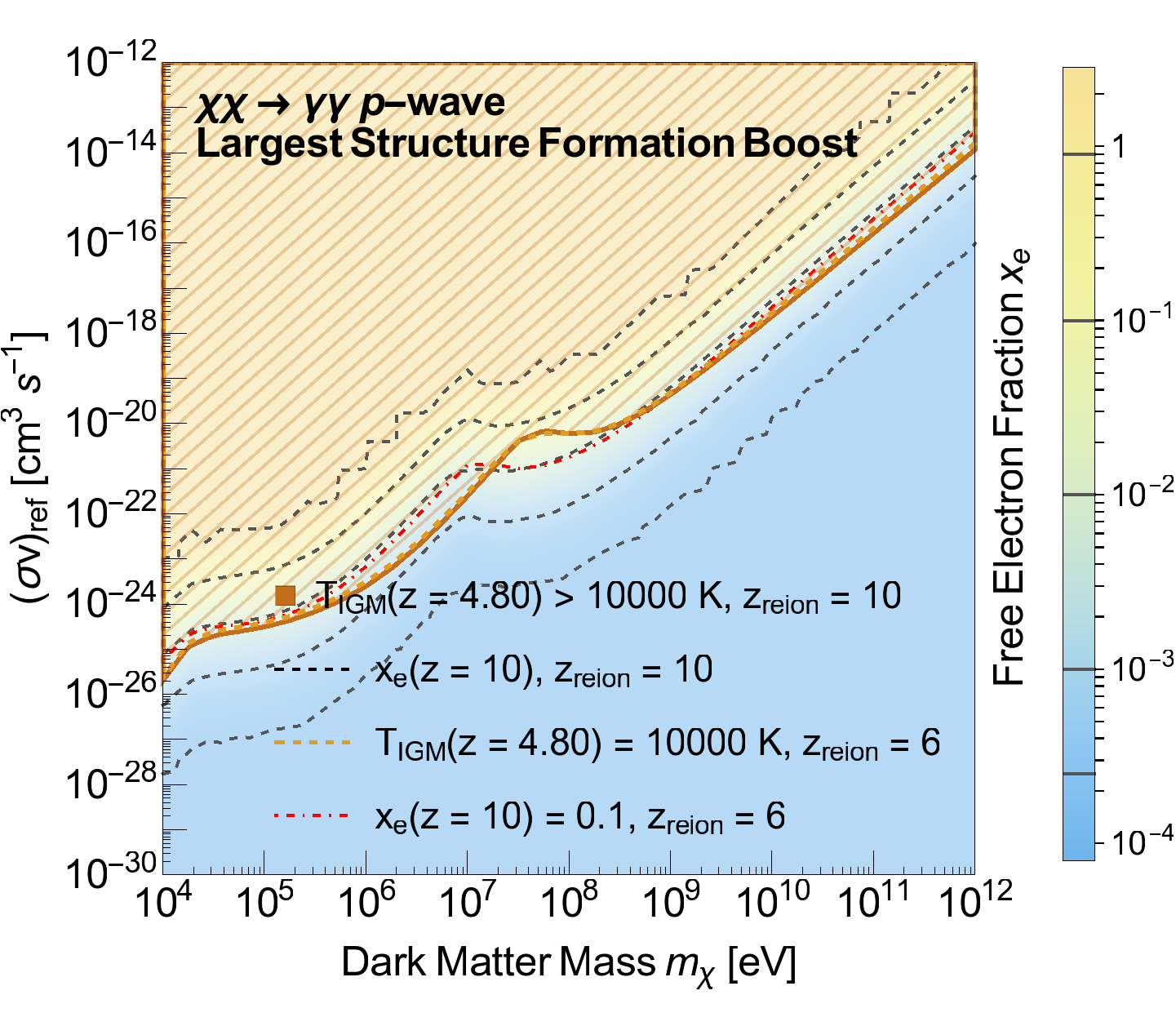}
	}
	\caption{\footnotesize{DM contribution to reionization for $\chi \chi \to e^+e^-$ (left) and $\chi \chi \to \gamma \gamma$ (right) $p$-wave annihilation assuming a different reionization scenario. The color density plot shows the DM contribution to $x_e$ just prior to reionization at $z = 10$, with contours (black, dashed) shown for a contribution to $x_e(z = 10) = $ 0.025\%, 0.1\%, 1\%, 10\% and 90\% respectively. The regions ruled out by the benchmark $T_{\text{IGM}}$ constraint $T_{\text{IGM}}(z=4.80) < \SI{10000}{K}$ assuming reionization at $z = 10$. The red, dot-dashed contour shows $x_e(z = 6) = 10\%$ and the dashed, bold orange contour shows $T_{\text{IGM}}(z=4.80) = \SI{10000}{K}$, both assuming reionization at $z = 6$, for comparison: the region above the IGM temperature contour is ruled out in this case. Note that the 10\% line for reionization at $z = 6$ lies close to the 1\% line for reionization at $z = 10$ in both cases.}}
	\label{fig:xeConstraintsReionSysPlot_pWave}
\end{figure*}

Next, we present our results assuming different reionization redshifts in Figure \ref{fig:xeConstraintsReionSysPlot_pWave}. These results show that the allowed region for $\chi \chi \to e^+e^-$ is shifted upward in cross-section, since a larger cross-section is required to reionize the universe at an earlier redshift, while $T_{\text{IGM}}$ actually becomes less constraining as the IGM temperature now has more time to decrease after reionization. This suggests that the region that permits significant reionization is relatively independent of the reionization condition. The same is not true for the case of $\chi \chi \to \gamma \gamma$: the IGM temperature constraints remain fairly similar, but since we are now extracting $x_e$ at a higher redshift, the overall contribution to $x_e$ by DM decreases. With reionization at $z = 10$, for the $\gamma \gamma$ channel, there is no allowable $m_\chi$ where the contribution to $x_e$ prior to reionization exceeds 10\%.

So far, there is still a range of DM masses with appropriate cross-sections that can reionize the universe at at least the 10\% level through $p$-wave annihilations into $e^+e^-$ ($ m_\chi \lesssim \SI{100}{\mega \eV}$, $(\sigma v)_{\text{ref}} \sim 10^{-24}$ - $10^{-23} \SI{}{ cm^3 s^{-1}}$), and into $\gamma \gamma$ ($m_\chi \sim \SI{100}{\mega \eV}$, $(\sigma v)_{\text{ref}} \sim 10^{-21}$ - $10^{-20} \SI{}{ cm^3 s^{-1}}$) with reionization at $z = 6$. We turn our attention now to two further bounds on $(\sigma v)_{\text{ref}}$ that are relevant to these regions in parameter space. 

First, we consider the cross-section constraints from the CMB power spectrum measurements. Although the results shown in Figure \ref{fig:excludedXSec} are bounds on $\langle \sigma v \rangle$ for $s$-wave annihilation, they also serve as an estimate for the bound on $\langle \sigma v \rangle = (\sigma v)_\text{ref} v^2/v_{\text{ref}}^2$ in the case of $p$-wave annihilations, since the results are only sensitive to the rate of energy deposition into ionization of the IGM during the cosmic dark ages. The main difference with $p$-wave annihilations is that the bound now depends on $v^2$ after recombination and during the cosmic dark ages. $v^2$ is strongly dependent on the primordial ``coldness'' of DM, which in turn depends on the nature of the DM particles, i.e. mass and kinetic decoupling temperature. While DM is coupled to photons, $v^2 \sim 3 T_\gamma/m_\chi$, whereas after decoupling, $v^2 \propto (1+z)^2$. Taking the limit $L(m_\chi)$ on $\langle \sigma v \rangle$ set by the CMB spectrum at a particular DM mass $m_\chi$ as shown in Figure \ref{fig:excludedXSec},
\begin{alignat}{1}
	(\sigma v)_\text{ref} \lesssim 3.7 L(m_\chi) \left(\frac{m_\chi}{\SI{1}{MeV}} \right)^2 \left( \frac{x_{\text{kd}}}{10^{-4}} \right) \left(\frac{\SI{1}{eV}}{T_\gamma} \right)^2,
\end{alignat}
where $x_{\text{kd}} \equiv T_{\text{kd}}/m_\chi$. $T_\gamma$ is some representative CMB temperature after recombination such that the CMB power spectrum is most sensitive to energy injections at the redshift $z$ corresponding to $T_\gamma$ ($z \sim 600$ in the $s$-wave case). 

In the case of $\chi \chi \to e^+e^-$, in the region of parameter space where a significant contribution to reionization can be made, the CMB bounds can rule out these regions if $x_{\text{kd}} \lesssim 10^{-2} - 10^{-1}$ for $m_\chi \sim \SI{1}{MeV}$ and $x_{\text{kd}} \lesssim 10^{-6}$ for $m_\chi \sim \SI{100}{MeV}$ (we have set $T_\gamma = \SI{0.14}{eV}$ as a representative value), while for 100 MeV DM annihilating $\chi \chi \to \gamma \gamma$, we have $x_{\text{kd}} \sim 10^{-3} - 10^{-2}$. Thus for the CMB bounds to exclude these regions, we would need to have $T_\mathrm{kd} \lesssim 100$ keV, and in some cases it would need to be much lower (at the sub-keV scale).

Values of $T_{\text{kd}}$ higher than these bounds are consistent (and expected) in a large variety of DM models, e.g. $T_{\text{kd}} \sim \SI{}{MeV} (m_\chi/\SI{}{GeV})^{2/3}$ for neutralino DM \cite{Chen2001}, and $T_{\text{kd}} \sim \SI{2.02}{MeV} (m_\chi/\SI{}{GeV})^{3/4}$ for DM-lepton interactions of the form $(1/\Lambda^2)(\bar{X} X)( \bar{l} l)$ for some interaction mass scale $\Lambda$, giving rise to $p$-wave suppressed cross-sections \cite{Shoemaker2013,Diamanti2014}. In general, $T_\mathrm{kd}$ below the scale of the electron mass is unusual, as the only relativistic species available to maintain kinetic equilibrium are photons and neutrinos.\footnote{Models such as neutrinophilic DM \cite{Shoemaker2013,VandenAarssen2012} can, however, exhibit such a behavior.} The CMB bounds therefore place few constraints on our parameter space for $p$-wave annihilation, in stark contrast to the $s$-wave case. 

\begin{figure*}[t!]
	\subfigure{
		\includegraphics[scale=0.59]{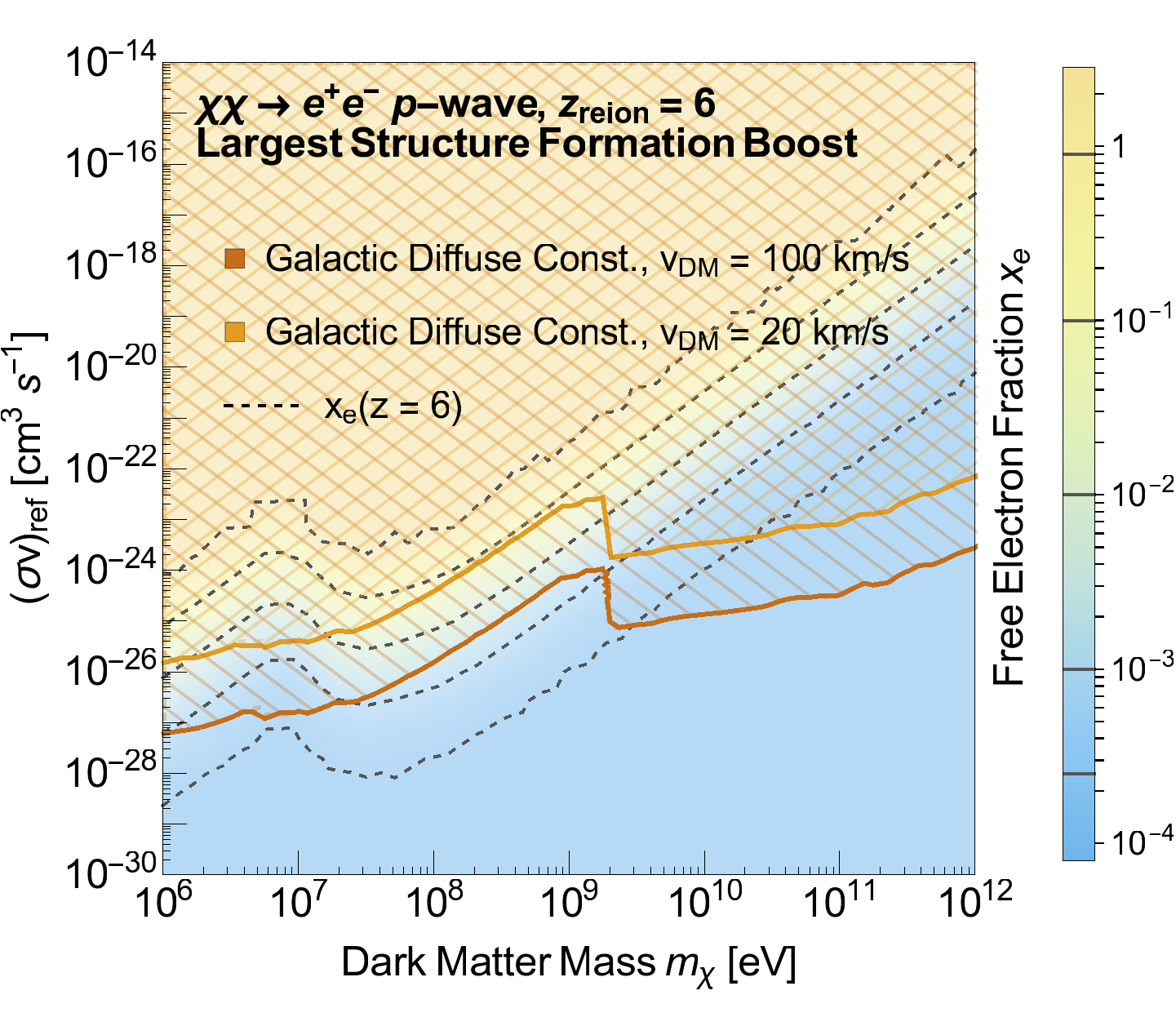}
	}
	\subfigure{
		\includegraphics[scale=0.59]{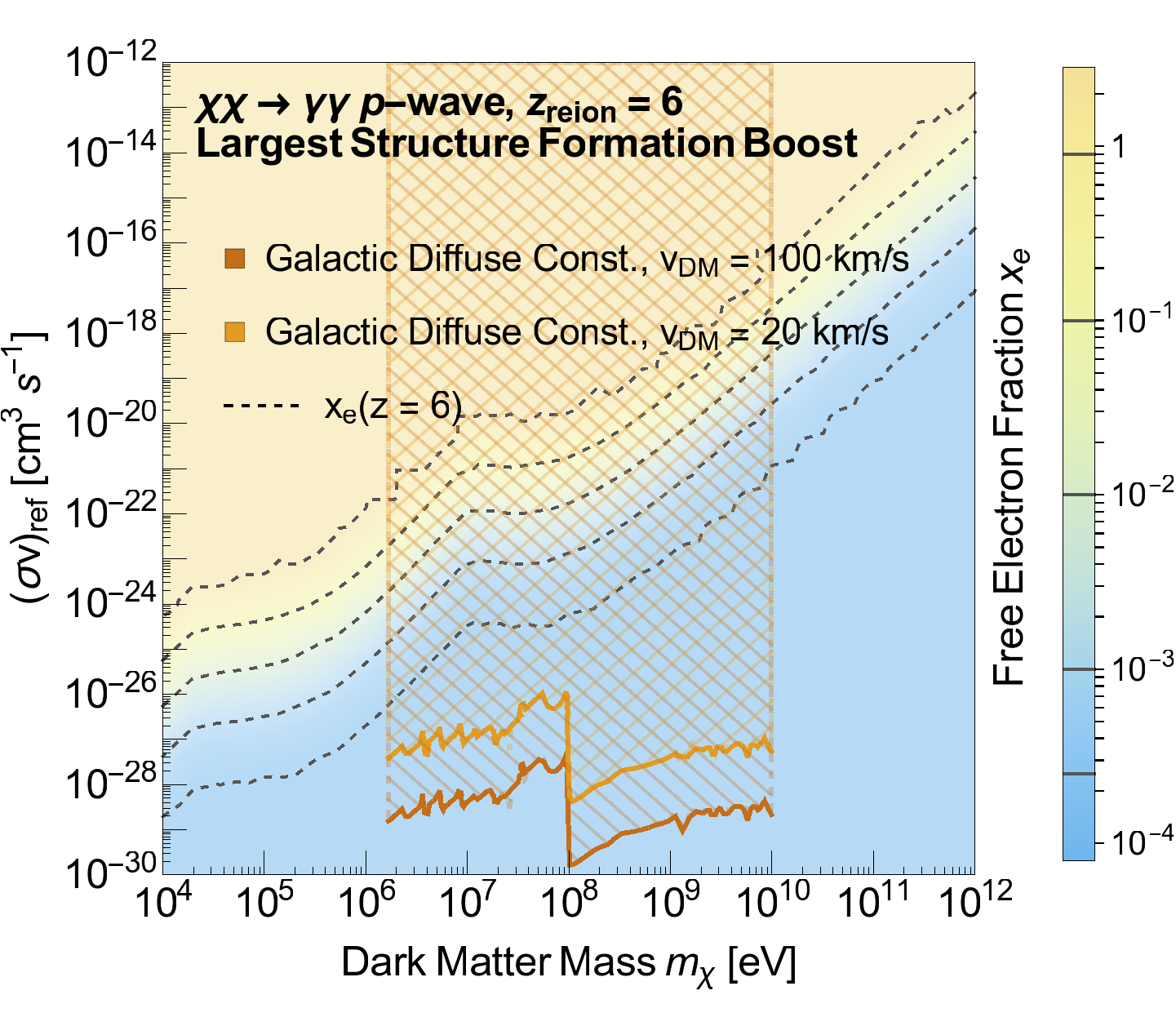}
	}
	\caption{\footnotesize{DM contribution to reionization for $\chi \chi \to e^+e^-$ (left) and $\chi \chi \to \gamma \gamma$ (right) $p$-wave annihilation, together with limits from the galactic diffuse background. The color density plot shows the DM contribution to $x_e$ just prior to reionization at $z = 6$, with contours (black, dashed) shown for a contribution to $x_e(z = 6) = $ 0.025\%, 0.1\%, 1\%, 10\% and 90\% respectively. These constraints are dependent on the dispersion velocity $v_{\text{DM}}$: we show the constraints obtained assuming that $v_{\text{DM}} = \SI{100}{km s^{-1}}$ (red hatched region) and \SI{20}{km s^{-1}} (orange hatched region). The $\chi \chi \to e^+e^-$ constraints are obtained from \cite{Essig2013,Massari2015}, while the $\chi \chi \to \gamma \gamma$ limits are from \cite{Boddy2015,Albert2014}.}}
	\label{fig:xeConstraintsGalacticPlot_pWave}
\end{figure*}

Next, we look at $p$-wave constraints from gamma ray flux measurements of the galactic diffuse background. The derived constraints from the galactic diffuse background are shown in Figure \ref{fig:xeConstraintsGalacticPlot_pWave}. For $\chi \chi \to e^+ e^-$, final state radiation produced as part of the annihilation process in the Milky Way halo produces gamma ray photons that can be measured by these experiments, placing an upper bound on the rate of $p$-wave annihilation into $e^+e^-$ for DM masses of up to \SI{10}{GeV} in the Milky Way. Constraints derived in \cite{Essig2013}  from a combination of data from INTEGRAL, COMPTEL and Fermi set a limit of $\langle \sigma v \rangle \lesssim 10^{-27}\SI{}{cm^3 s^{-1}}$ for $m_\chi \lesssim \SI{100}{\mega \eV}$. This was derived assuming an NFW profile, which is a relatively conservative choice for these experiments: the constraints fluctuate by a factor of a few if different DM halo profiles are chosen. All of the measured photon flux is conservatively attributed to DM annihilation in the galaxy halo only, without accounting for extragalactic DM annihilation or other more conventional sources like inverse Compton scattering off starlight or synchrotron radiation. 

The translation of these velocity-averaged cross-section bounds to constraints on $(\sigma v)_{\text{ref}}$ depends on the velocity dispersion $v_{\text{DM}}$ around the solar circle. Given a measured photon flux, a larger $v_{\text{DM}}$ would place a stronger constraint on $(\sigma v)_{\text{ref}}$, since the photon flux is proportional to the annihilation rate, which is in turn proportional to $(\sigma v)_{\text{ref}} v_{\text{DM}}^2$ in a $p$-wave process. Because of this, the constrained $(\sigma v)_{\text{ref}}$ is proportional to $1/v_{\text{DM}}^2$. However, in order for some region of parameter space with more than a 10\% contribution to reionization from DM to be allowed, the dispersion velocity in the solar circle needs to satisfy $v_{\text{DM}} < \SI{20}{km s^{-1}}$, which is significantly smaller than the local velocity of the solar circle and is hence unrealistic \cite{Cerdeno:2010jj}. 

Similar results hold for $\chi \chi \to \gamma \gamma$, where searches for sharp spectral features such as lines or boxes in the galactic diffuse gamma ray background place strong bounds on the annihilation cross section of this process. By requiring the number of counts from $\chi \chi \to \gamma \gamma$ in each energy bin in the spectrum to not exceed the measured number of counts by $2 \sigma$, the gamma ray spectrum from COMPTEL and EGRET can be used to set an upper limit of $\langle \sigma v \rangle \lesssim 10^{-27} \SI{}{cm^3 s^{-1}}$ for $m_\chi \sim \SI{100}{MeV}$ \cite{Boddy2015}, with a similar analysis using Fermi data \cite{Albert2014} giving a limit of $\langle \sigma v \rangle \lesssim 10^{-29} \SI{}{cm^3 s^{-1}}$ for $m_\chi \gtrsim \SI{100}{MeV}$. This means that the dispersion velocity required for a 10\% contribution to reionization is $v_{\text{DM}} \sim \SI{0.1}{km s^{-1}}$, which is once again unrealistic.

Although we have freely used the constraints for $\langle \sigma v \rangle$ to directly set constraints on $(\sigma v)_{\text{ref}}$, some caution must be taken when doing so. The contribution of DM annihilations to the observed photon flux measured by a detector is due to annihilations all along the line-of-sight. In order to set constraints on DM annihilation from gamma ray flux measurements, the appropriate function of the DM density and velocity must therefore be averaged along the line-of-sight. $\langle \sigma v \rangle$ bounds are frequently set by averaging over the DM density, but without taking into account the velocity dispersion of the Milky Way halo. Without performing this average, $\langle \sigma v \rangle$ bounds are implicitly assumed to be for $s$-wave processes only. 

However, as we demonstrate in Appendix \ref{app:JFactor}, averaging over the velocity dispersion as well as the density appears to change the $\langle \sigma v \rangle$ bounds for $p$-wave annihilation by less than a factor of 2 under many different assumptions. These bounds would need to relax by at least 2 orders of magnitude for $\chi \chi \to e^+e^-$ and 4 orders of magnitude for $\chi \chi \to \gamma \gamma$ to allow any significant contribution to reionization at all.

Overall, the possible contribution of $p$-wave DM annihilation to reionization appears to be constrained to the $<10\%$ level across all of the masses and injection species considered here. At $m_\chi \gtrsim \SI{10}{GeV}$, this contribution is limited by $T_\text{IGM}$ measurements, while for $m_\chi \lesssim \SI{10}{GeV}$, any allowed parameter space with more than 10\% contribution to reionization after accounting for $T_{\text{IGM}}$ appears to be ruled out by observations of the galactic diffuse emission gamma ray spectrum.

\subsection{Decay}

\begin{figure*}[t!]
	\subfigure{
		\includegraphics[scale=0.59]{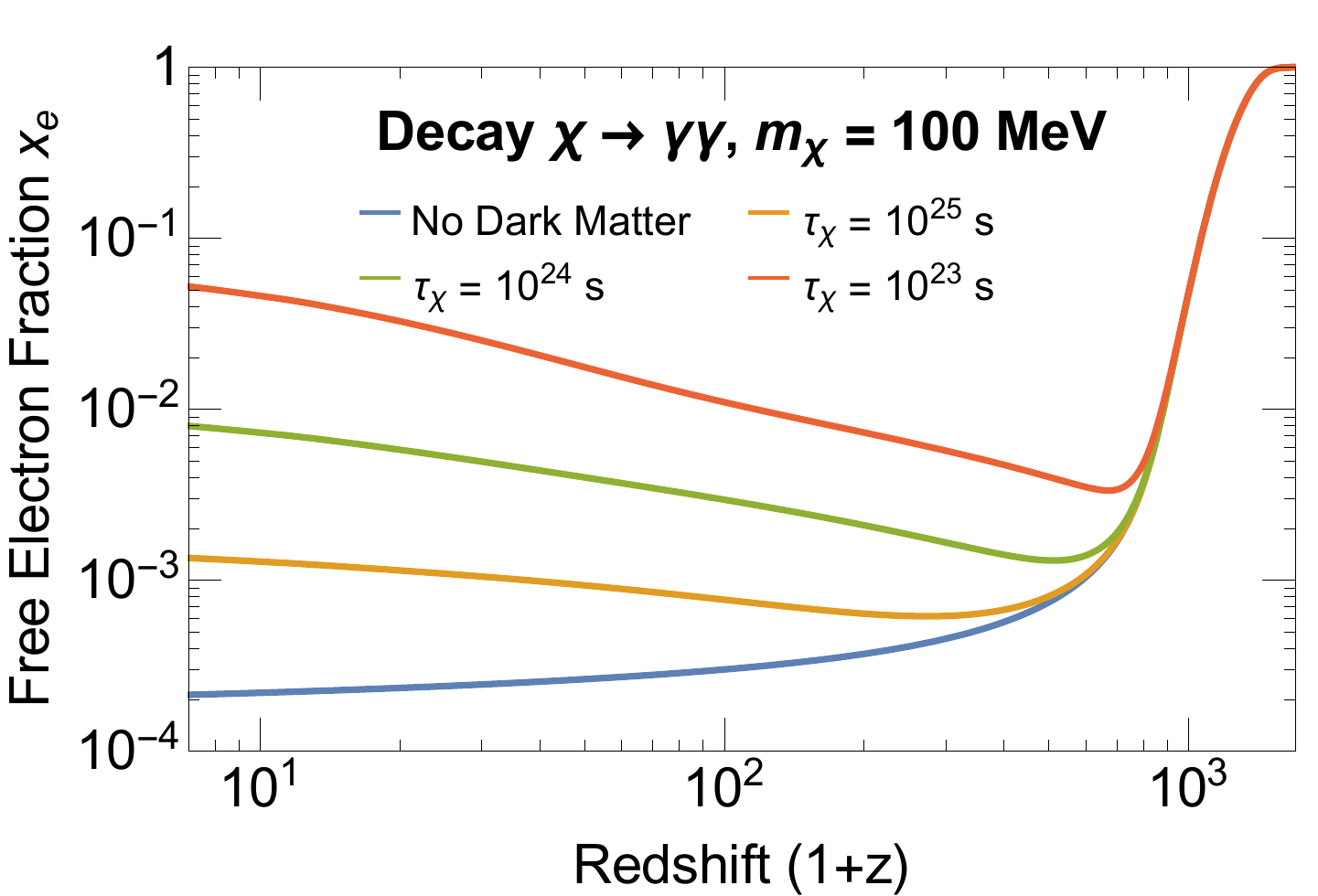}
	}
	\subfigure{
		\includegraphics[scale=0.58]{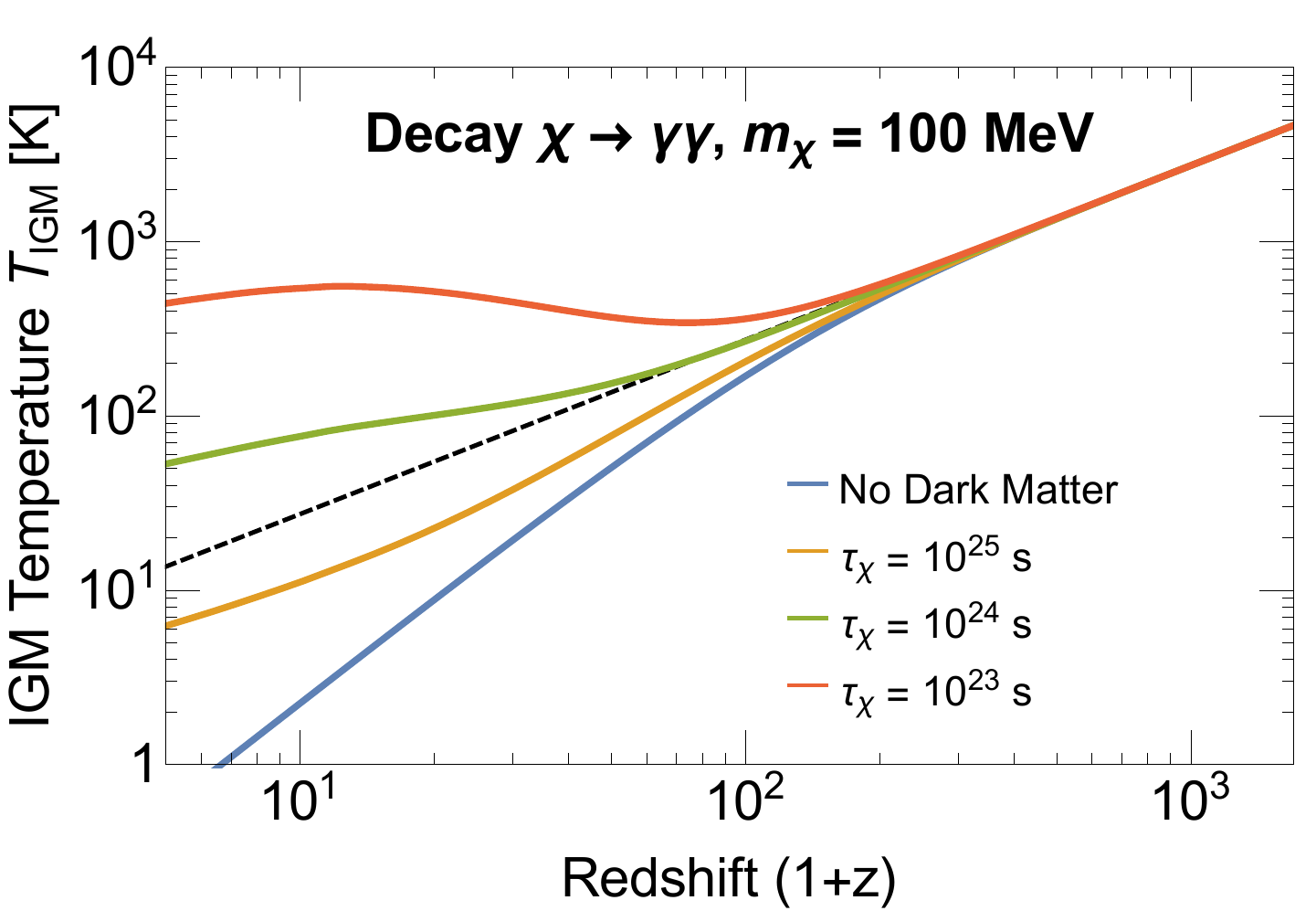}
	}
	\caption{\footnotesize{Integrated free electron fraction $x_e$ and IGM temperature $T_{\text{IGM}}$ for $\chi \to \gamma \gamma$ decays ($m_\chi = \SI{100}{\mega \eV}$) with (from bottom to top): no DM, $\tau_\chi =  10^{25}$ s, $10^{24}$ s and $10^{23}$\SI{}{ s} respectively. The CMB temperature is shown as a dashed line for reference. No reionization is assumed. }}
	\label{fig:freeEleFracDecay}
\end{figure*}

Figure \ref{fig:freeEleFracDecay} shows $x_e(z)$ and $T_{\text{IGM}}(z)$ for $m_\chi = \SI{100}{MeV}$ DM undergoing $\chi \to \gamma \gamma$ decays (each photon now has an energy of \SI{50}{MeV}) with various representative decay lifetimes, which are typical for other masses and decay modes. Compared to $s$-wave annihilation, the energy injection rate in decays is not dependent on structure formation, and the $(1+z)^3$ redshift dependence for decays (compared to $(1+z)^6$ for $s$-wave annihilation) means that the energy injection is less weighted toward earlier redshifts. This leads to a steady rise in $x_e$ from immediately before recombination to the present day. 

\begin{figure*}[t!]
	\subfigure{
		\includegraphics[scale=0.59]{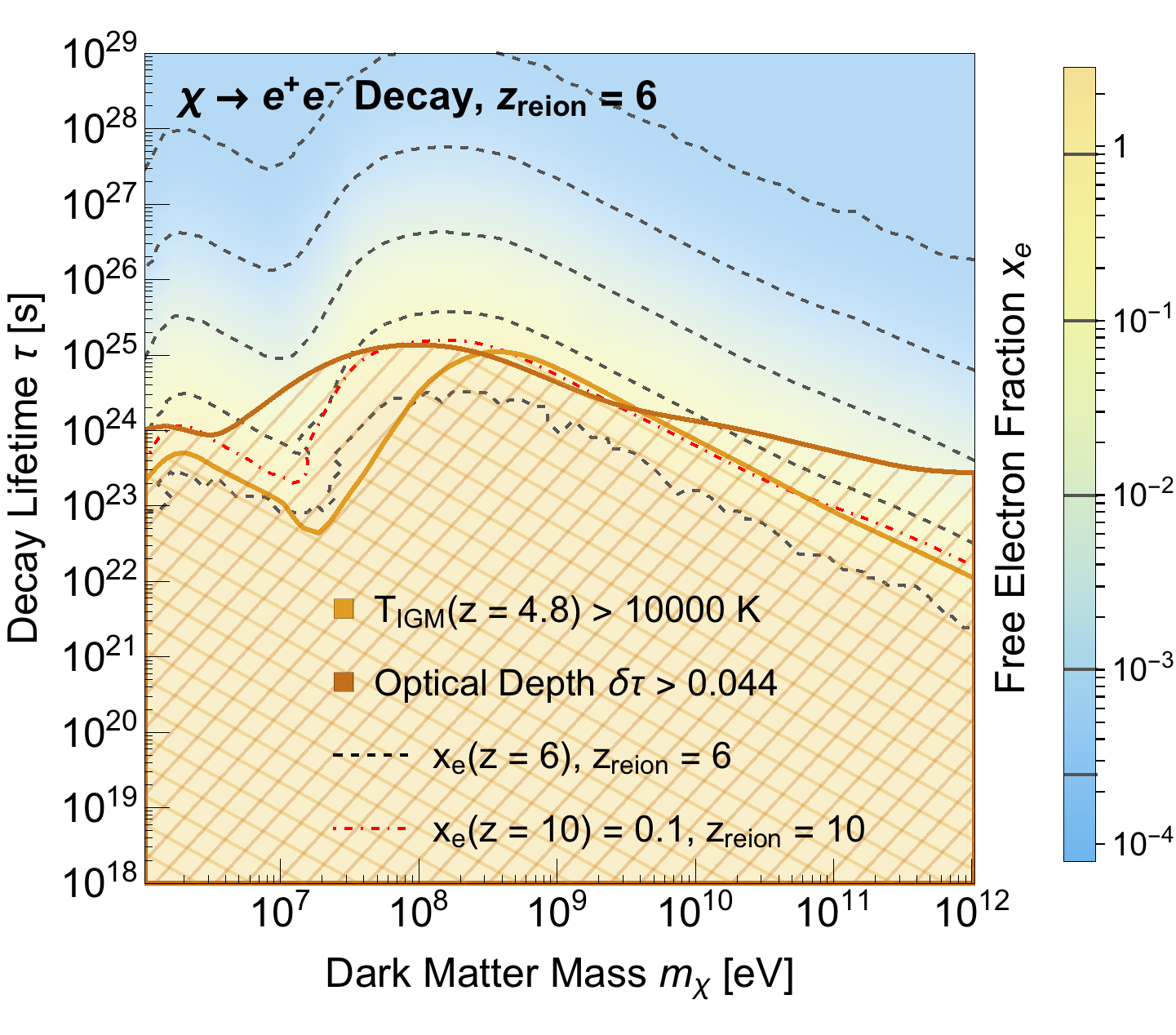}
	}
	\subfigure{
		\includegraphics[scale=0.59]{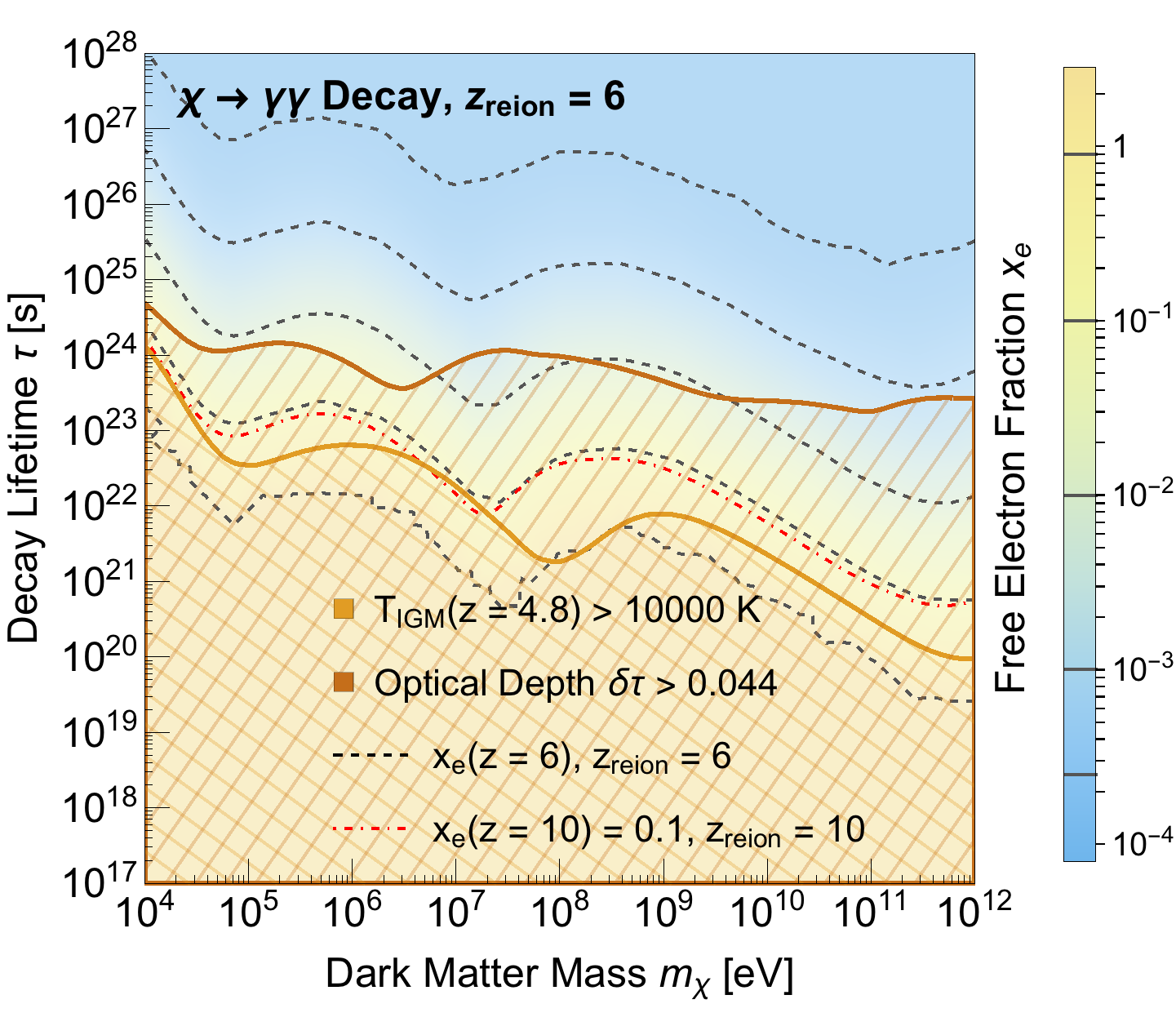}
	}
	\caption{\footnotesize{DM contribution to reionization for $\chi \to e^+e^-$ (left) and $\chi \to \gamma \gamma$ (right) decays, benchmark scenario. The hatched regions correspond to parameter space ruled out by the optical depth (red) and the IGM temperature constraint $T_{\text{IGM}}(z=4.80) < \SI{10000}{K}$ (orange) respectively. The color density plot shows the DM contribution to $x_e$ just prior to reionization at $z = 6$, with contours (black, dashed) shown for a contribution to $x_e(z = 6) = $ 0.025\%, 0.1\%, 1\%, 10\% and 90\% respectively. We have also shown $x_e(z = 10) = 10\%$ when reionization occurs at $z = 10$ (red, dot-dashed contour). The optical depth limits are similar in both reionization scenarios, while the $T_{\text{IGM}}$ limits are similar between \SI{100}{MeV} and \SI{10}{GeV}, where they are more constraining than the optical depth limits.}}
	\label{fig:xeConstraintsPlot_decay}
\end{figure*}

Optical depth constraints play an important role in placing bounds on the decay lifetime: with no structure formation boost, the only way for significant ionization at low redshifts to occur is for $x_e$ to be relatively high throughout the cosmic dark ages, contributing significantly to the optical depth. Figure \ref{fig:xeConstraintsPlot_decay} shows the region of the ($\tau_\chi$,$m_\chi$) parameter space where DM can contribute significantly to reionization, as well as the constraints on the decay lifetime coming from IGM temperature and the optical depth. Significant reionization occurs for relatively longer decay lifetimes for masses where $f_{\text{H ion.}}(z)$ is large at low redshifts. However, both optical depth and IGM temperature constraints rule out large parts of the allowed parameter space for $\chi \to e^+e^-$ and all of the parameter space for $\chi \to \gamma \gamma$ at the 10\% level of contribution to reionization, with the $T_{\text{IGM}}$ bounds being more effective than optical depth for the $m_\chi \sim \SI{100}{MeV} - \SI{10}{GeV}$ range for $\chi \chi \to e^+e^-$. 

Figure \ref{fig:xeConstraintsPlot_decay} also shows the same results after considering different reionization conditions. Once again, the optical depth constraints change very little with respect to reionization redshift, while the $T_{\text{IGM}}$ constraints are very similar in both reionization scenarios in the region where they are stronger than the optical depth, and we can hence simply compare the $x_e$ contributions with the $\delta \tau$ and $T_{\text{IGM}}$ constraints at $z_{\text{reion}} = 6$. As before, earlier reionization makes it more difficult for DM to contribute to $x_e$ just prior to reionization. For $\chi \to e^+e^-$, almost all decay lifetimes and masses which previously resulted in a 10\% contribution to reionization now result in a contribution below 10\% when the redshift of reionization is changed to $z = 10$, while the results for $z = 6$ and $z = 10$ for $\chi \to \gamma \gamma$ are similar. 

\begin{figure}[t!]
	\centering
	\includegraphics[scale=0.6]{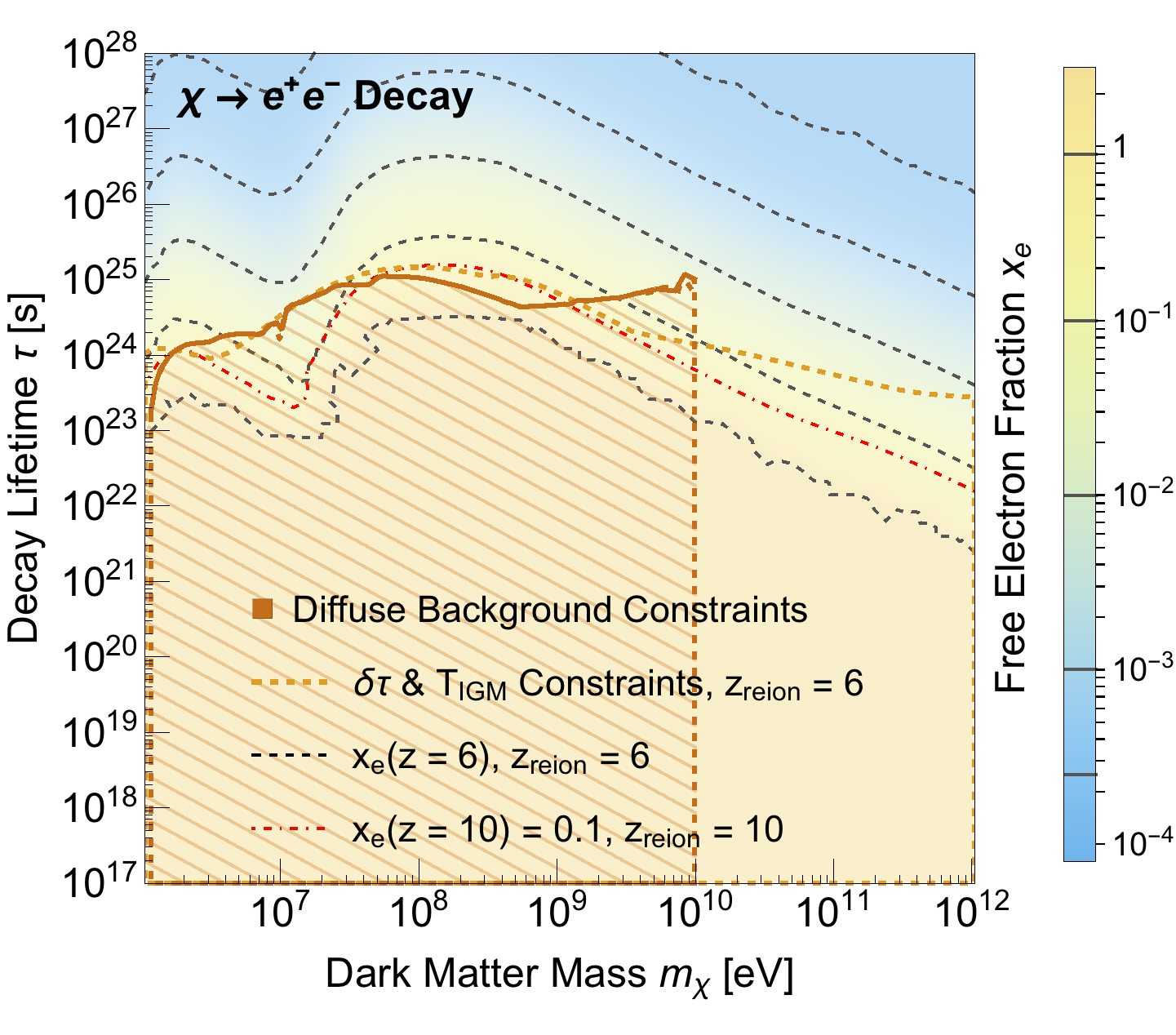}
	\caption{\footnotesize{DM contribution to reionization for $\chi \to e^+e^-$ decays, benchmark scenario, including constraints from the galactic diffuse background (red contour, hatched) derived from \cite{Essig2013}. The color density plot shows the DM contribution to $x_e$ just prior to reionization at $z = 6$, with contours (black, dashed) shown for a contribution to $x_e(z = 6) = $ 0.025\%, 0.1\%, 1\%, 10\% and 90\% respectively. We have also shown $x_e(z = 10) = 10\%$ when reionization occurs at $z = 10$ (red, dot-dashed contour) for comparison. The combined constraint from both optical depth $\delta \tau < 0.044$ and IGM temperature $T_{\text{IGM}}(z = 4.8) < \SI{10000}{K}$ (orange, dashed contour) is shown as well, with regions below this contour ruled out. These limits are almost identical in either reionization scenario.}}
	\label{fig:xeConstraintsGalacticPlot_electron_decay}
\end{figure}

Nevertheless, a contribution to $x_e$ just prior to reionization at more than the 10\% level still remains possible for $\chi \to e^+e^-$ at a DM mass of $m_\chi \sim \SI{100}{MeV} - \SI{10}{GeV}$, $\tau_\chi \sim 10^{24} - 10^{25} \SI{}{s}$, as well as $m_\chi \sim \SI{1}{MeV}$, $\tau_\chi \sim 10^{24} \SI{}{s}$ in the benchmark reionization scenario. As with $p$-wave annihilation, the galactic diffuse background provides an additional constraint on the decay lifetime. These constraints are derived in a similar way to the $p$-wave case, i.e. by conservatively assuming that all of the diffuse gamma ray background comes from FSR from the DM decay. However, unlike with $p$-wave annihilation, the diffuse background constraints are of the same order as the optical depth bounds that we have set here. Figure \ref{fig:xeConstraintsGalacticPlot_electron_decay} shows these constraints superimposed on Figure \ref{fig:xeConstraintsPlot_decay}, showing that none of the experimental constraints are able to rule out the possibility of a more than 10\% contribution to $x_e$ prior to reionization in the $m_\chi \sim \SI{10}{} - \SI{100}{MeV}$, $\tau_\chi \sim 10^{25}\SI{}{s}$ and $m_\chi \sim \SI{1}{MeV}$, $\tau_\chi \sim 10^{24} \SI{}{s}$ regions of parameter space. This conclusion still holds true for a different redshift of reionization for $m_\chi \sim \SI{100}{MeV}$. 

\begin{figure*}[t!]
	\subfigure{
		\includegraphics[scale=0.59]{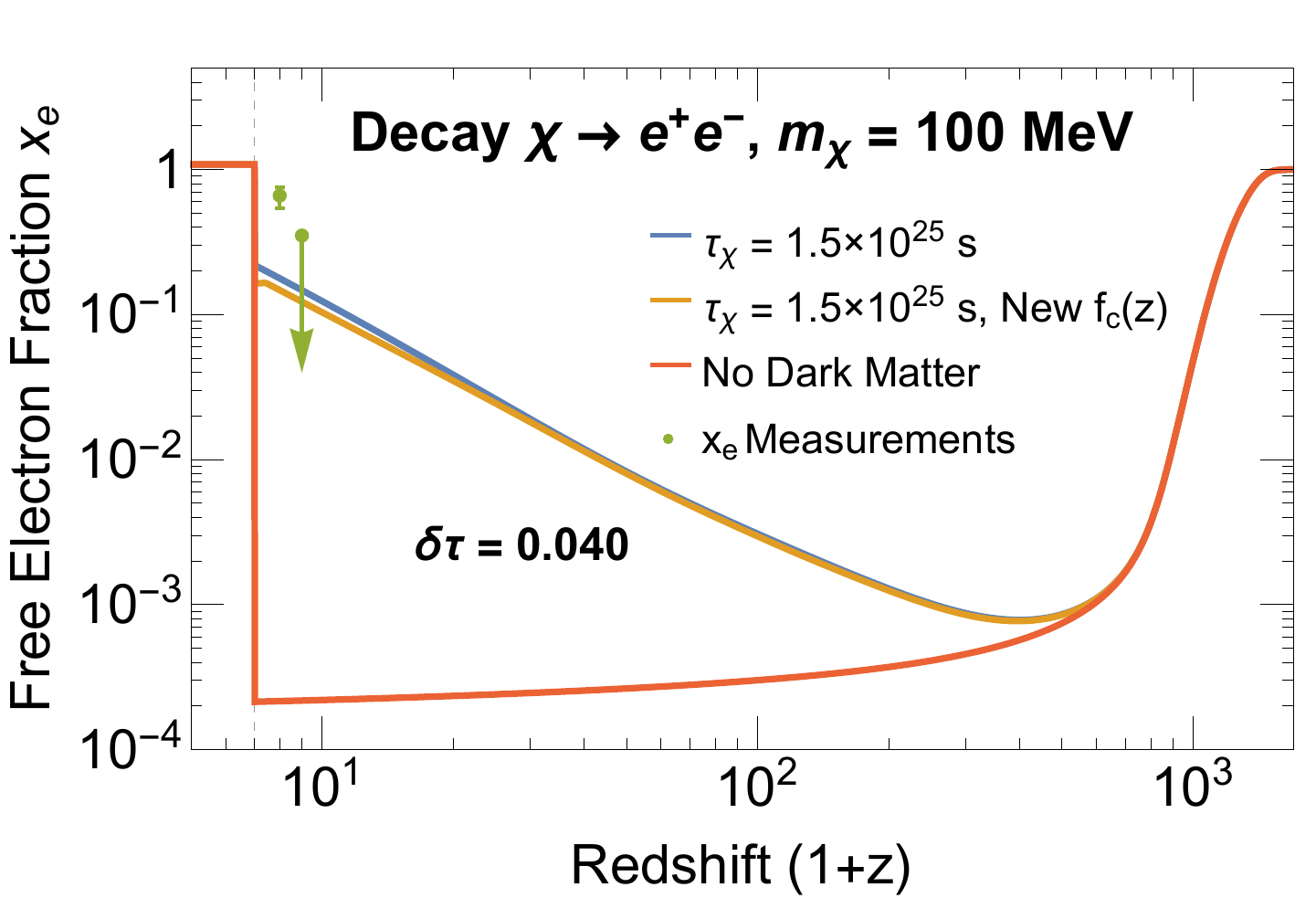}
	}
	\subfigure{
		\includegraphics[scale=0.58]{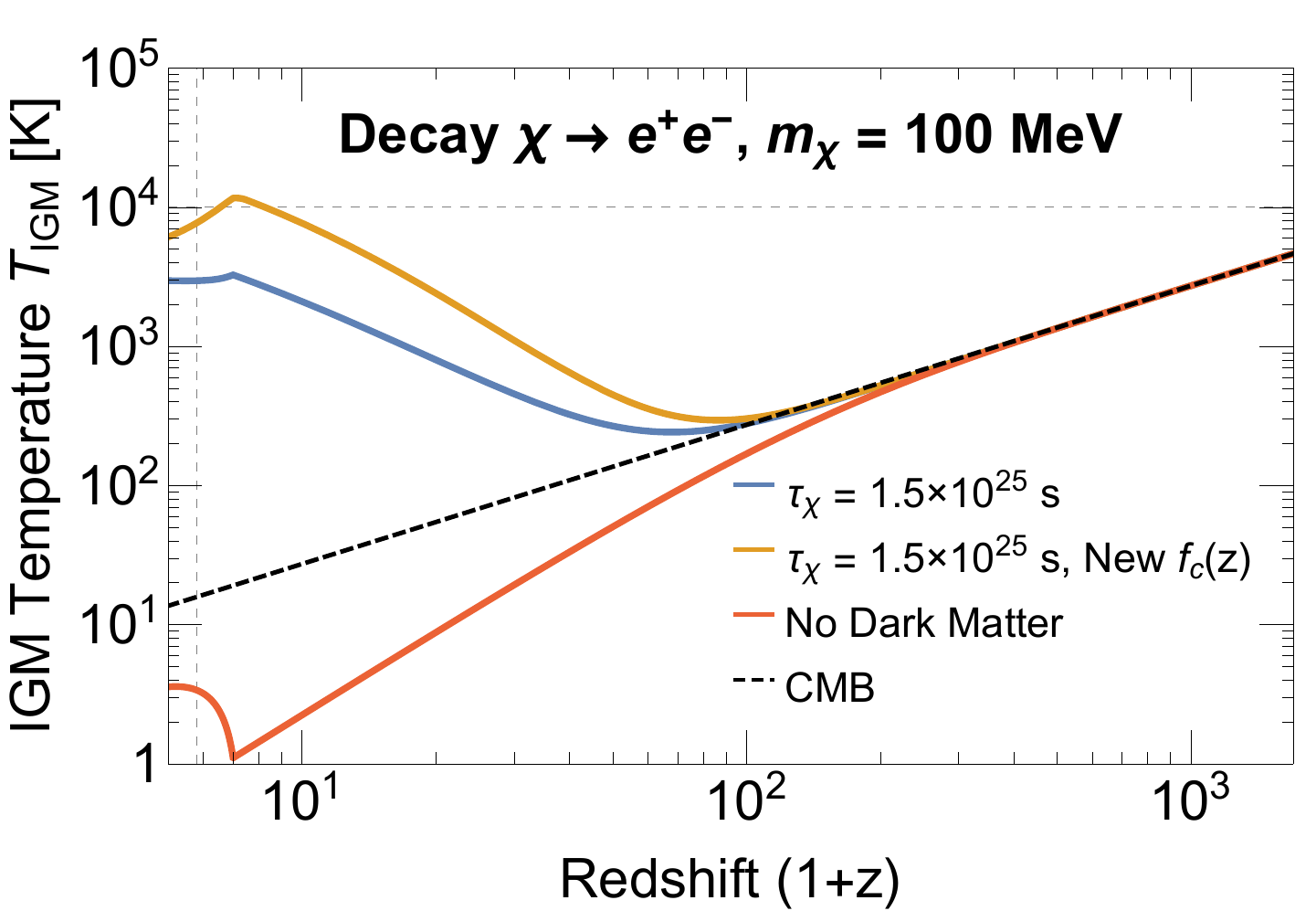}
	}
	\caption{\footnotesize{Integrated free electron fraction $x_e$ and IGM temperature $T_{\text{IGM}}$ for $\chi \to e^+e^-$ decays ($m_\chi = \SI{100}{\mega \eV}$) with: (red) no DM; (blue) $\tau_\chi = \SI{1.5e25}{s}$ with the default $f_c(z)$; (orange) $  \SI{1.5e25}{s}$ with $f_c(z)$ computed using $x_e(z)$ obtained from the default $f_c(z)$ shown in blue. The green points and error bars show the observational limits for $x_e$ near reionization \cite{Schenker2014}. The CMB temperature (bold, dashed line) and $T_{\text{IGM}}(z = 4.8) = \SI{10000}{K}$ (dashed line) are shown for reference. Reionization at $z = 6$ is assumed.}}
	\label{fig:freeEleFracDecayAllowedRegion}
\end{figure*}

The blue curve in Figure \ref{fig:freeEleFracDecayAllowedRegion} shows $x_e(z)$ and $T_{\text{IGM}}(z)$ assuming reionization at $z = 6$, with $m_\chi = \SI{100}{MeV}$ and $\tau_\chi = \SI{1.5e25}{s}$, parameters which lie in one of the allowed regions found above. Reionization at $z = 6$ causes the behavior of $T_{\text{IGM}}$ to change abruptly due to the instantaneous change of $x_e$. Just before reionization, $x_e(z = 6) \sim 0.2$, with the integrated optical depth being $\delta \tau = 0.040$, which lies within the allowed limit. $T_{\text{IGM}}(z = 4.8)$ lies below the lower limit of the $T_{\text{IGM}}$ constraint, but as we have previously explained, $T_{\text{IGM}}$ is always underestimated with the default ionization history. 

We have also performed the integration of $x_e(z)$ and $T_{\text{IGM}}(z)$ with $f_c(z)$ derived from the ionization history that we obtained above. Since $f_c(z)$ as calculated from the default ionization history overestimates $x_e(z)$, using this new $f_c(z)$ ensures that the allowed regions are not ruled out by a more accurate estimate of $x_e(z)$. The result is also shown in orange in Figure \ref{fig:freeEleFracDecayAllowedRegion}. As we expect, this more accurate $f_c(z)$ increases $T_{\text{IGM}}(z)$ and decreases $x_e(z)$ slightly. The contribution to reionization remains the same, while still staying consistent with the $T_{\text{IGM}}(z = 4.8)$ and the optical depth bounds. 

Figure \ref{fig:freeEleFracDecayAllowedRegion} also shows two measurements of $x_e$ from just before reionization obtained by \cite{Schenker2014}, corresponding to
\begin{alignat}{1}
	x_e(z=7) &= 0.66^{+0.12}_{-0.09}, \nonumber \\
	x_e(z=8) &< 0.35.
\label{eqn:schenkerxe}
\end{alignat}
The ionization history for $m_\chi = \SI{100}{MeV}$ and $\tau_\chi = \SI{1.5e25}{s}$ is consistent with the bound from $z=8$, and can be made consistent with the $z=7$ bound with the addition of other sources of ionization between these two redshifts.  

In summary, optical depth constraints as well as bounds from the galactic diffuse background constraints rule out reionization from $\chi \to \gamma \gamma$ and almost rules out reionization from $\chi \to e^+e^-$ at the 10\% level, except for $m_\chi \sim \SI{10}{} - \SI{100}{MeV}$, $\tau_\chi \sim 10^{25} \SI{}{s}$ and $m_\chi \sim \SI{1}{MeV}$, $\tau_\chi \sim 10^{24} \SI{}{s}$. The former region remains viable even under the different reionization scenarios considered here. 

\section{Conclusion}
\label{sec:Conclusion}

We have studied the potential impact of $s$-wave annihilation, $p$-wave annihilation and decay of DM to $e^+e^-$ and $\gamma \gamma$ on the process of reionization. Using the latest calculations for the fraction of the energy deposition rate in channel $c$ to the energy injection rate at redshift $z$, $f_c(z)$, we have determined the free electron fraction $x_e$ and IGM temperature $T_{\text{IGM}}$ as a function of redshift. We have extended the $f_c(z)$ calculation from $1+z = 10$ down to $1+z = 4$ by assuming three different reionization scenarios and determining the total amount of energy deposited as ionization of HeII, IGM heating and continuum photons once reionization occurs. 

We have also considered multiple detailed structure formation models in order to accurately calculate the $s$-wave and $p$-wave annihilation rates. This modeling accounts for the formation of DM haloes and their subhaloes, with abundance and internal properties that are consistent with current cosmological simulations. It also considers the uncertainties at the smallest scales (corresponding to low-mass haloes, $<10^8$~M$_\odot$, devoid of gas and stars) that cannot be resolved in current simulations in a full cosmological setting, but that are very relevant in predicting the annihilation rate in the case of $s$-wave self-annihilation. This is particularly important at low redshifts: at $z\sim10$, the uncertainty in $\rho_{\rm eff}^2$ is $\sim5$ for the case of $s$-wave self-annihilation (see Figure \ref{fig_rho_eff}). On the other hand, for $p$-wave self-annihilation, the uncertainties in the unresolved regime are irrelevant since the signal is dominated by massive haloes (see Figure \ref{fig_rho_eff_pwave}).

The integrated free electron fraction $x_e(z)$ and IGM mean temperature $T_{\text{IGM}}(z)$ were both computed using a pair of coupled differential equations derived from a three-level atom model, modified to include energy injection from DM. This simplified model agrees well with \texttt{RECFAST}, and enables us to compute these two quantities and set constraints across a large range of annihilation cross-sections/decay lifetimes and DM masses $m_\chi$. For each process, we obtained constraints for different assumptions on the redshift of reionization, structure formation prescriptions as well as $T_{\text{IGM}}$ constraints to check the robustness of the constraints. 

For $s$-wave annihilation, constraints from measurements on the CMB power spectrum and on the integrated optical depth $\tau$ rule out any possibility of DM contributing significantly to reionization, with the CMB power spectrum constraints on $\langle \sigma v \rangle$ being approximately an order of magnitude stronger at a given $m_\chi$. The maximum allowed value of $\langle \sigma v \rangle$ can at most contribute to 2\% of $x_e$ at reionization for $\chi \chi \to e^+e^-$, and less than 0.1\% for $\chi \chi \to \gamma \gamma$. These results are largely independent of reionization redshift and structure formation prescription. 

In the case of $p$-wave annihilation, the velocity suppression at early times greatly relaxes the CMB constraints compared to $s$-wave annihilation, since the former are mainly dependent on the cross-section immediately after recombination. However, the sudden increase in energy deposition once structure formation becomes important leads to a sharp rise in $T_\text{IGM}$, making astrophysical measurements of $T_{\text{IGM}}$ at redshifts $z \sim 4$ to 6 important. The most optimistic assumptions appear to allow for significant contributions to reionization, but much of the allowed parameter space is ruled out with the stricter $T_{\text{IGM}}$ constraint and earlier reionization. The sole exception to this is in the channel $\chi \chi \to e^+e^-$ with $m_\chi$ between \SI{1}{MeV} and \SI{100}{MeV}, but this region is in turn ruled out by constraints from the photon flux from the galactic diffuse background emission. Overall, we find that only a $\sim 0.1\%$ contribution to $x_e$ at reionization is permitted for $p$-wave annihilation dominantly to $e^+ e^-$ pairs; for annihilation dominantly to photons, a $\sim 5\%$ contribution is possible.

Finally, for DM decay, optical depth constraints rule out any large contribution from decays into $\gamma \gamma$, with the strongest bounds occurring for heavier DM (a contribution to $x_e$ at the $\sim 10\%$ level is viable for the lightest DM we consider, around 10 keV). Contributions at the 20-40\% level from decays into $e^+e^-$ are possible for $m_\chi \sim \SI{10}{} - \SI{100}{MeV}$, $\tau_\chi \sim 10^{25} \SI{}{s}$ and $m_\chi \sim \SI{1}{MeV}$, $\tau_\chi \sim 10^{24} \SI{}{s}$, with this result being independent of our assumptions on the redshift of reionization.

\begin{figure*}[t!]
	\subfigure{
		\includegraphics[scale=0.59]{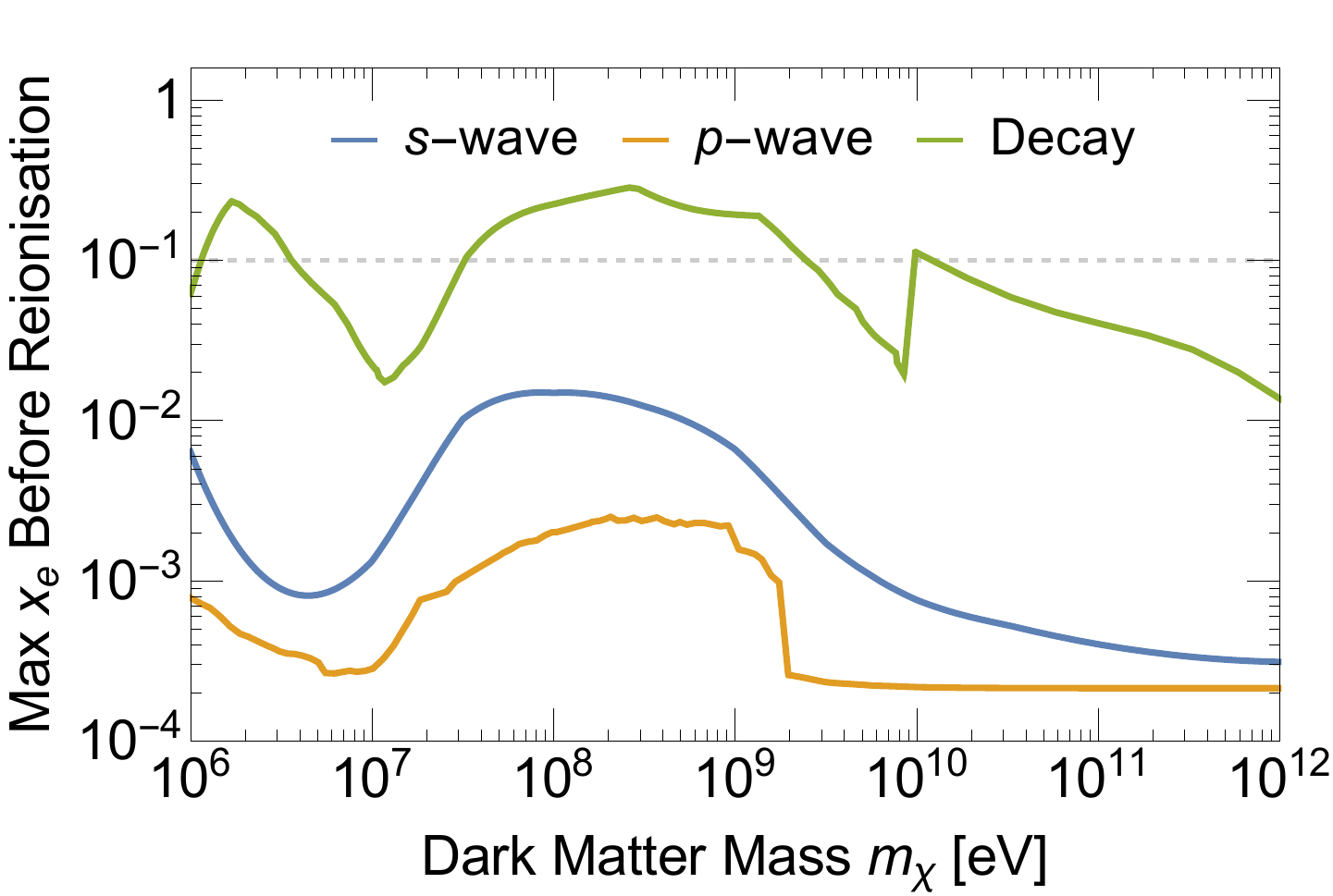}
	}
	\subfigure{
		\includegraphics[scale=0.59]{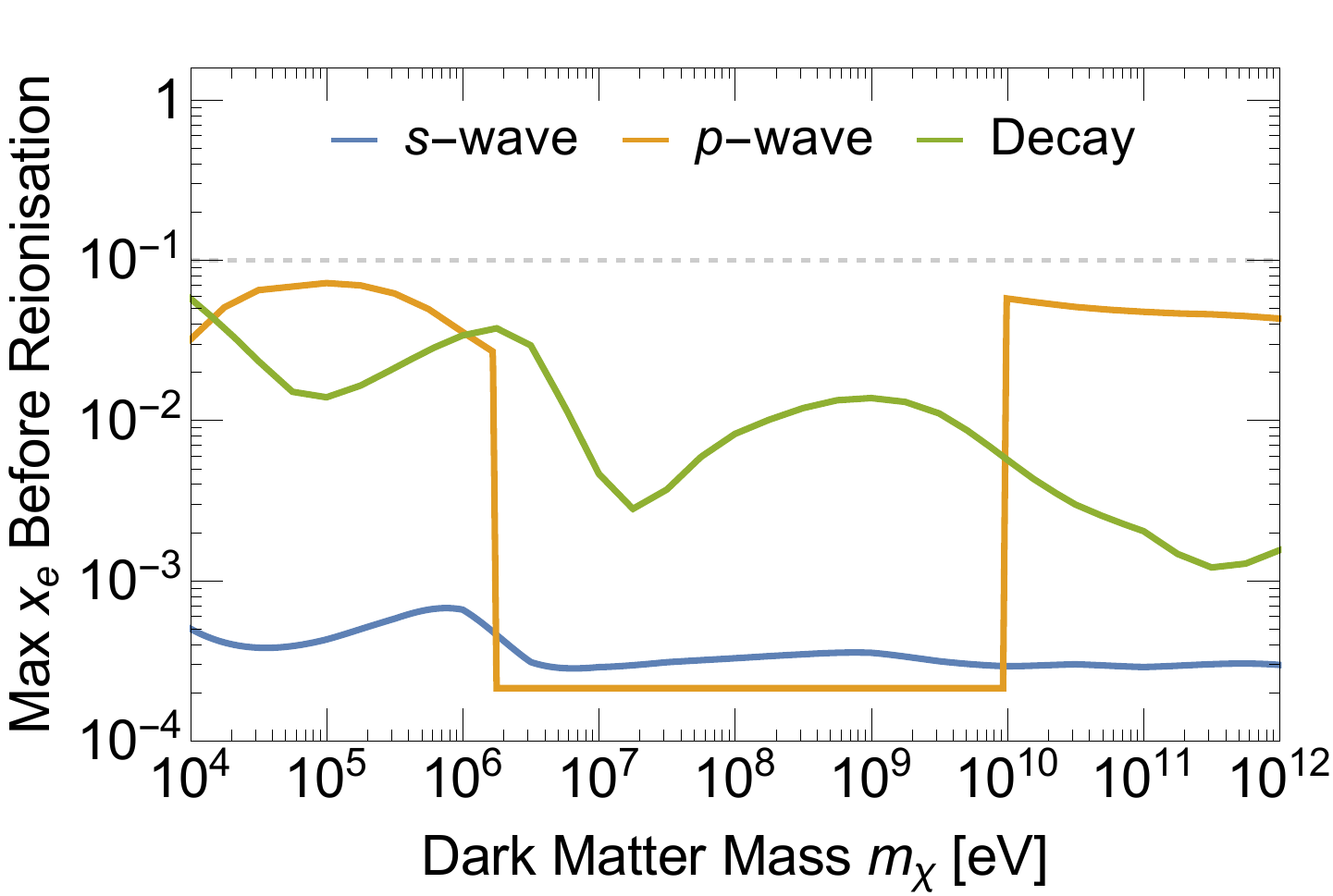}
	}
	\caption{\footnotesize{The maximum free electron fraction $x_e$ just prior to reionization consistent with all constraints used in this paper for $s$-wave annihilations (blue), $p$-wave annihilations (yellow) and decays (green) into $e^+e^-$ (left) and $\gamma \gamma$ (right). }}
	\label{fig:xeMaxConstraints}
\end{figure*}

Overall, we find that DM is mostly unable to contribute more than 10\% of the free electron fraction after reionization across most of the DM processes and annihilation or decay products considered in this paper, even after allowing for different structure formation prescriptions, reionization scenarios and choice of constraint. The one exception to this is found in $\chi \chi \to e^+e^-$, with a possible contribution of up to 40\% near $m_\chi = \SI{100}{MeV}$. Figure \ref{fig:xeMaxConstraints} summarizes the maximum $x_e$ achievable prior to reionization that is consistent with all of the constraints considered in this paper.

With potential input from 21 cm tomography and improved measurements of the IGM at large redshift and the CMB, we expect our understanding of the process of reionization and the end of the cosmic dark ages to improve dramatically in the near future. These future results may be sensitive to a contribution to reionization by DM at well below the 10\% level, and may serve as a good probe of the properties of DM.\footnote{See \cite{Lopez-Honorez2016} for recent work in understanding the impact of DM annihilations on the 21 cm signal, using methods that are similar to those used here.} The continued relevance of DM to reionization and vice-versa serves as strong motivation to improve on the results developed here. Future work may include new ways to calculate $f_c(z)$ at $1+z \leq 10$ with greater accuracy by taking into account the ionization and thermal history of the universe near reionization, as well as understanding the potential impact of DM annihilation products on the haloes in which they are generated, building on results from \cite{Schon2014}. 

\section{Acknowledgments}

The Dark Cosmology Centre is funded by the DNRF. JZ is supported by the EU under a Marie Curie International Incoming Fellowship, contract PIIF-GA-2013-62772. TS and HL are supported by the U.S. Department of Energy under grant Contract Numbers DE$-$SC00012567 and DE$-$SC0013999. The authors would like to thank Jens Chluba, Rouven Essig, Dan Hooper, Katie Mack, Lina Necib, Nicholas Rodd, Sergio Palomares Ruiz, Aaron Vincent and Chih-Liang Wu for helpful comments and discussions.

\appendix

\section{Additional Constraints}
\label{app:additionalConstraints}

\begin{figure*}[t!]
	\subfigure{
		\includegraphics[scale=0.59]{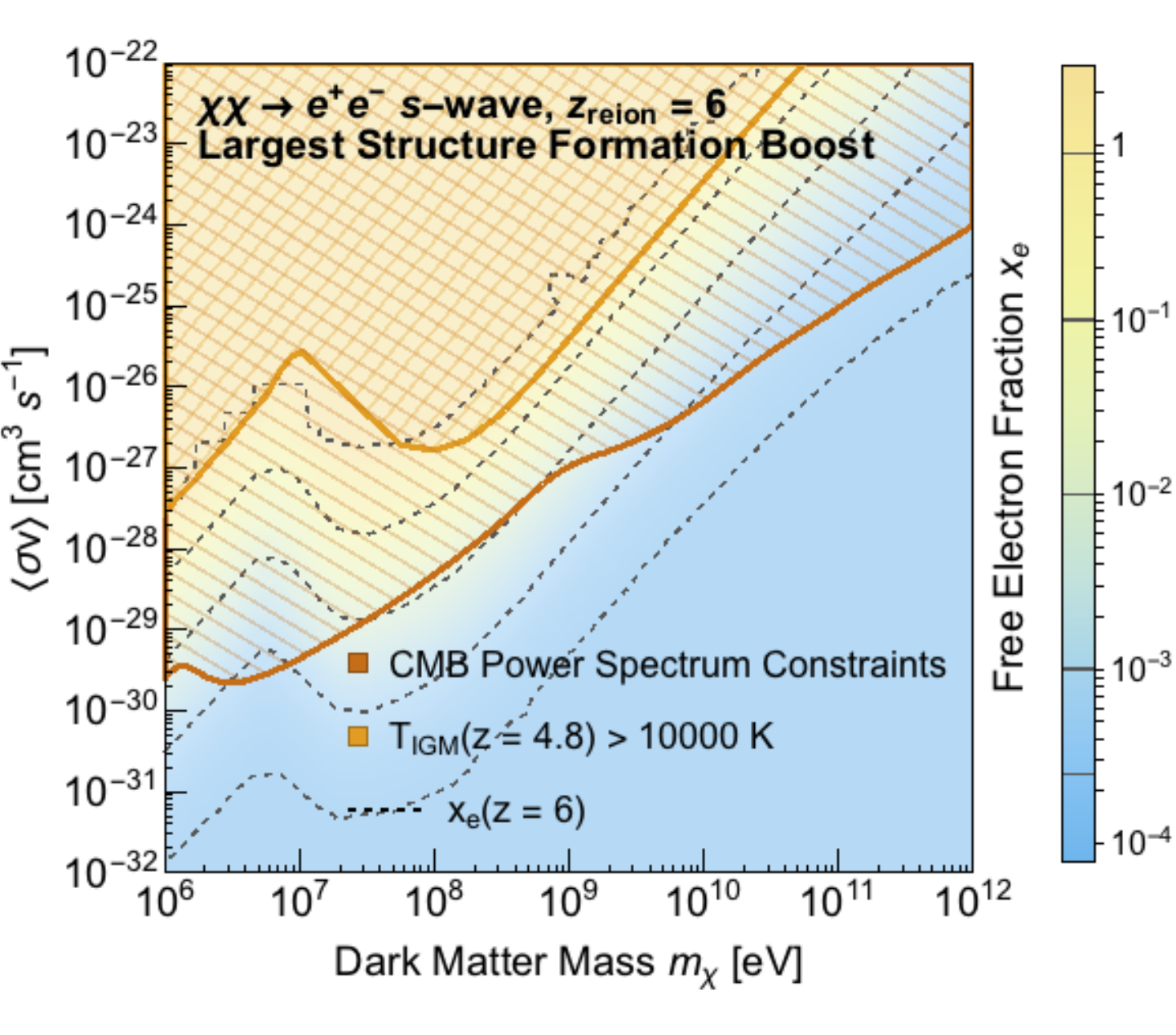}
	}
	\subfigure{
		\includegraphics[scale=0.59]{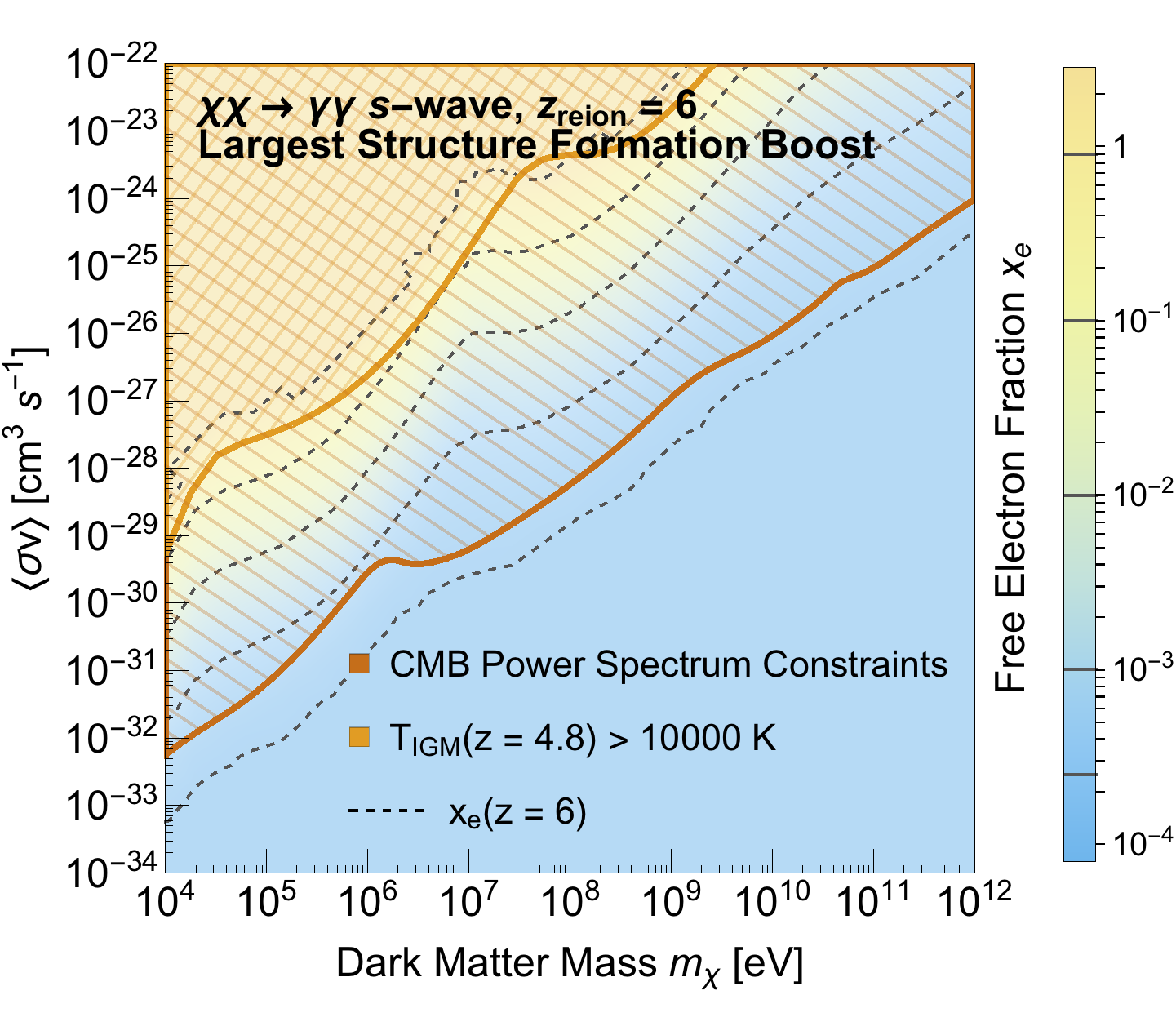}
	}
	\caption{\footnotesize{DM contribution to reionization for $\chi \chi \to e^+e^-$ (left) and $\chi \chi \to \gamma \gamma$ (right) $s$-wave annihilation, benchmark scenario. The hatched regions correspond to parameter space ruled out by the CMB power spectrum constraints as measured by Planck (red) and $T_{\text{IGM}}(z = 4.8) < \SI{10000}{K}$ (orange) respectively. The color density plot shows the DM contribution to $x_e$ just prior to reionization at $z = 6$, with contours (black, dashed) shown for a contribution to $x_e(z = 6) = $ 0.025\%, 0.1\%, 1\%, 10\% and 90\% respectively.}}
	\label{fig:xeConstraintsTIGMPlot_sWave}
\end{figure*}

Figure \ref{fig:xeConstraintsTIGMPlot_sWave} shows the free electron fraction just prior to reionization $x_e(z=6)$ for the benchmark scenario of both $\chi \chi \to e^+e^-$ and $\chi \chi \to \gamma \gamma$ $s$-wave annihilations, as well as the excluded cross-sections due to constraints from the CMB power spectrum as measured by Planck and from the $T_{\text{IGM}}(z=4.8)$ constraints. The $T_{\text{IGM}}$ bounds alone can almost rule out a 10\% contribution from $\chi \chi \to e^+e^-$ above a mass of approximately \SI{1}{GeV}, but are weaker for $\chi \chi \to \gamma \gamma$, since less energy goes into heating for this process. However, if the structure formation boost factor has been underestimated in our paper, these bounds will become stronger. This effectively sets a limit on how large the boost can be.

Throughout this paper, we have obtained the limits on the contribution to reionization from DM in the case of $s$- and $p$-wave annihilation by considering the processes $\chi \chi \to e^+e^-$ and $\chi \chi \to \gamma \gamma$ with each annihilation product having fixed, identical total energy $E = m_\chi$. This allowed us to set limits on $\langle \sigma v \rangle$ or $(\sigma v)_{\text{ref}}$ as a function of $m_\chi$. However, the constraints that we set here extend beyond these two annihilation scenarios. The energy injection rate from annihilations is set only by the quantity $\langle \sigma v \rangle/m_\chi$, and is independent of the annihilation products produced; only the energy deposition rate is dependent on the species and energies of the annihilation products. 

Thus, if we were to recast the $\langle \sigma v \rangle - m_\chi$ parameter space in Figures \ref{fig:xeConstraintsPlot_sWave} and \ref{fig:xeConstraintsPlot_pWave} as a $\langle \sigma v \rangle/m_\chi - m_\chi$ parameter space, the latter parameter actually corresponds to the injection energy of the annihilation products, which is not necessarily equal to the DM mass. 

\begin{figure*}[t!]
	\subfigure{
		\includegraphics[scale=0.59]{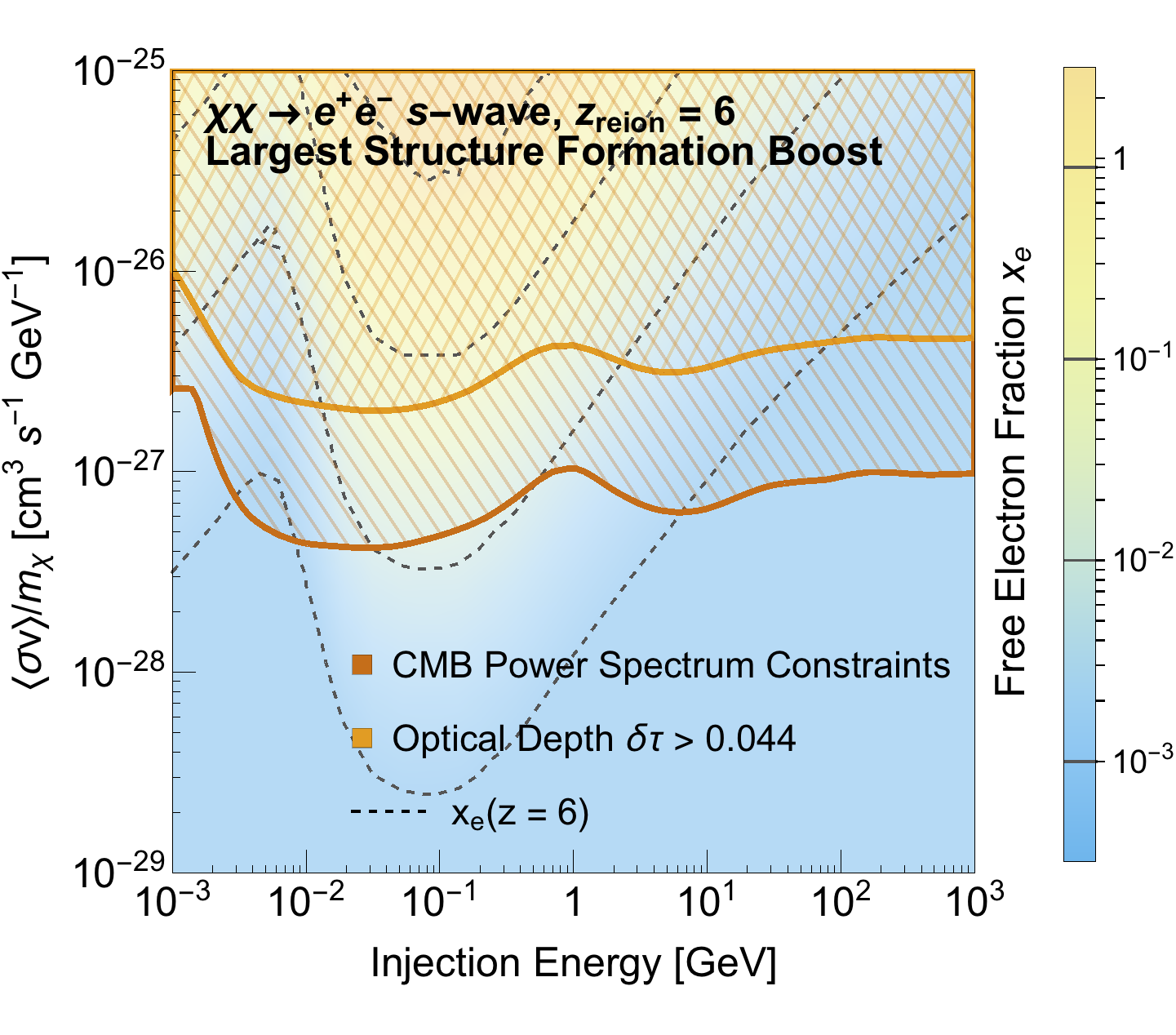}
	}
	\subfigure{
		\includegraphics[scale=0.59]{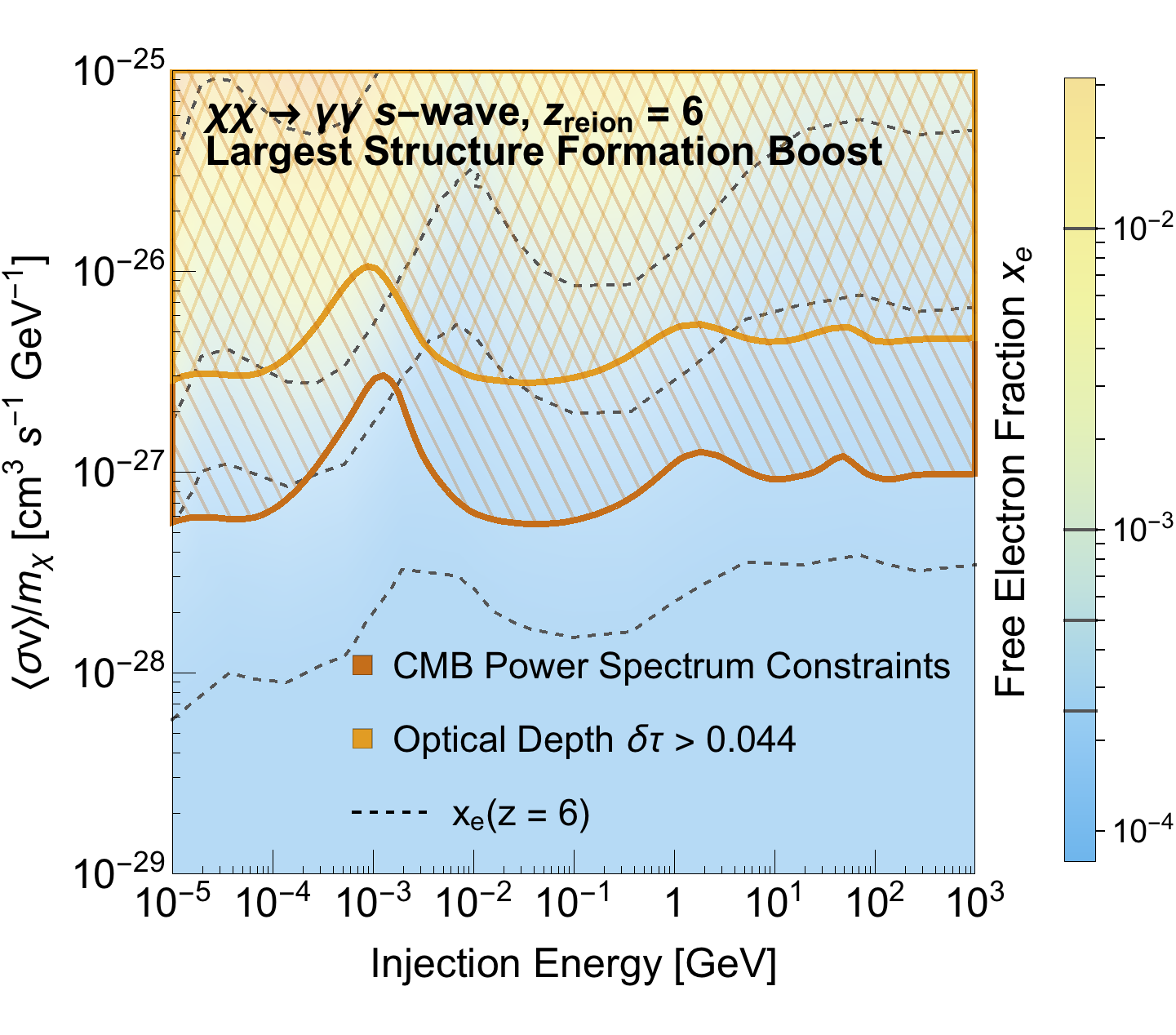}
	}
	\caption{\footnotesize{DM contribution to reionization for $\chi \chi \to e^+e^-$ (left) and $\chi \chi \to \gamma \gamma$ (right) $s$-wave annihilation, plotted as a function of $\langle \sigma v \rangle / m_\chi$ and injection energy, benchmark scenario. The hatched regions correspond to parameter space ruled out by the CMB power spectrum constraints as measured by Planck (red) and optical depth constraints (orange) respectively. The color plot indicates the DM contribution to $x_e$, with contours drawn for a contribution of 0.025\%, 0.1\%, 1\%, 10\% and 90\% respectively. }}
	\label{fig:xeConstraintsPlotSigmavOverMChi_sWave}
\end{figure*}

\begin{figure*}[t!]
	\subfigure{
		\includegraphics[scale=0.59]{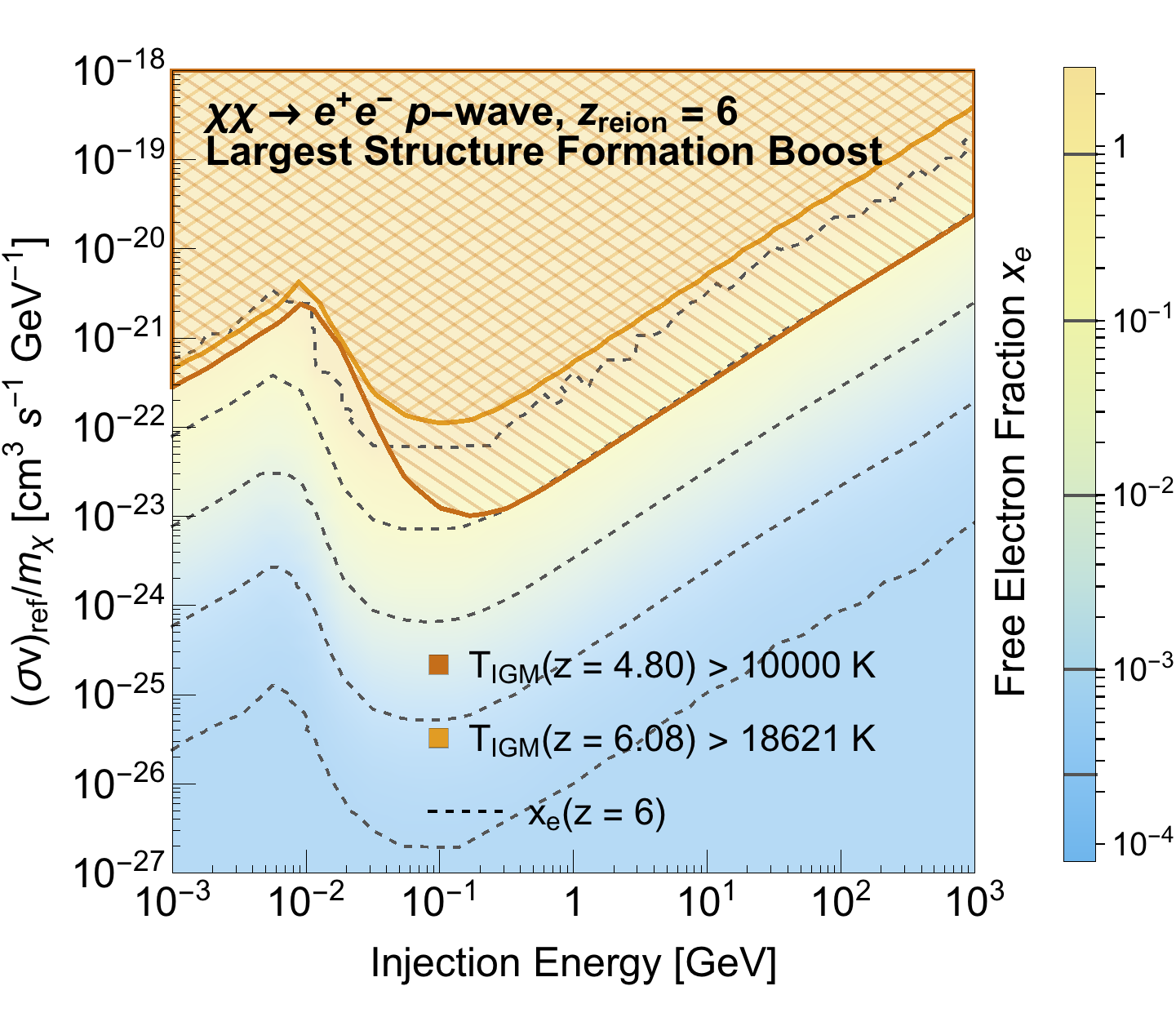}
	}
	\subfigure{
		\includegraphics[scale=0.59]{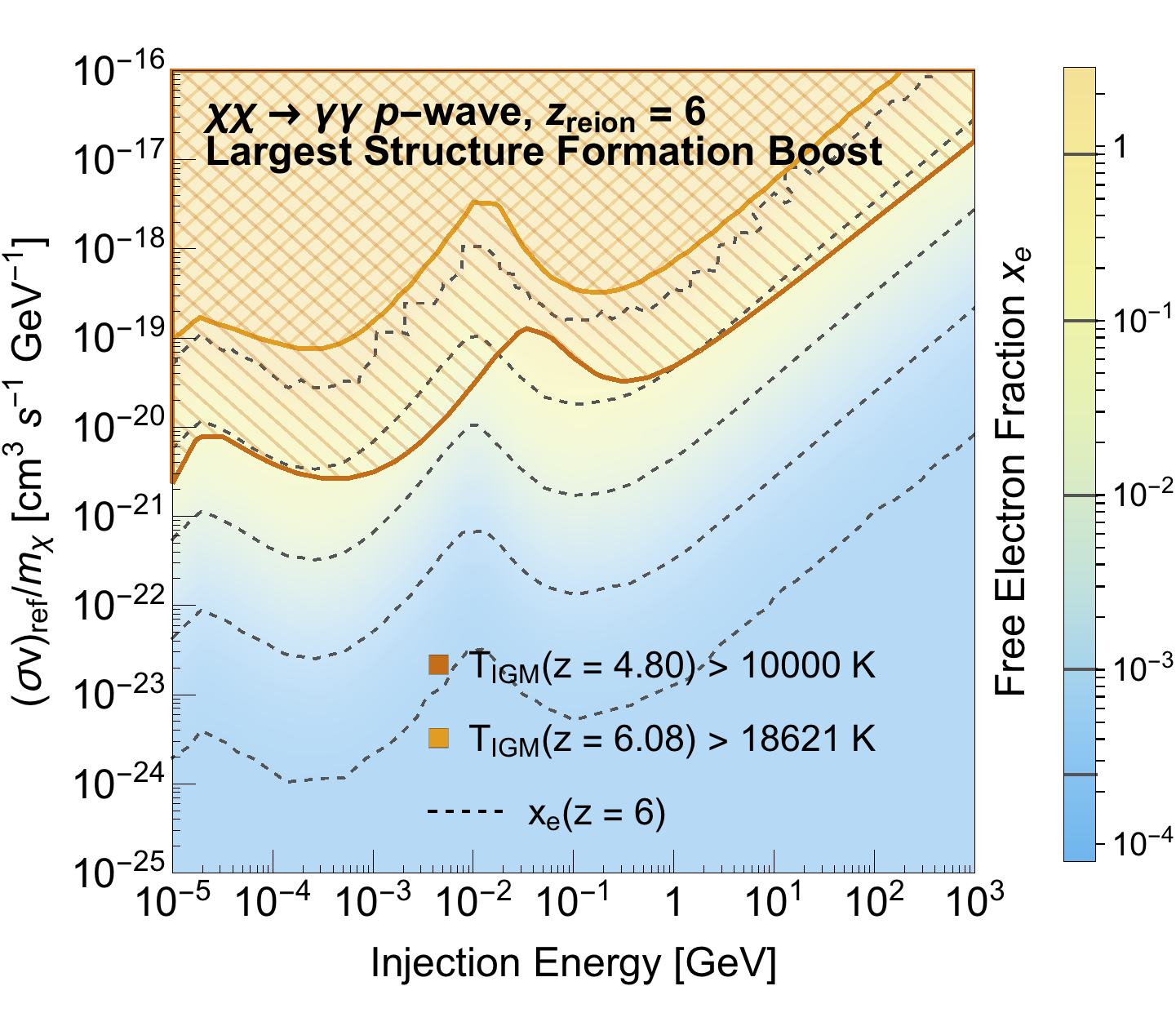}
	}
	\caption{\footnotesize{DM contribution to reionization for $\chi \chi \to e^+e^-$ (left) and $\chi \chi \to \gamma \gamma$ (right) $p$-wave annihilation, plotted as a function of $\langle \sigma v \rangle / m_\chi$ and injection energy, benchmark scenario. The hatched regions correspond to parameter space ruled out by the CMB power spectrum constraints as measured by Planck (red) and optical depth constraints (orange) respectively. The color plot indicates the DM contribution to $x_e$, with contours drawn for a contribution of 0.025\%, 0.1\%, 1\%, 10\% and 90\% respectively. }}
	\label{fig:xeConstraintsPlotSigmavOverMChi_pWave}
\end{figure*}

 Figures \ref{fig:xeConstraintsPlotSigmavOverMChi_sWave} and \ref{fig:xeConstraintsPlotSigmavOverMChi_pWave} present the same set of constraints and results for $x_e(z = 6)$ as a function of $\langle \sigma v \rangle/m_\chi$ or $(\sigma v)_{\text{ref}}/m_\chi$ and the injection energy of the $s$- or $p$-wave annihilation products, which in general can be very different from $m_\chi$. Table \ref{tab:Constraints} gives the $s$-wave CMB power spectrum constraints and the $p$-wave $T_{\text{IGM}}(z = 4.80) > \SI{10000}{K}$ constraints in table form for the convenience of the reader. For any arbitrary annihilation process, the total contribution to $x_e$ prior to reionization is strictly less than the highest contribution to $x_e$ possible among the different particles with different energies produced from the annihilation. This implies that for a given injection rate, the only dependence on the spectrum of the annihilation products enters through $f_c(z)$, and as a result, the CMB power spectrum constraints are relatively insensitive to the details of the injection spectrum from DM annihilations \cite{Elor2015a}.  

\renewcommand{\arraystretch}{1.5} 
\begin{table*}
\begin{tabularx}{500pt}{c >{\centering\arraybackslash}X >{\centering\arraybackslash}X >{\centering\arraybackslash}X >{\centering\arraybackslash}X} 
	\toprule
	\multirow{3}{*}{$\log_{10}[m_\chi (\SI{}{GeV})]$} & \multicolumn{2}{c}{$s$-wave} & \multicolumn{2}{c}{$p$-wave} \\ \cline{2-5}
	& \multicolumn{2}{c}{$\log_{10} \left[ \langle \sigma v \rangle/m_\chi (\SI{}{cm^3 s^{-1} GeV^{-1}}) \right]$} & \multicolumn{2}{c}{$\log_{10} \left[ (\sigma v)_{\text{ref}}/m_\chi (\SI{}{cm^3 s^{-1} GeV^{-1}}) \right]$ } \\ \cline{2-3} \cline{4-5}
	& $\chi \chi \to e^+e^-$ & $\chi \chi \to \gamma \gamma$ & $\chi \chi \to e^+e^-$ & $\chi \chi \to \gamma \gamma$ \\ \colrule
	-5.00 & & -27.2502 & & -20.6327 \\
	-4.75 & & -27.2243 & & -20.1114 \\
	-4.50 & & -27.2311 & & -20.1027 \\ 
	-4.25 & & -27.2326 & & -20.2672 \\ 
	-4.00 & & -27.1866 & & -20.4146 \\
	-3.75 & & -27.0830 & & -20.5190 \\
	-3.50 & & -26.9280 & & -20.5746 \\
	-3.25 & & -26.7415 & & -20.5746 \\
	-3.00 & -26.5871 & -26.5424 & -21.5524 & -20.5075 \\
	-2.75 & -26.7722 & -26.6038 & -21.3538 & -20.3684 \\
	-2.50 & -27.1549 & -26.9224 & -21.1154 & -20.1486 \\
	-2.25 & -27.3000 & -27.1003 & -20.8725 & -19.8619 \\
	-2.00 & -27.3572 & -27.2023 & -20.5468 & -19.5262 \\
	-1.75 & -27.3727 & -27.2421 & -21.0758 & -19.1676 \\
	-1.50 & -27.3787 & -27.2574 & -21.8876 & -18.8817 \\
	-1.25 & -27.3611 & -27.2570 & -22.5907 & -18.9666 \\
	-1.00 & -27.3186 & -27.2409 & -22.9054 & -19.2229 \\
	-0.75 & -27.2587 & -27.2056 & -23.0043 & -19.4243 \\
	-0.50 & -27.1635 & -27.1489 & -22.9120 & -19.4912 \\
	-0.25 & -27.0370 & -27.0626 & -22.7140 & -19.4418 \\
	 0.00 & -26.9831 & -26.9568 & -22.4788 & -19.3185 \\
	 0.25 & -27.0701 & -26.9007 & -22.2346 & -19.1527 \\
	 0.50 & -27.1613 & -26.9332 & -21.9916 & -18.9624 \\
	 0.75 & -27.2024 & -27.0015 & -21.7520 & -18.7597 \\
	 1.00 & -27.1837 & -27.0369 & -21.5127 & -18.5503 \\
	 1.25 & -27.1212 & -27.0208 & -21.2700 & -18.3361 \\
	 1.50 & -27.0662 & -26.9702 & -21.0248 & -18.1182 \\
	 1.75 & -27.0467 & -26.9416 & -20.7816 & -17.8968 \\
	 2.00 & -27.0246 & -27.0247 & -20.5460 & -17.6747 \\
	 2.25 & -27.0014 & -27.0301 & -20.3158 & -17.4536 \\
	 2.50 & -27.0101 & -27.0116 & -20.0852 & -17.2340 \\
	 2.75 & -27.0139 & -27.0102 & -19.8505 & -17.0141 \\
	 3.00 & -27.0090 & -27.0089 & -19.6115 & -16.7924 \\
	 \botrule
\end{tabularx}
\caption{Tabulated $s$-wave CMB power spectrum constraints and $p$-wave $T_{\text{IGM}}(z = 4.80) > \SI{10000}{K}$ constraints.}
\label{tab:Constraints}
\end{table*}

\section{$p$-wave $J$-Factor}
\label{app:JFactor}

The photon flux per unit energy due to DM annihilations from DM in the galaxy is given by \cite{Essig2013}
\begin{alignat}{1}
	\frac{d\Phi}{dE} = \frac{1}{2} \frac{r_\odot}{4\pi} \frac{\rho_\odot^2}{m_\chi} \frac{\langle \sigma v \rangle_\odot}{m_\chi} \frac{dN_\gamma}{dE} J,
\end{alignat}
where $dN_\gamma/dE$ is the annihilation photon yield, and $r_\odot$ and $\rho_\odot$ are the distance from the Sun to the galactic center and the local DM density respectively. $J$ is a dimensionless factor that encapsulates the averaging of the DM density along the line-of-sight of the entire field of observation, and is given by
\begin{alignat}{1}
	J = \int d\Omega \frac{ds}{r_\odot} \left(\frac{\rho(s)}{\rho_\odot} \right)^2.
\end{alignat}
For $s$-wave annihilations, $J$ contains all of the dependence of the photon flux on the DM distribution in the galaxy. In $p$-wave annihilations, however, the rate of DM annihilations also depends on the velocity dispersion of DM, and thus both the density and the velocity of DM along each line-of-sight must be averaged. We should therefore replace $J$ with
\begin{alignat}{1}
	J_p = \int d\Omega \frac{ds}{r_\odot} \left(\frac{\rho(s)}{\rho_\odot} \right)^2 \frac{v^2(s)}{v_\odot^2},
\end{alignat}
and now $\langle \sigma v \rangle_\odot$ is explicitly the local annihilation cross-section due to the velocity dependence of $\langle \sigma v \rangle$. 

Previous studies have implicitly assumed that $J$ and $J_p$ are equal. To assess the significance of this assumption, we consider a pure NFW DM profile given by equation (\ref{rho_smooth}) with $\alpha = 1$, with a corresponding velocity dispersion profile given by the following relation \cite{Zavala2014}:
\begin{alignat}{1}
	\frac{\rho(r)}{\sigma_{\text{1D}}^3(r)} \propto r^{-1.9}.
\end{alignat}
where $\sigma_{\text{1D}}$ is the 1D velocity dispersion that we use as a proxy for $v$. The constant of proportionality of this equation is determined by setting $\rho(r_\odot) = \SI{0.3}{GeV cm^{-3}}$ and assuming a Maxwellian distribution of the dark matter particles in the halo with a peak value set equal to the rotation velocity of the Sun given by $v = \SI{220}{km s^{-1}}$. With these assumptions, we find a difference between $J_p$ and $J$ of about 5 - 10\%, after averaging over the solid angle within some typical galactic diffuse gamma-ray background survey regions. This result has also been confirmed using DM particle dispersion velocities as a function of radius \cite{Necib2016} derived from the Illustris $N$-body simulation \cite{Vogelsberger2014}, which models both DM and baryons. 

We have therefore assumed throughout our analysis that $J_p = J$, and anticipate an error of about 10\% in translating the $\langle \sigma v \rangle$ constraints assuming and $s$-wave distribution directly into constraints for $(\sigma v)_{\text{ref}}$ in $p$-wave annihilations. Since the $p$-wave constraints that we have used rule out regions of parameter space with a contribution to reionization exceeding 10\% by more than 2 orders of magnitude, we do not expect this assumption to change our conclusions in any significant way. 

\section{Contour Plots of $f_c(z)$}
\label{app:fz}

\begin{figure*}[t!]
	\begin{tabular}{cc}
	\subfigure{
		\includegraphics[scale=0.32]{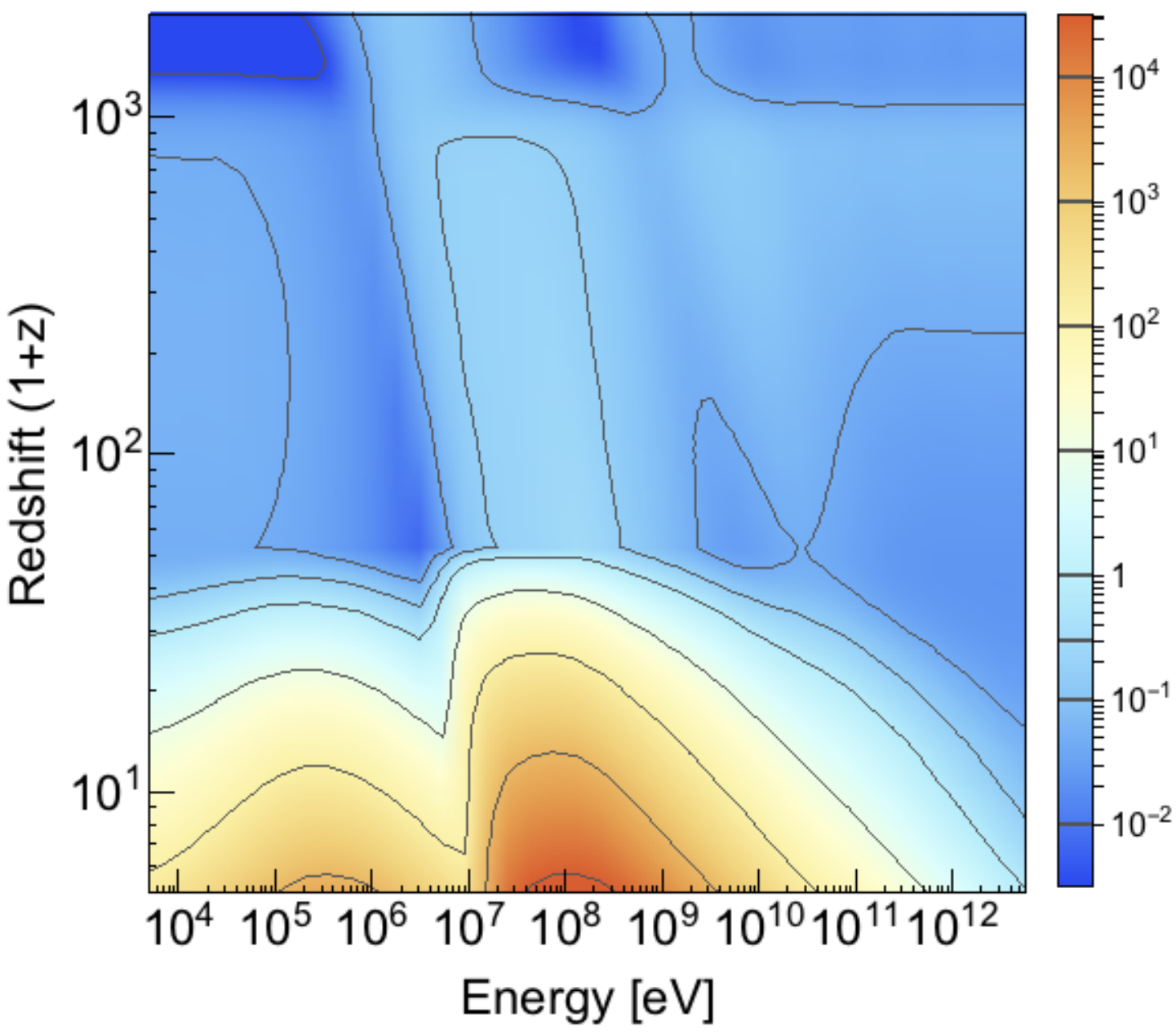}
	} & 
	\subfigure{
		\includegraphics[scale=0.32]{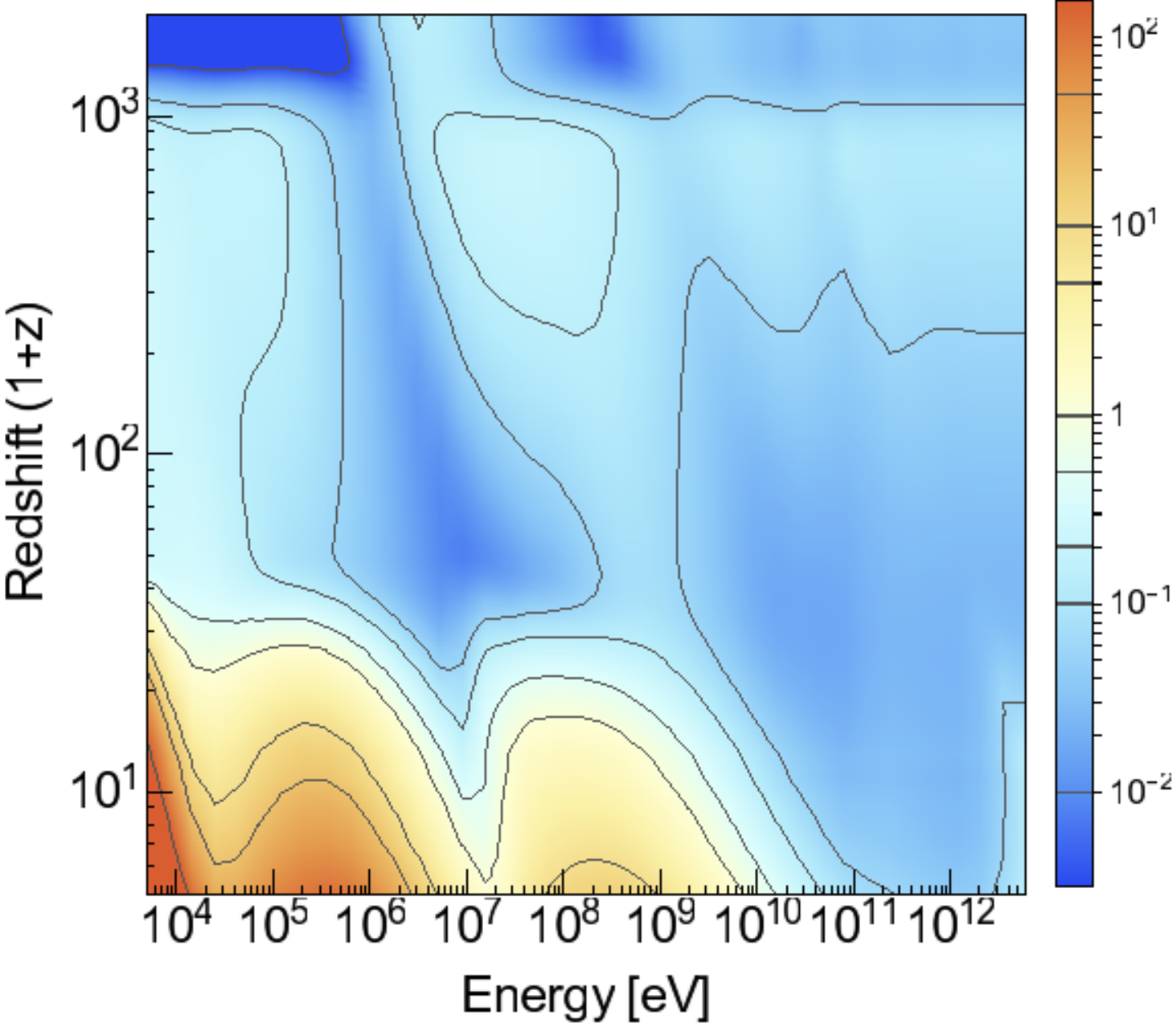}
	} \\
	\subfigure{
		\includegraphics[scale=0.32]{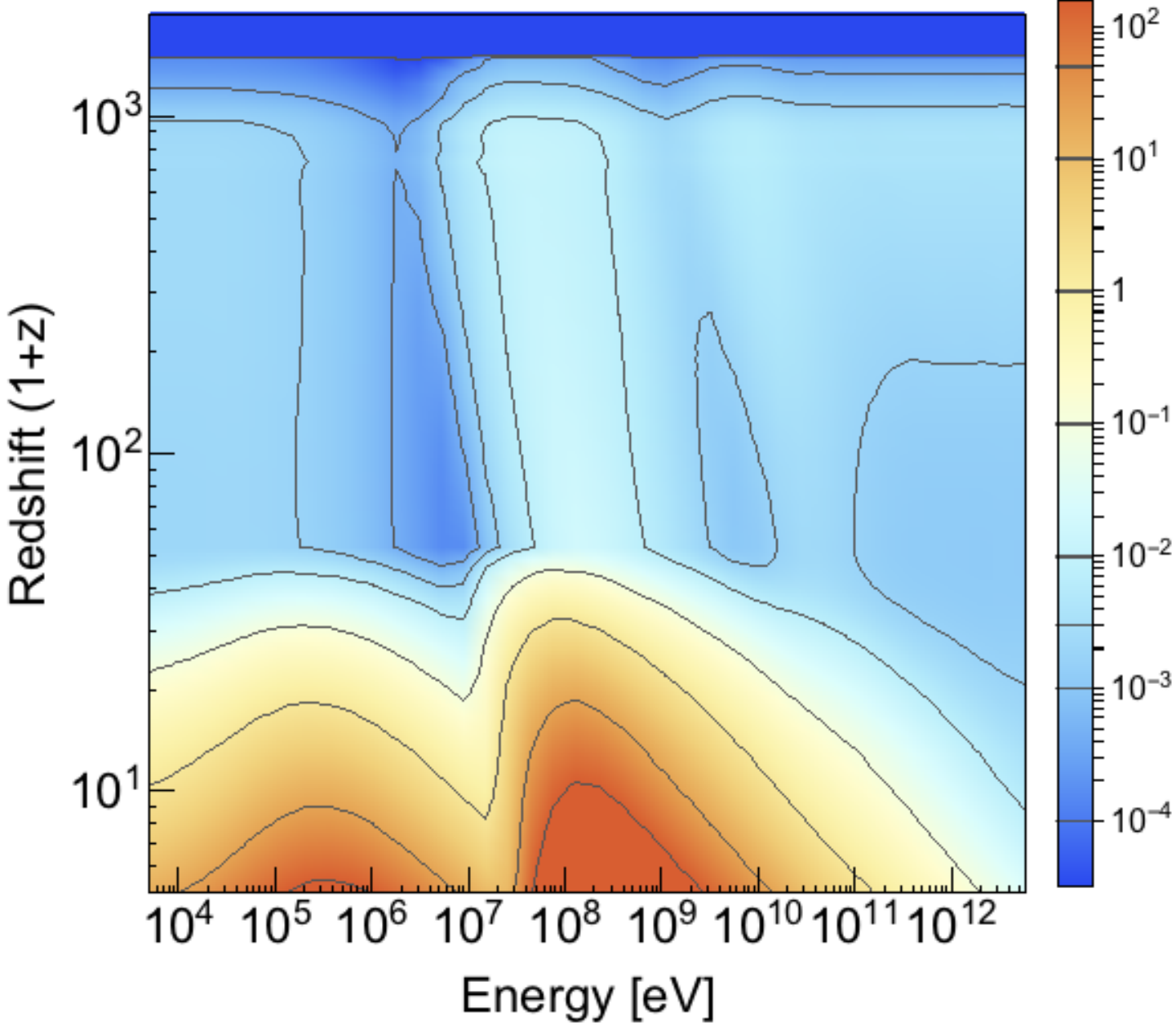}
	} & 
	\subfigure{
		\includegraphics[scale=0.32]{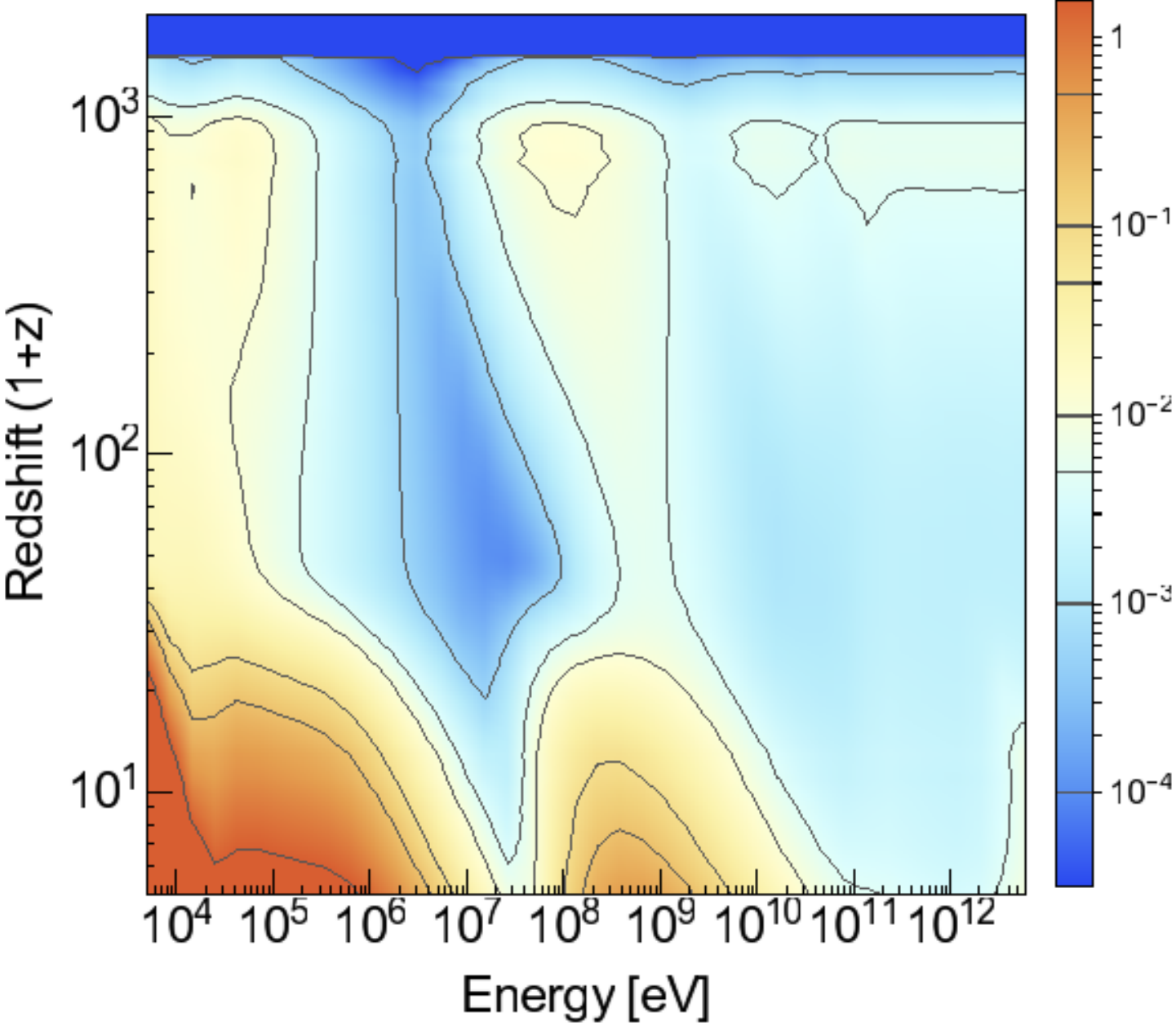}
	} \\
	\subfigure{
		\includegraphics[scale=0.32]{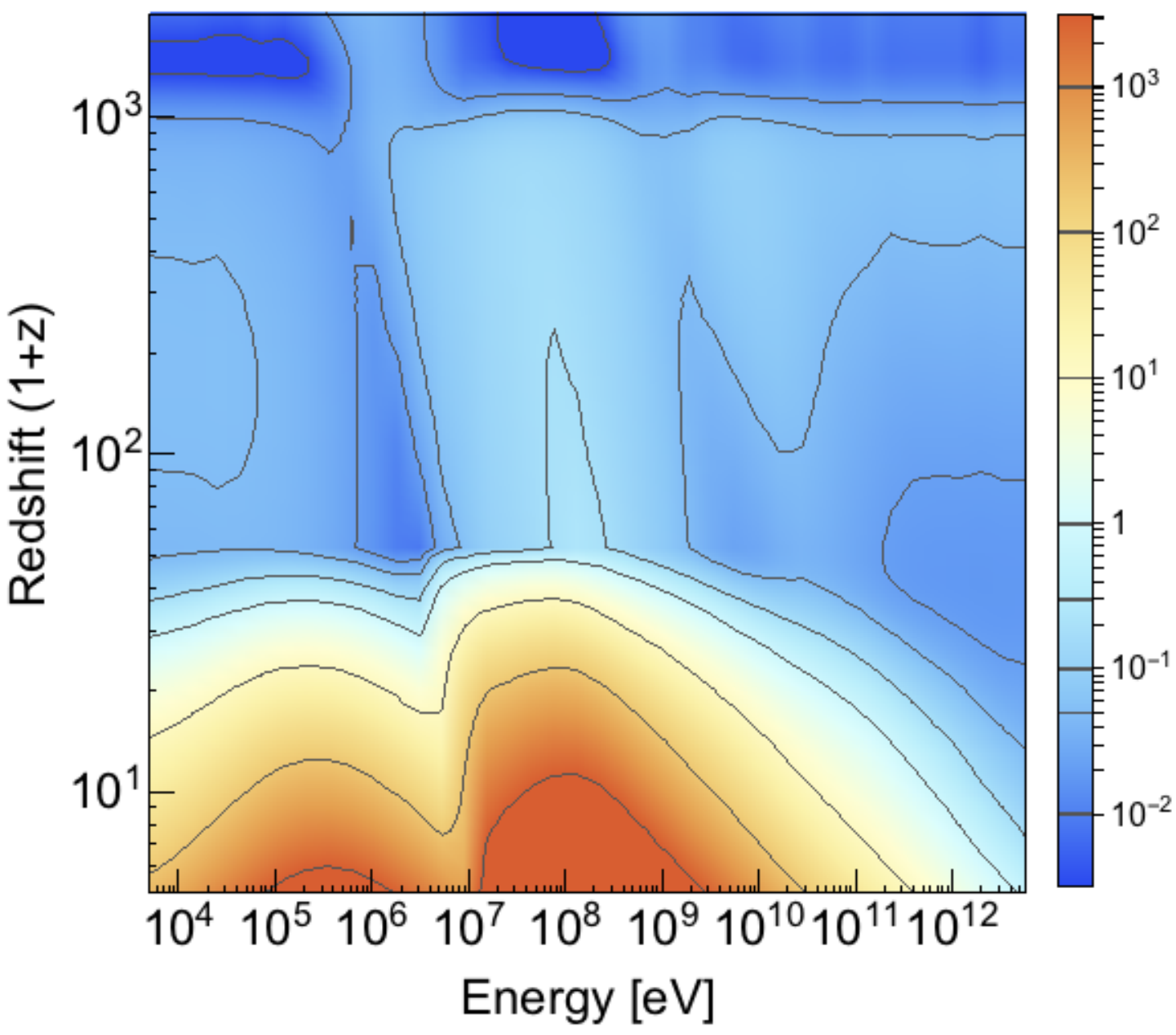}
	} & 
	\subfigure{
		\includegraphics[scale=0.32]{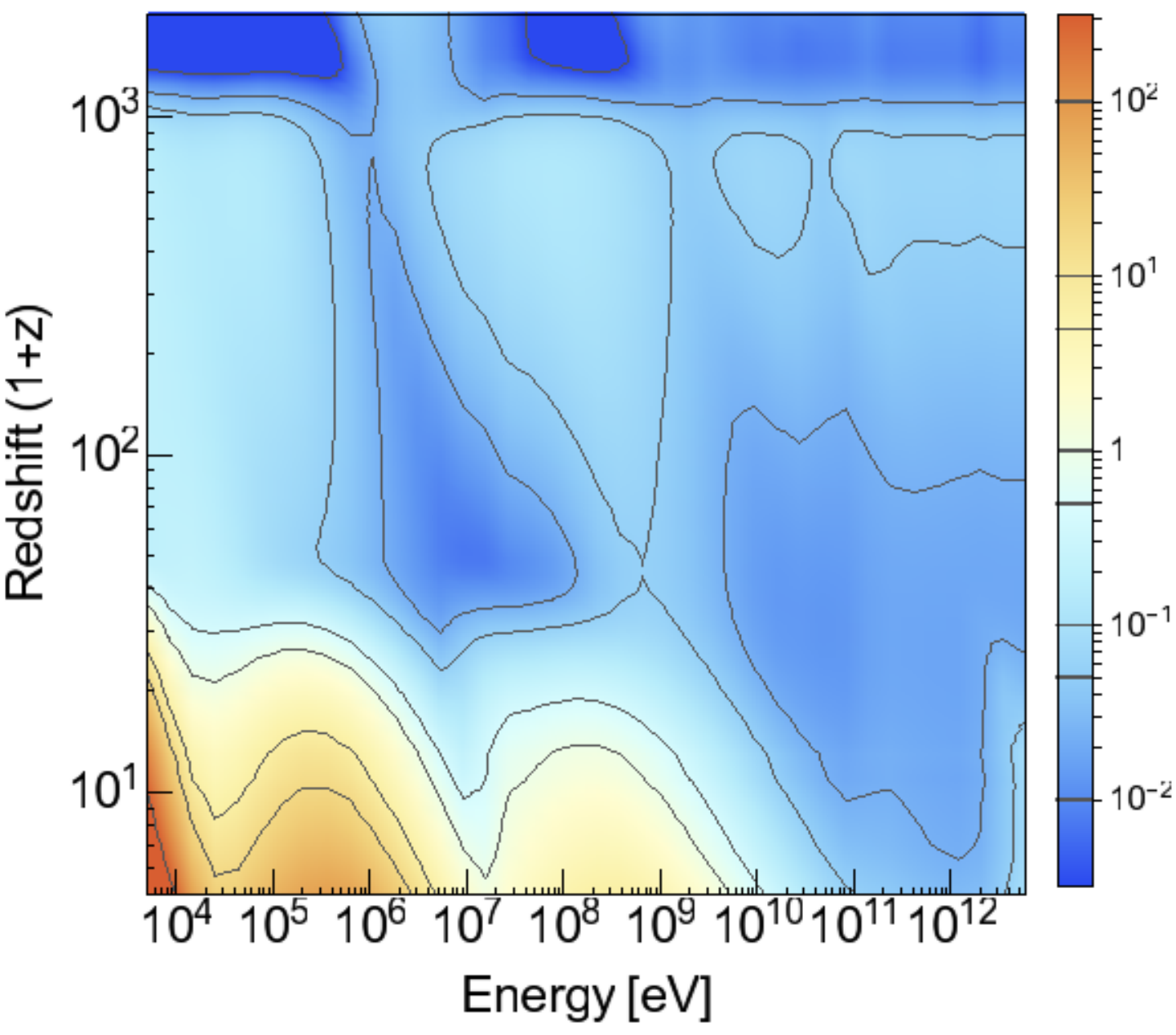}
	} \\
	\subfigure{
		\includegraphics[scale=0.32]{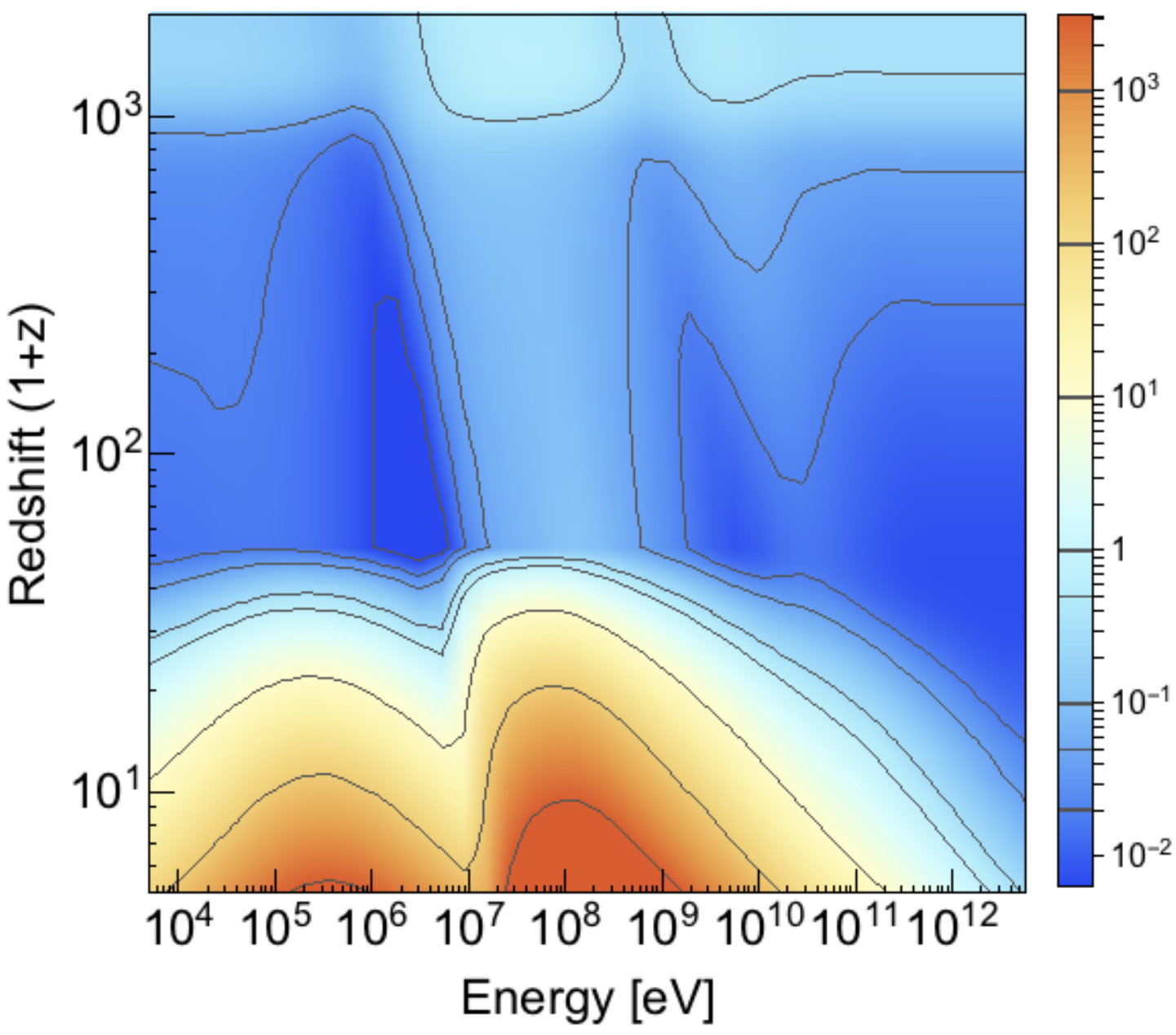}
	} & 
	\subfigure{
		\includegraphics[scale=0.32]{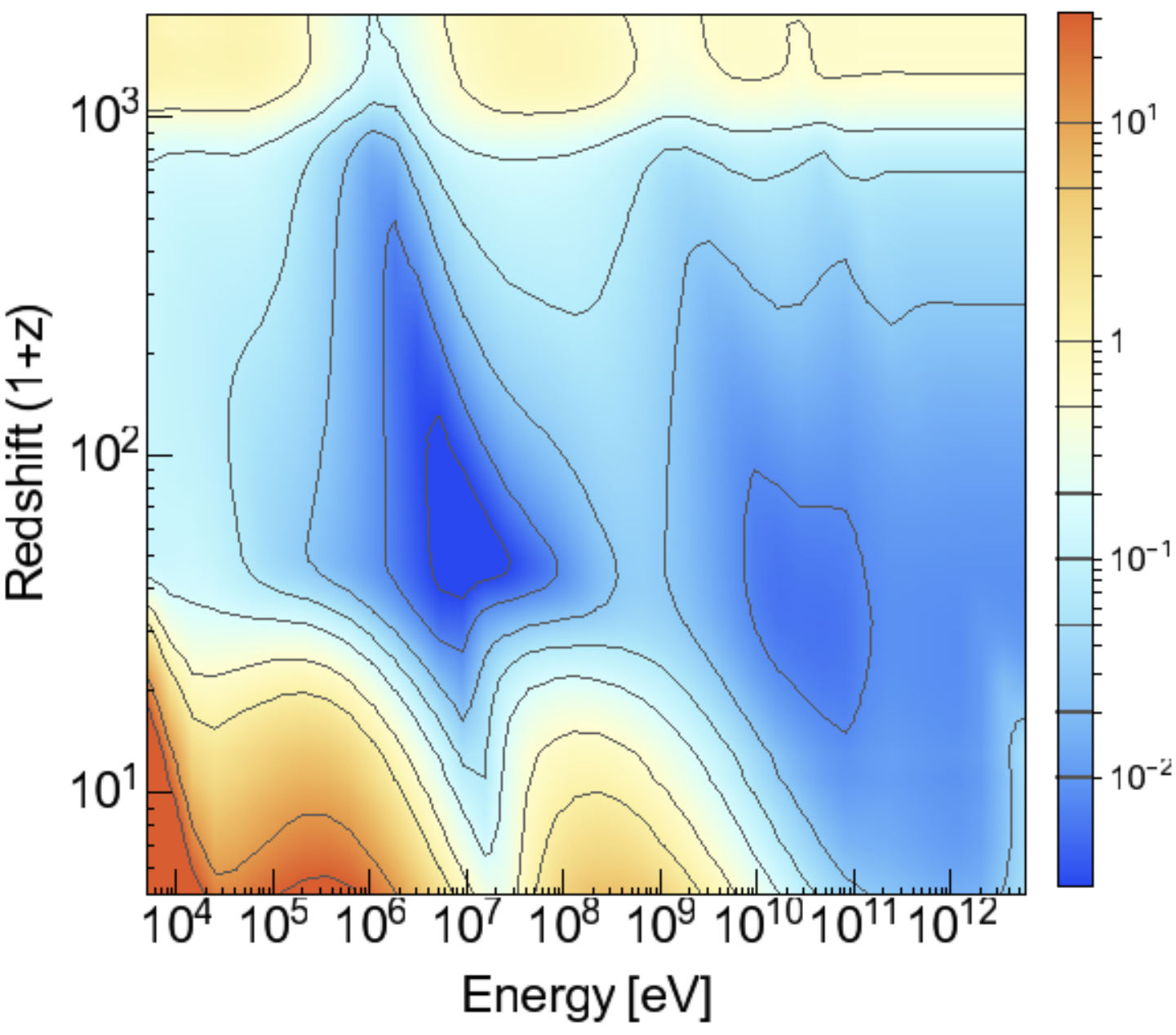}
	} \\
	\subfigure{
		\includegraphics[scale=0.32]{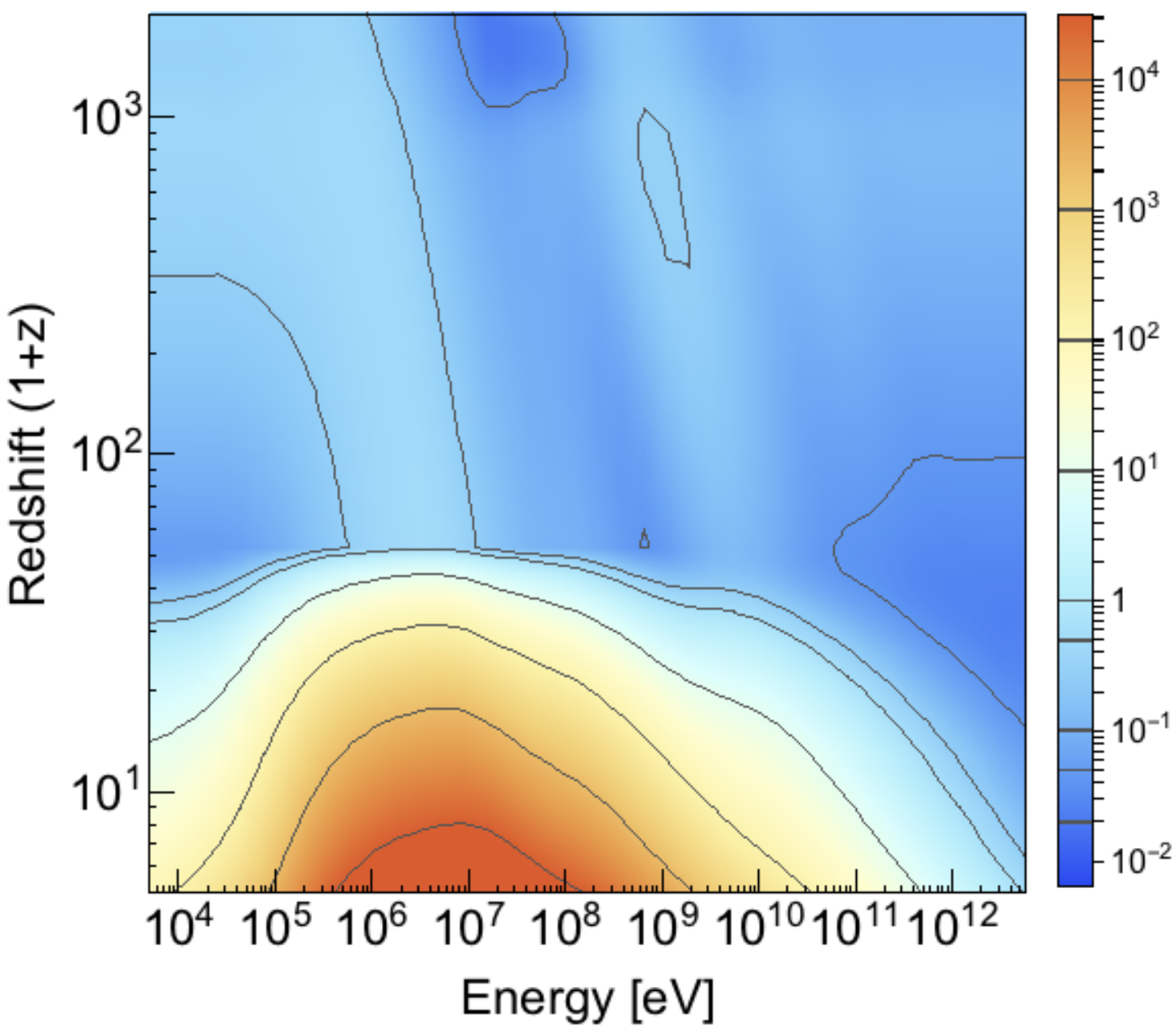}
	} & 
	\subfigure{
		\includegraphics[scale=0.32]{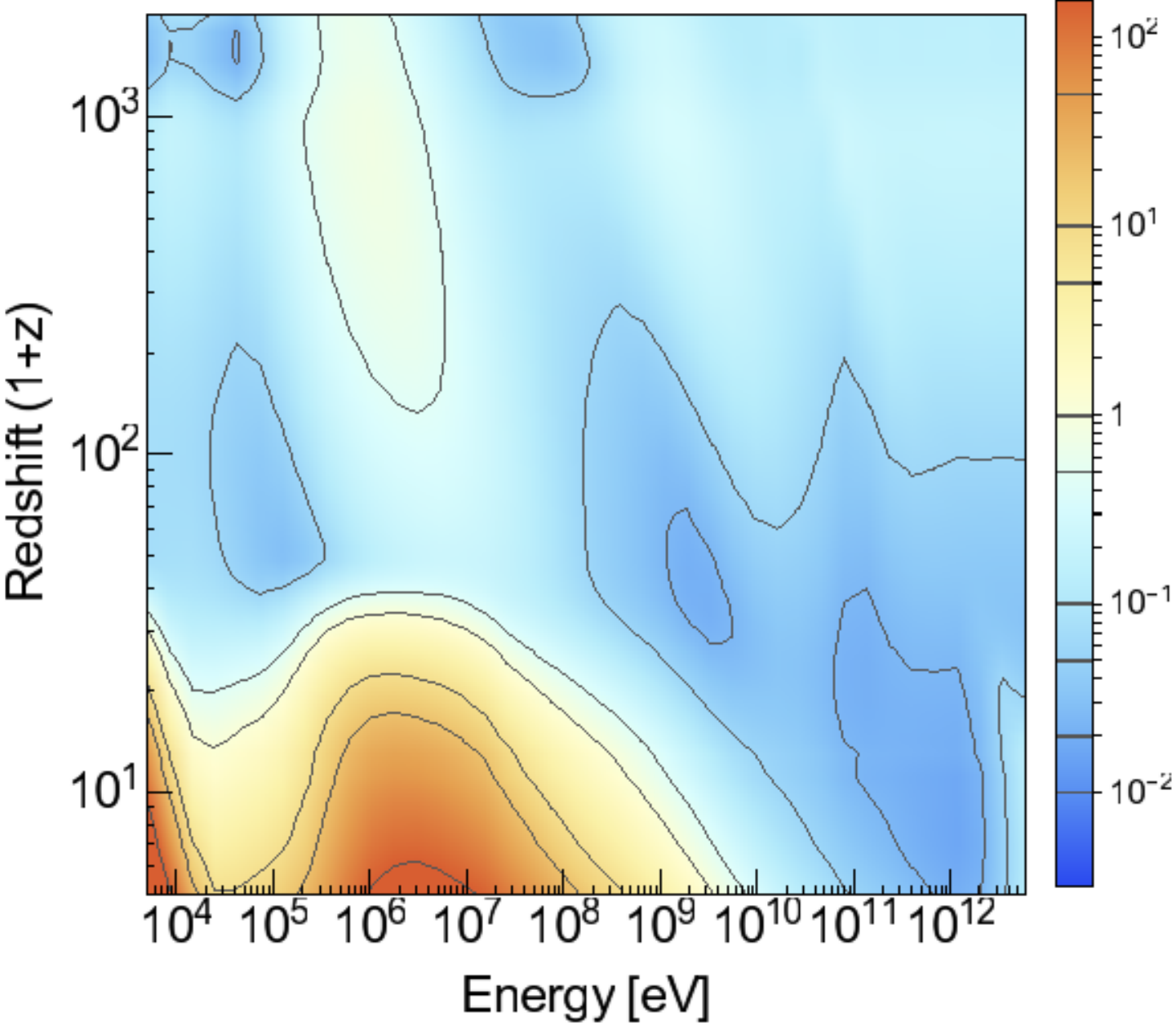}
	} \\
	\end{tabular}
	\caption{\footnotesize{Contour plots of $f_c(z)$ for $\chi \chi \to e^+ e^-$ (left) and $\chi \chi \to \gamma \gamma$ (right) $s$-wave annihilations into (from top to bottom) H ionization; He ionization; Lyman-$\alpha$; heating; and sub-10.2 eV continuum photons as a function of injection energy and redshift. Lines on the bar legend indicate the value of $f_c(z)$ at which contours are drawn. The structure formation prescription with the largest boost is used (see Figure \ref{fig_rho_eff}), and no reionization is assumed.  }}
	\label{fig:fz_sWave}
\end{figure*}

\begin{figure*}[t!]
	\begin{tabular}{cc}
	\subfigure{
		\includegraphics[scale=0.32]{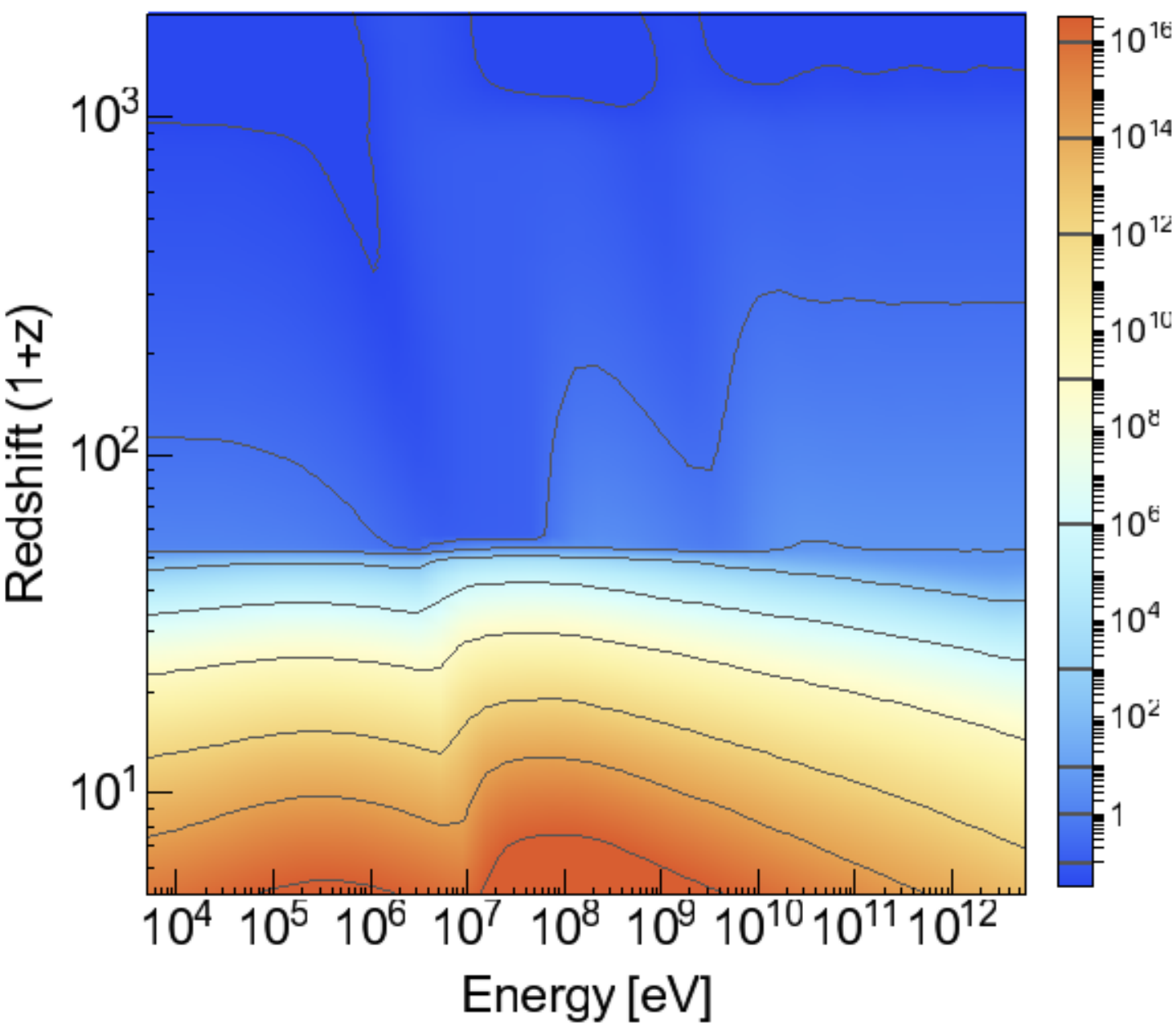}
	} & 
	\subfigure{
		\includegraphics[scale=0.32]{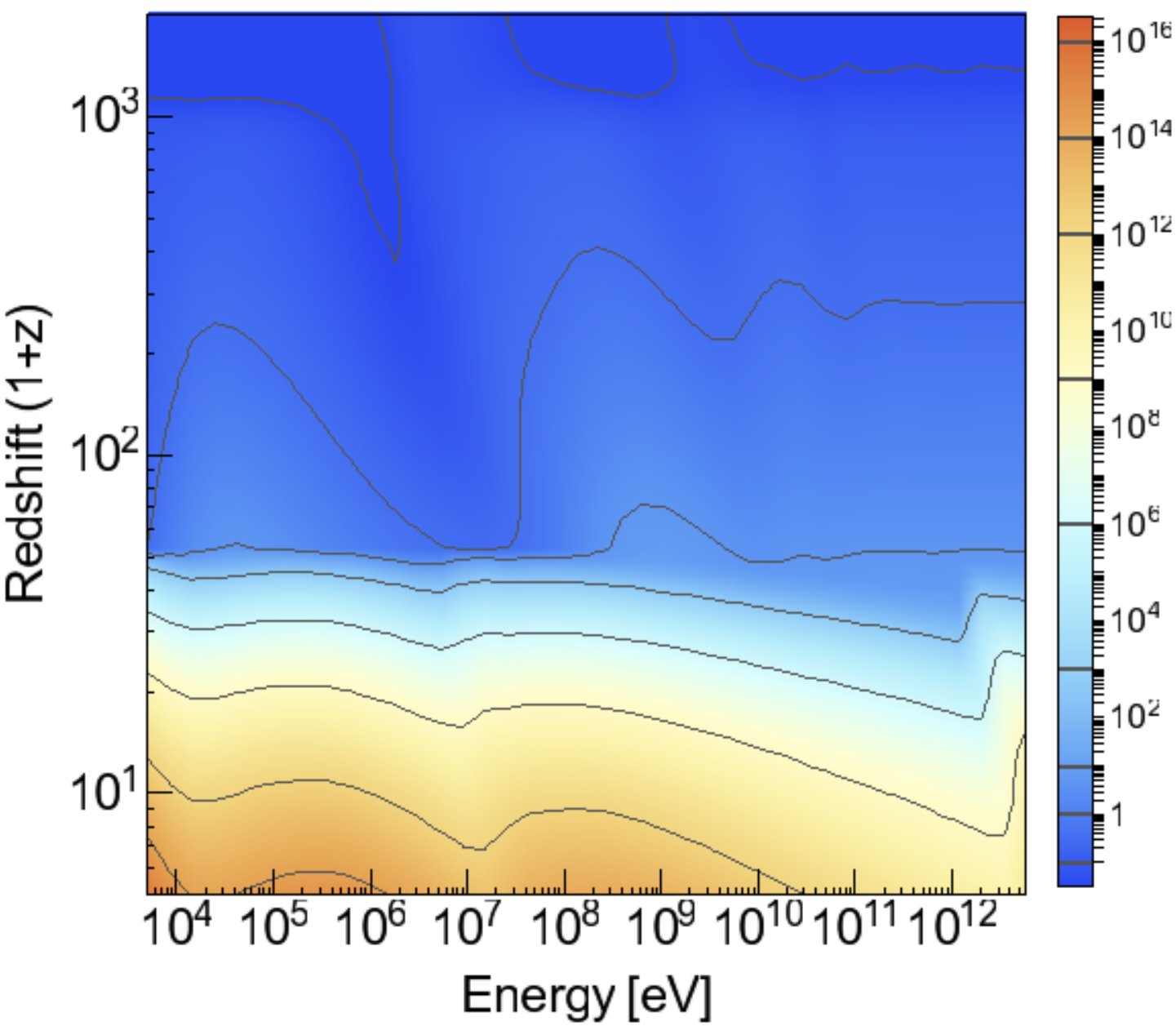}
	} \\
	\subfigure{
		\includegraphics[scale=0.32]{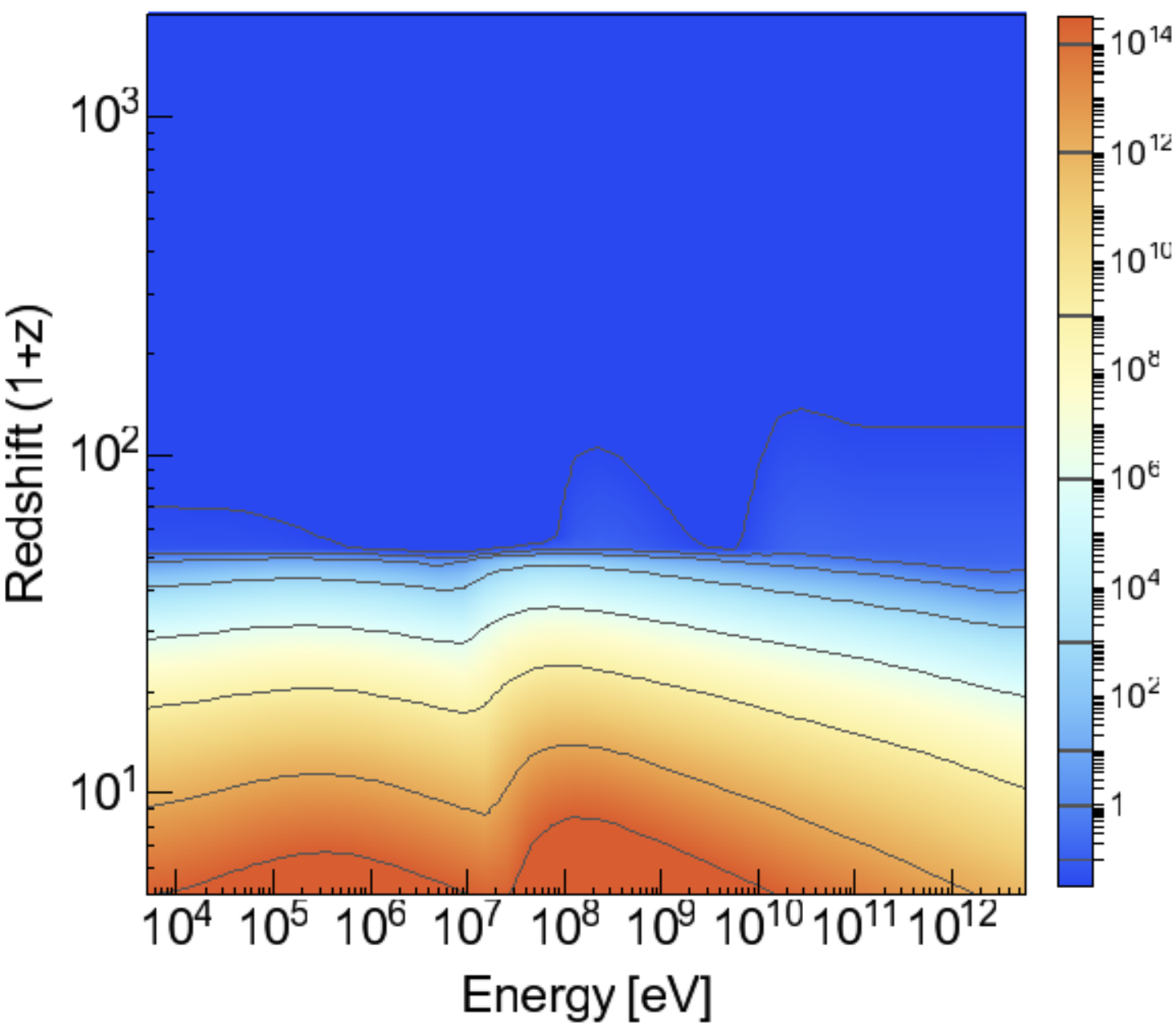}
	} & 
	\subfigure{
		\includegraphics[scale=0.32]{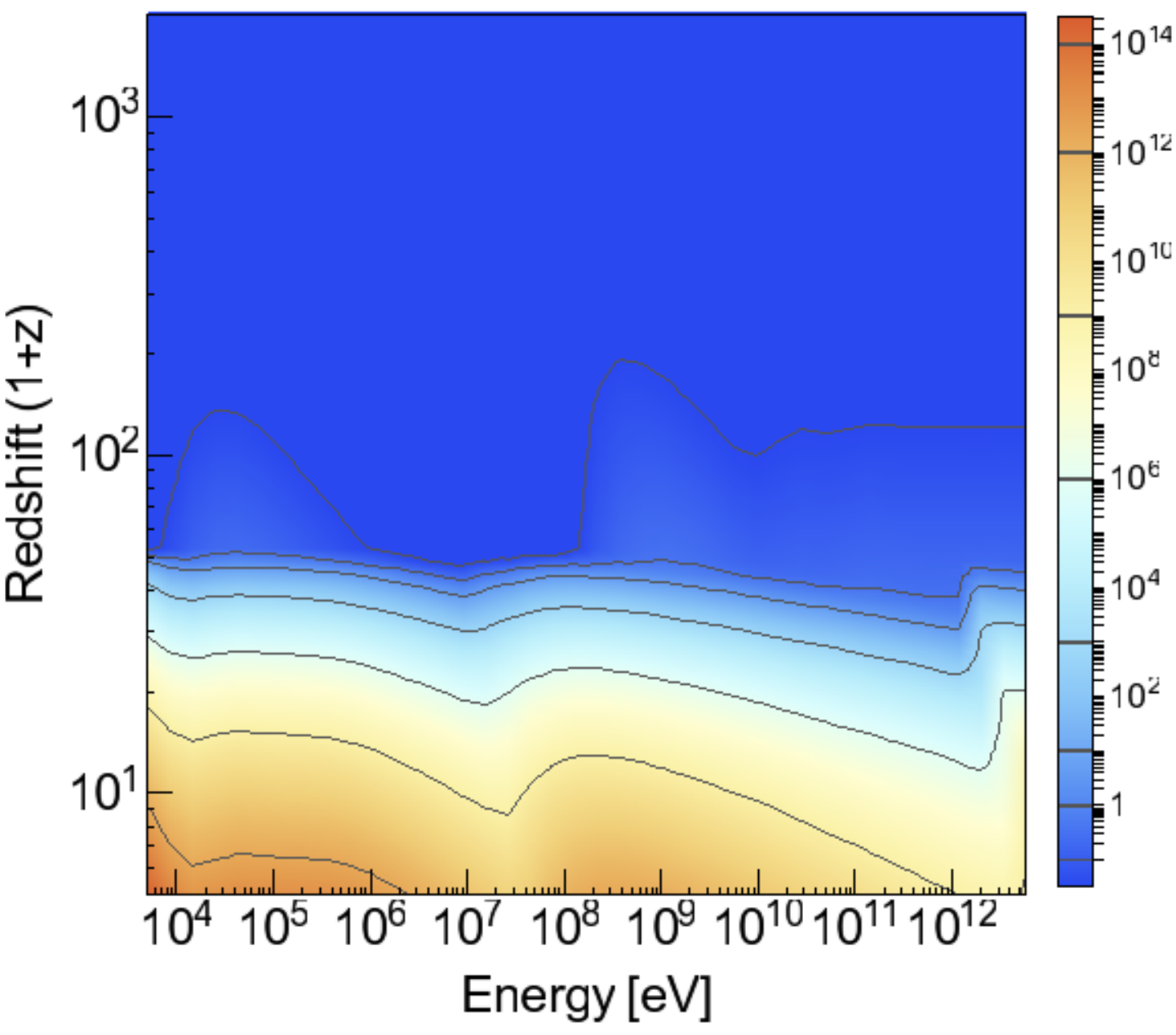}
	} \\
	\subfigure{
		\includegraphics[scale=0.32]{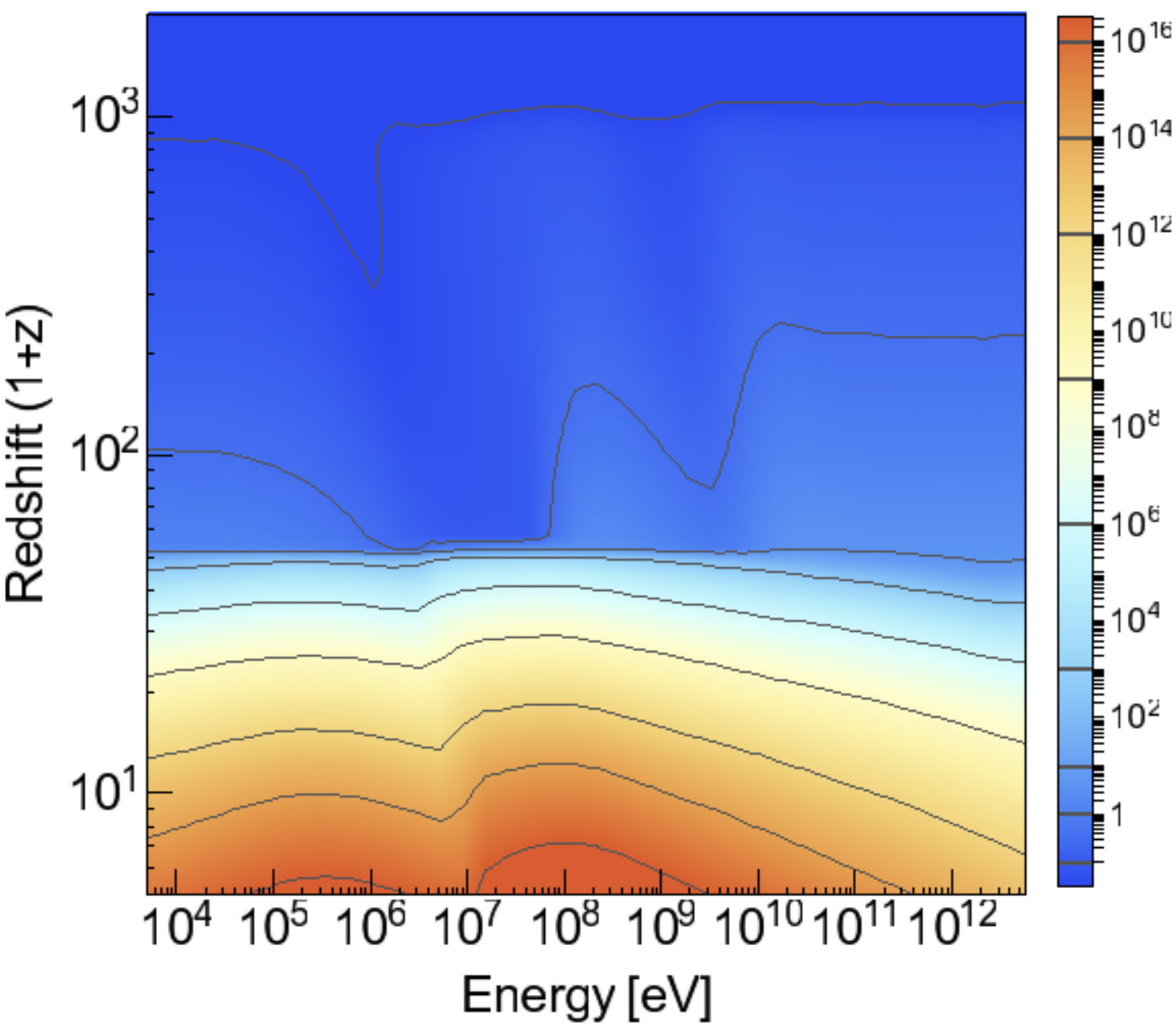}
	} & 
	\subfigure{
		\includegraphics[scale=0.32]{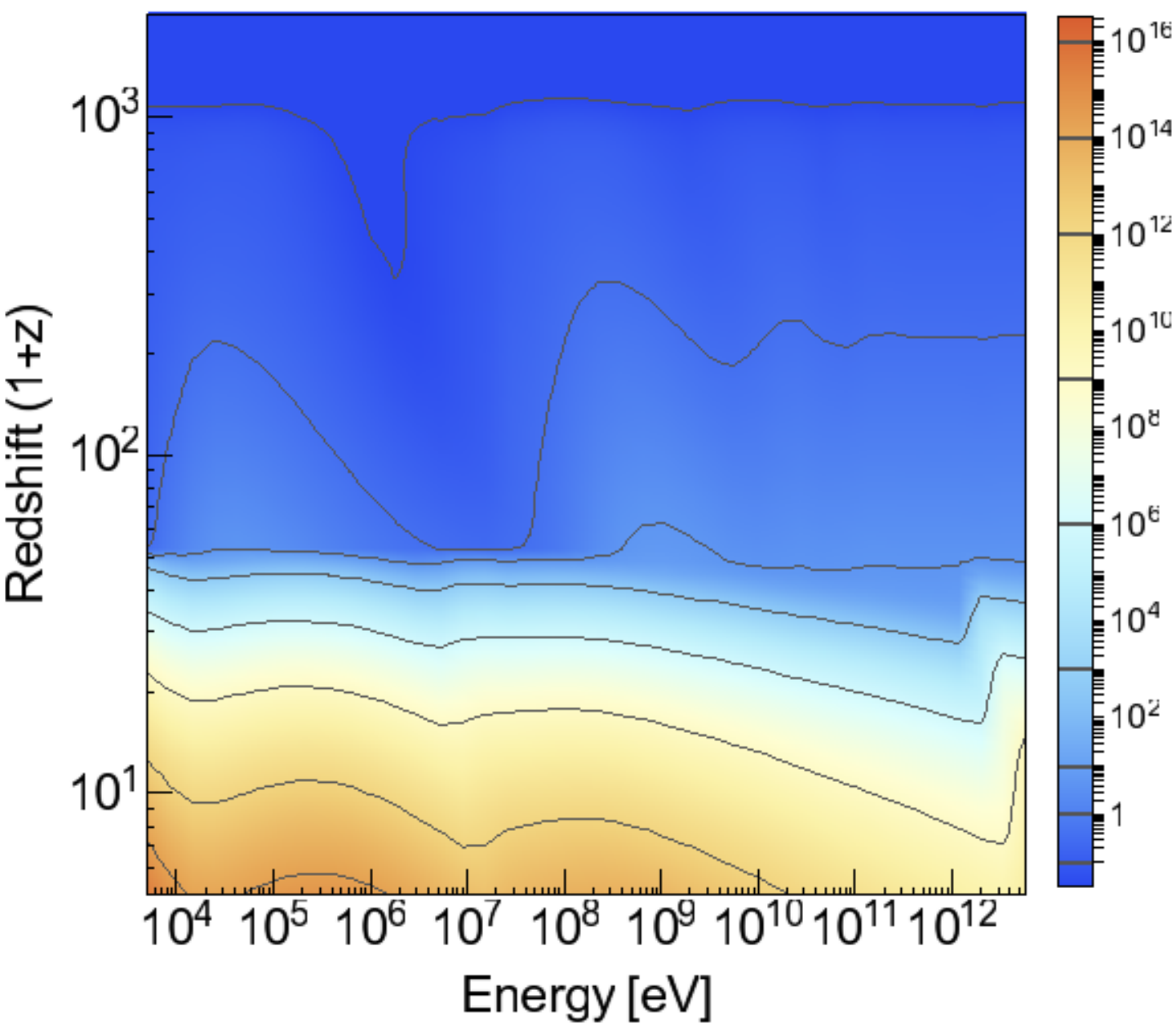}
	} \\
	\subfigure{
		\includegraphics[scale=0.32]{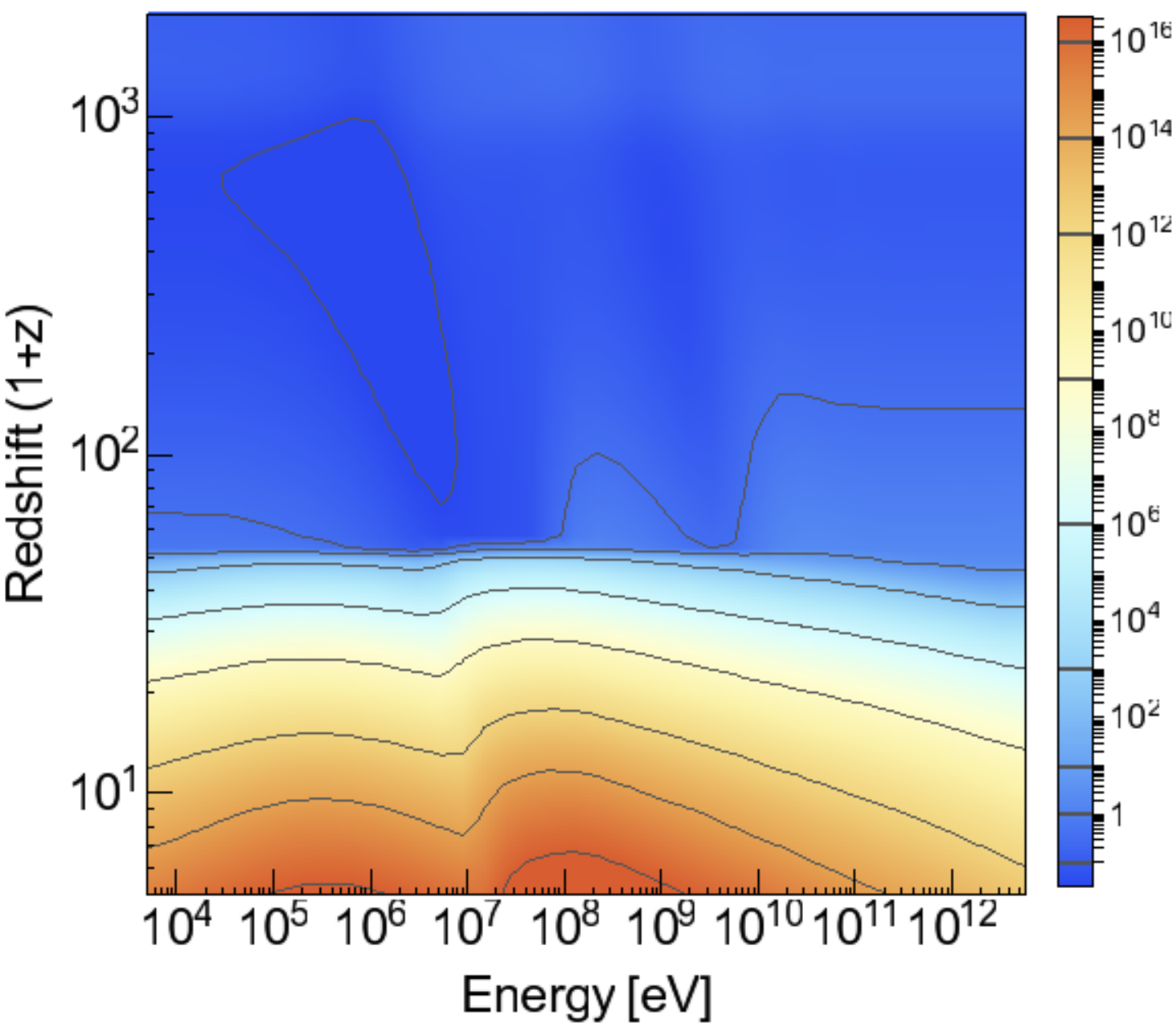}
	} & 
	\subfigure{
		\includegraphics[scale=0.32]{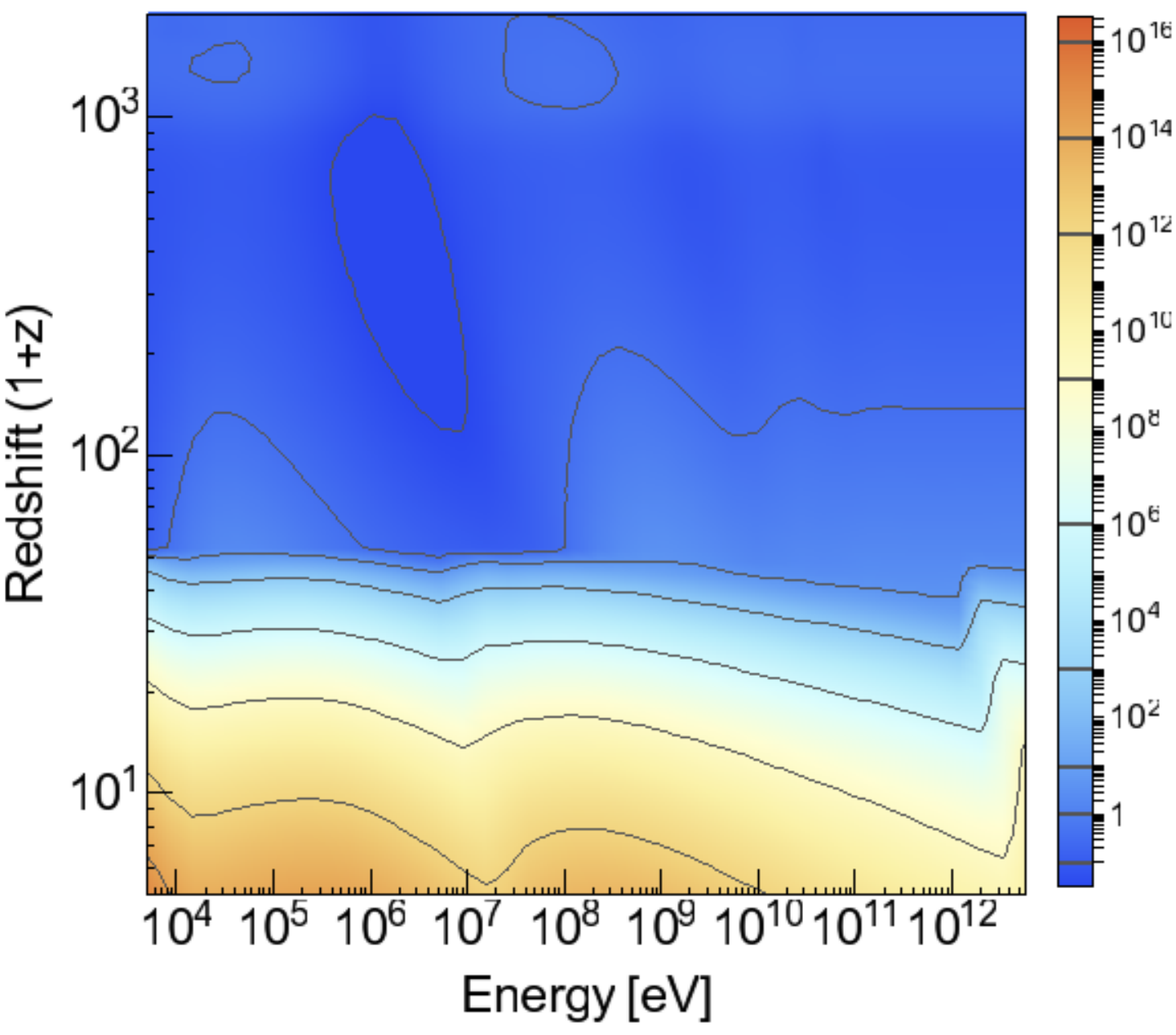}
	} \\
	\subfigure{
		\includegraphics[scale=0.32]{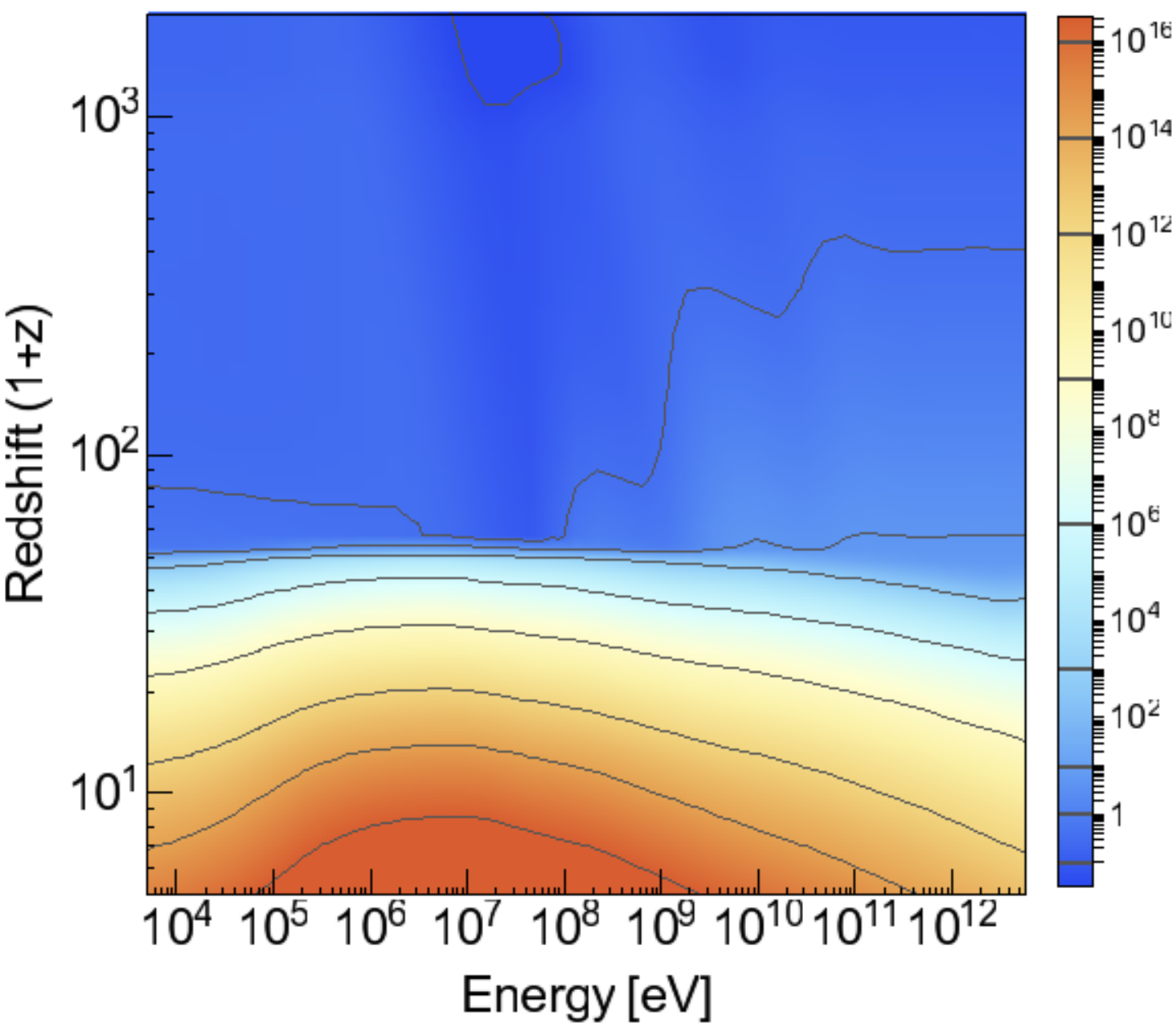}
	} & 
	\subfigure{
		\includegraphics[scale=0.32]{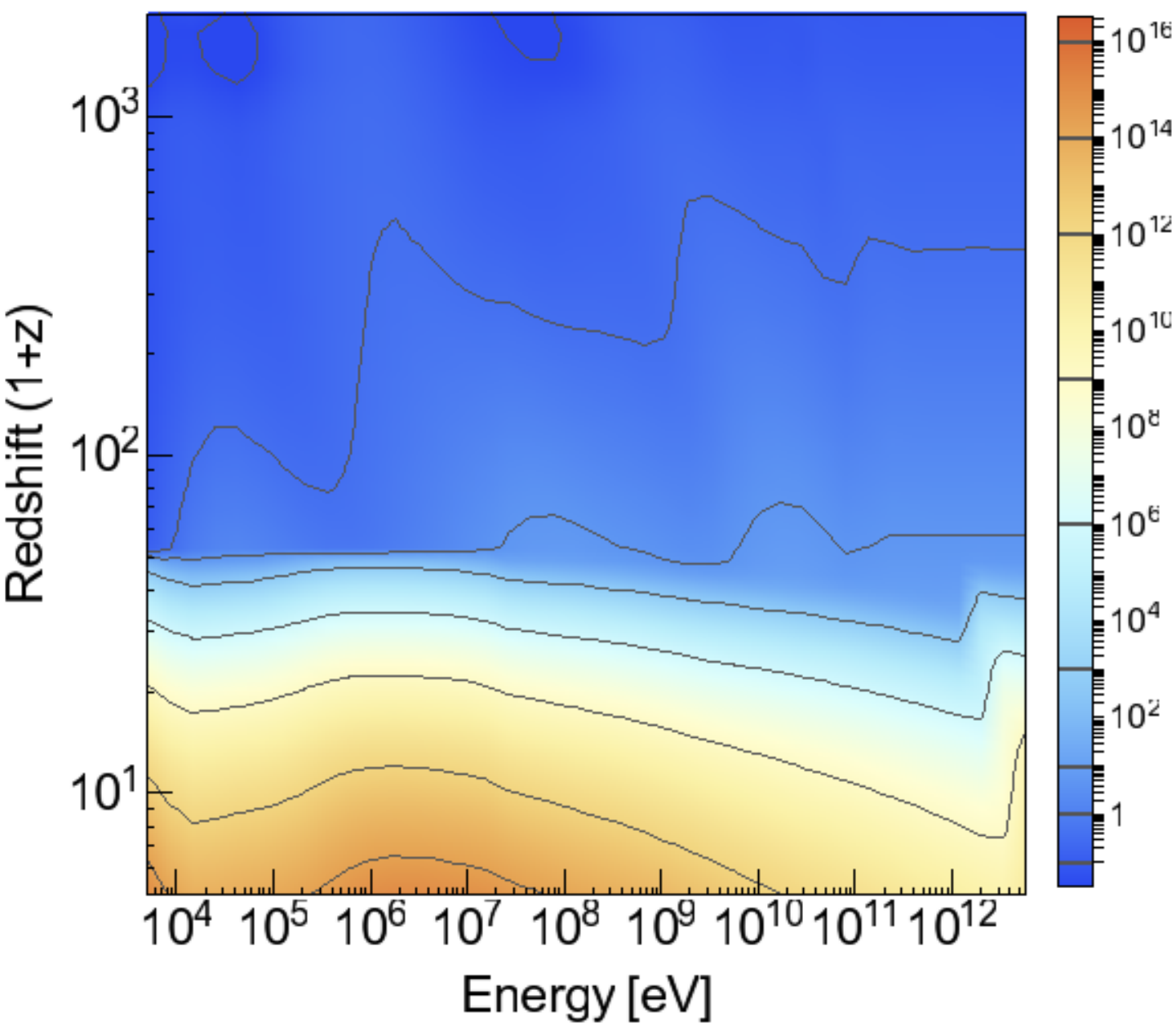}
	} \\
	\end{tabular}
	\caption{\footnotesize{Contour plots of $f_c(z)$ for $\chi \chi \to e^+ e^-$ (left) and $\chi \chi \to \gamma \gamma$ (right) $p$-wave annihilations into (from top to bottom) H ionization; He ionization; Lyman-$\alpha$; heating; and sub-10.2 eV continuum photons as a function of injection energy and redshift. Lines on the bar legend indicate the value of $f_c(z)$ at which contours are drawn. The structure formation prescription with the largest boost is used, and no reionization is assumed.  }}
	\label{fig:fz_pWave}
\end{figure*}

\begin{figure*}[t!]
	\begin{tabular}{cc}
	\subfigure{
		\includegraphics[scale=0.32]{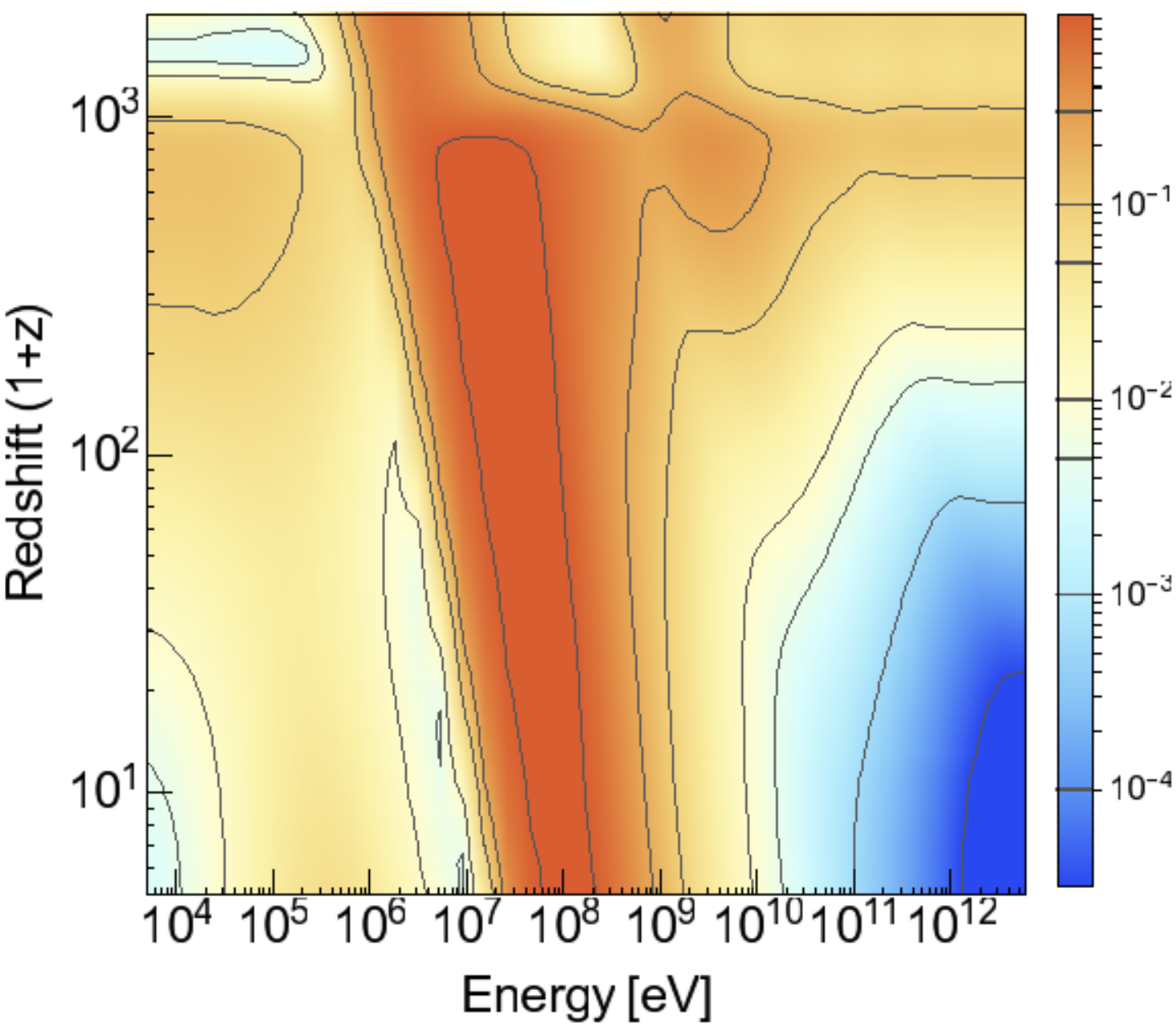}
	} & 
	\subfigure{
		\includegraphics[scale=0.32]{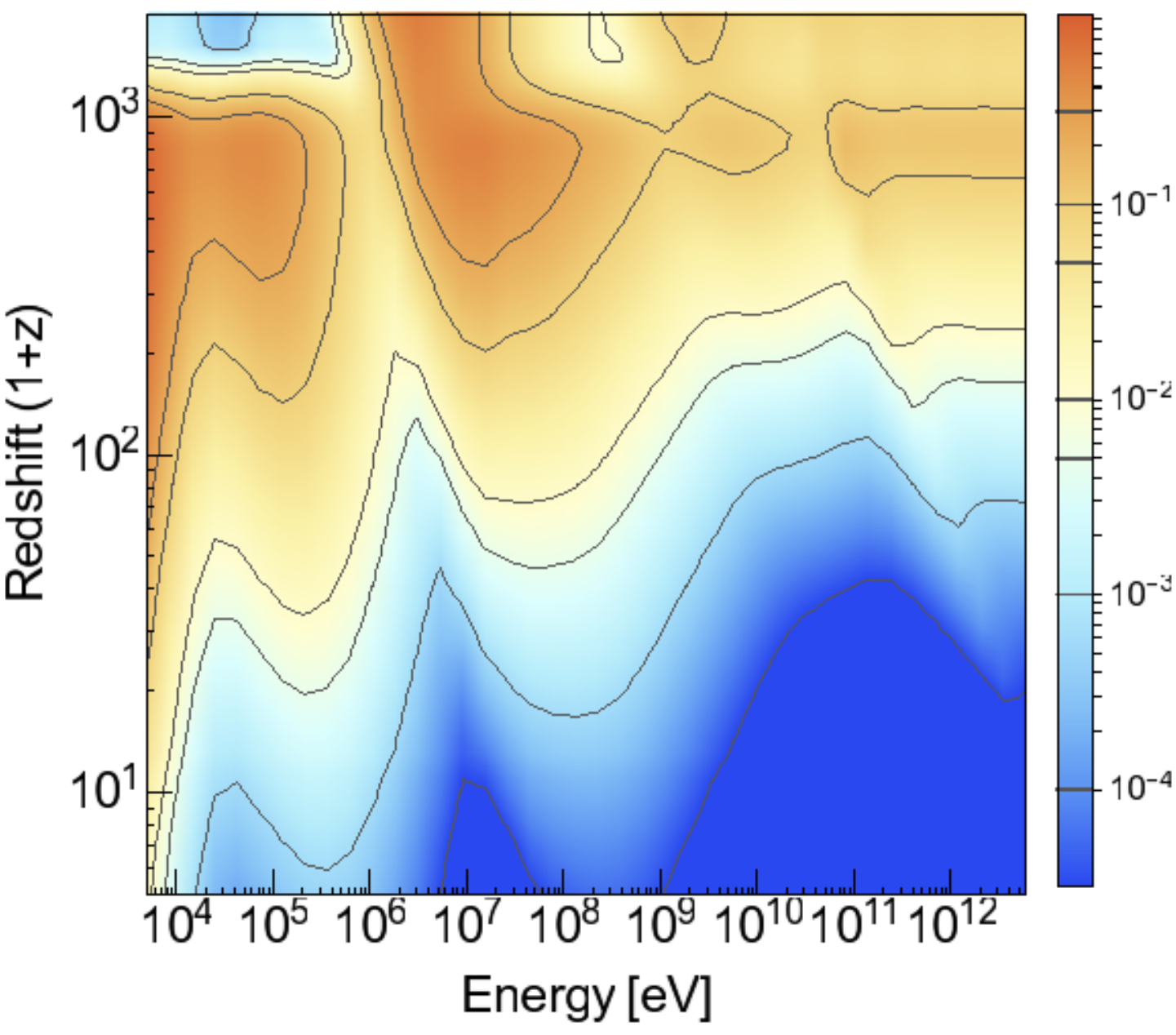}
	} \\
	\subfigure{
		\includegraphics[scale=0.32]{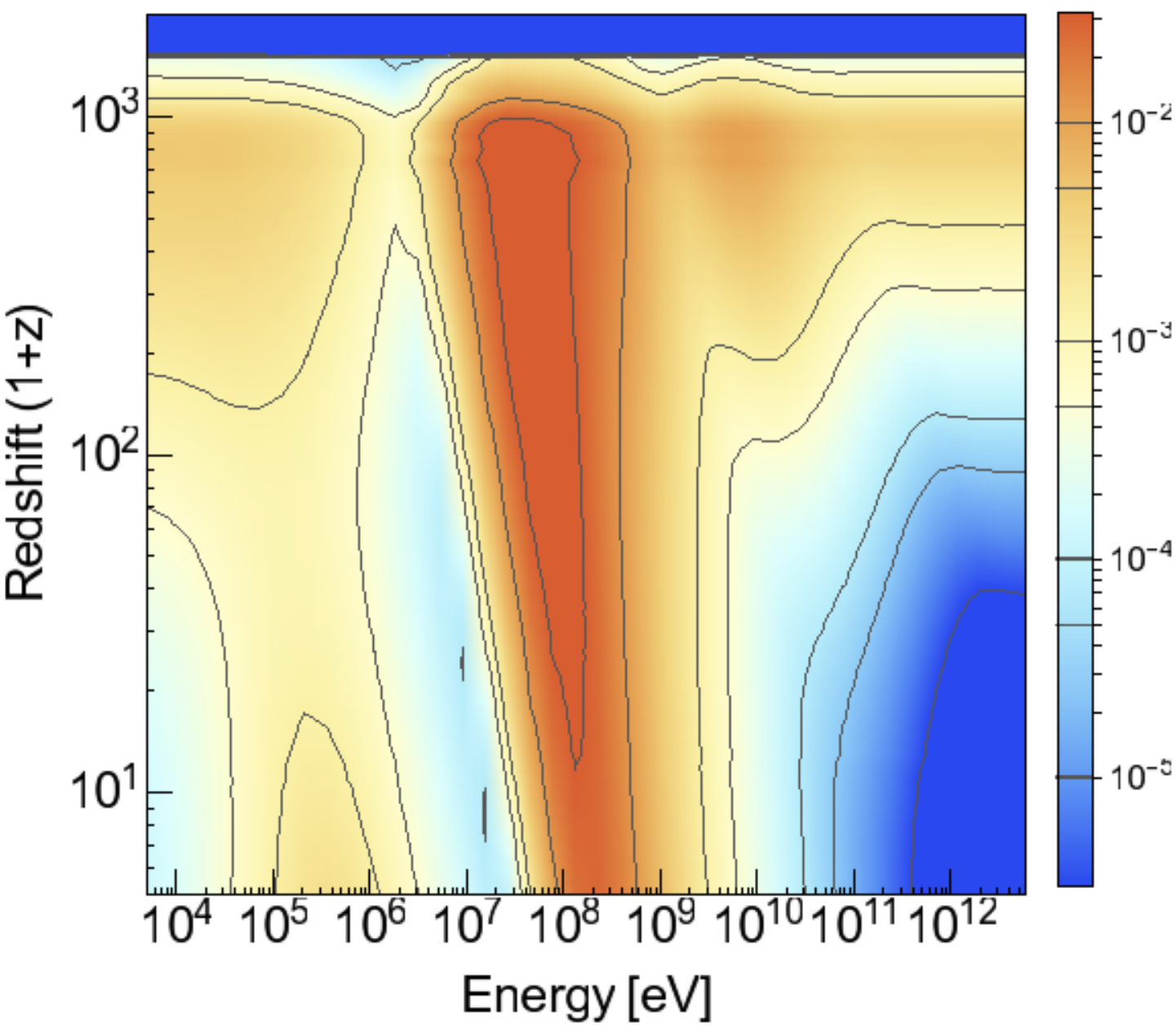}
	} & 
	\subfigure{
		\includegraphics[scale=0.32]{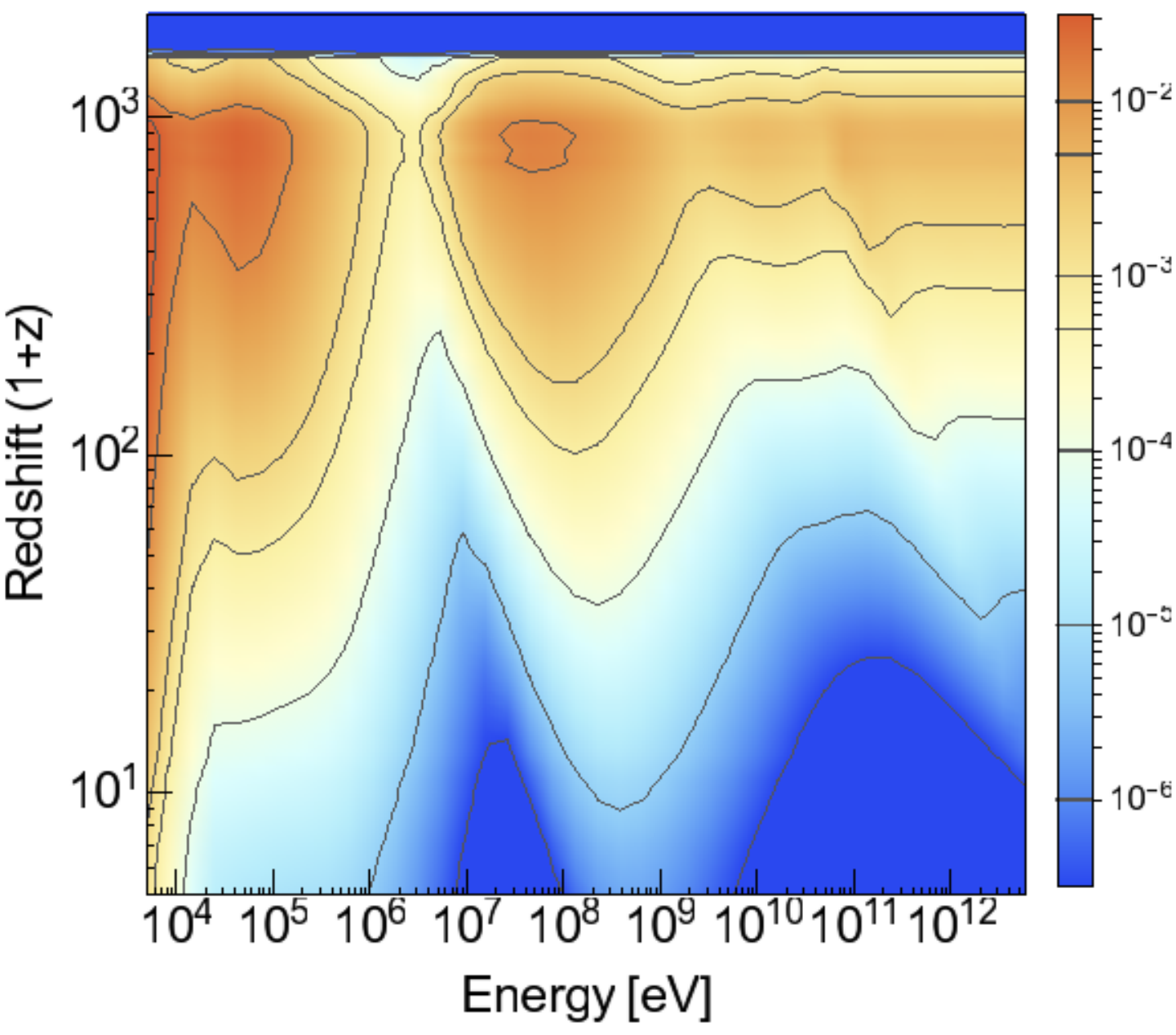}
	} \\
	\subfigure{
		\includegraphics[scale=0.32]{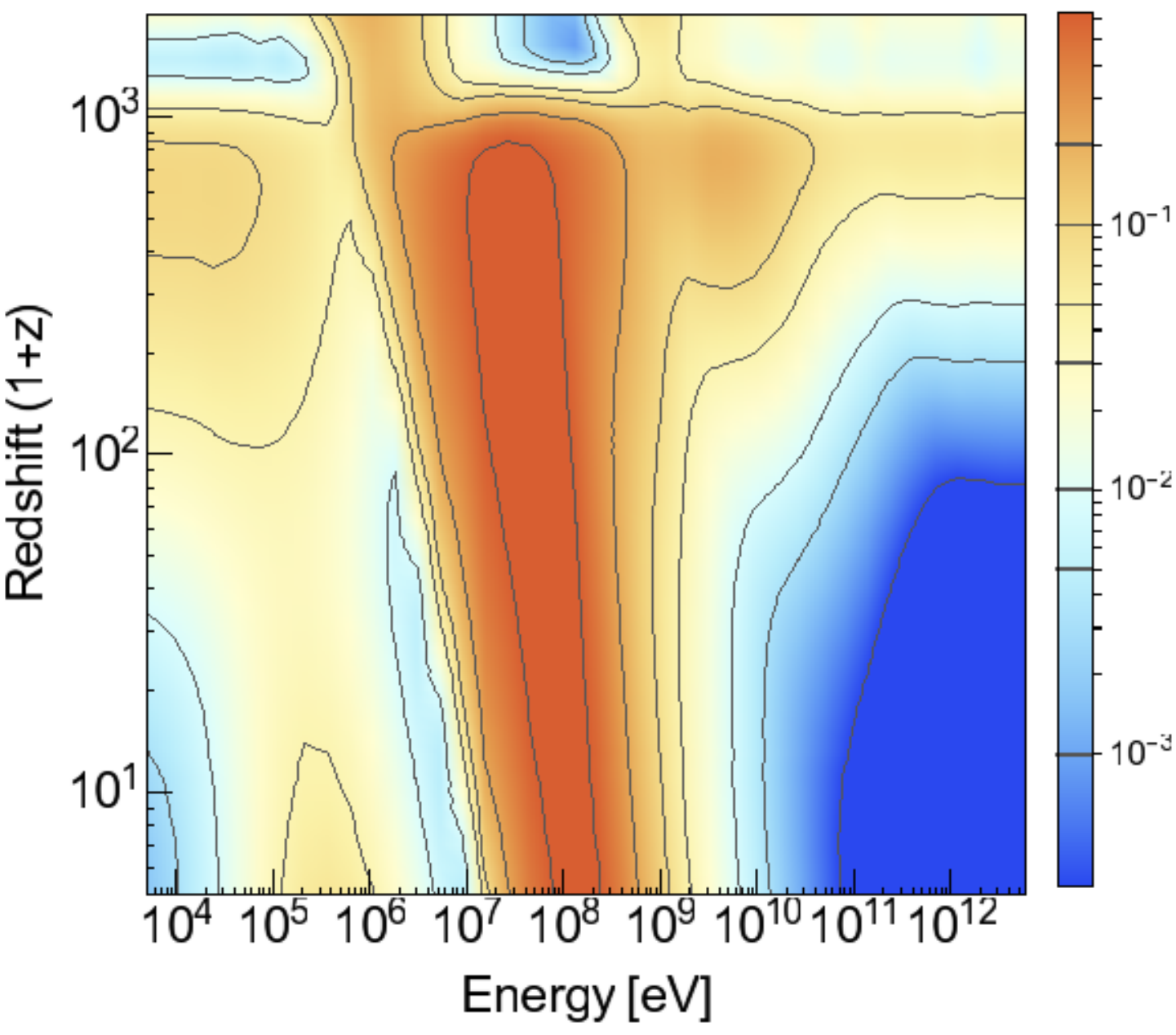}
	} & 
	\subfigure{
		\includegraphics[scale=0.32]{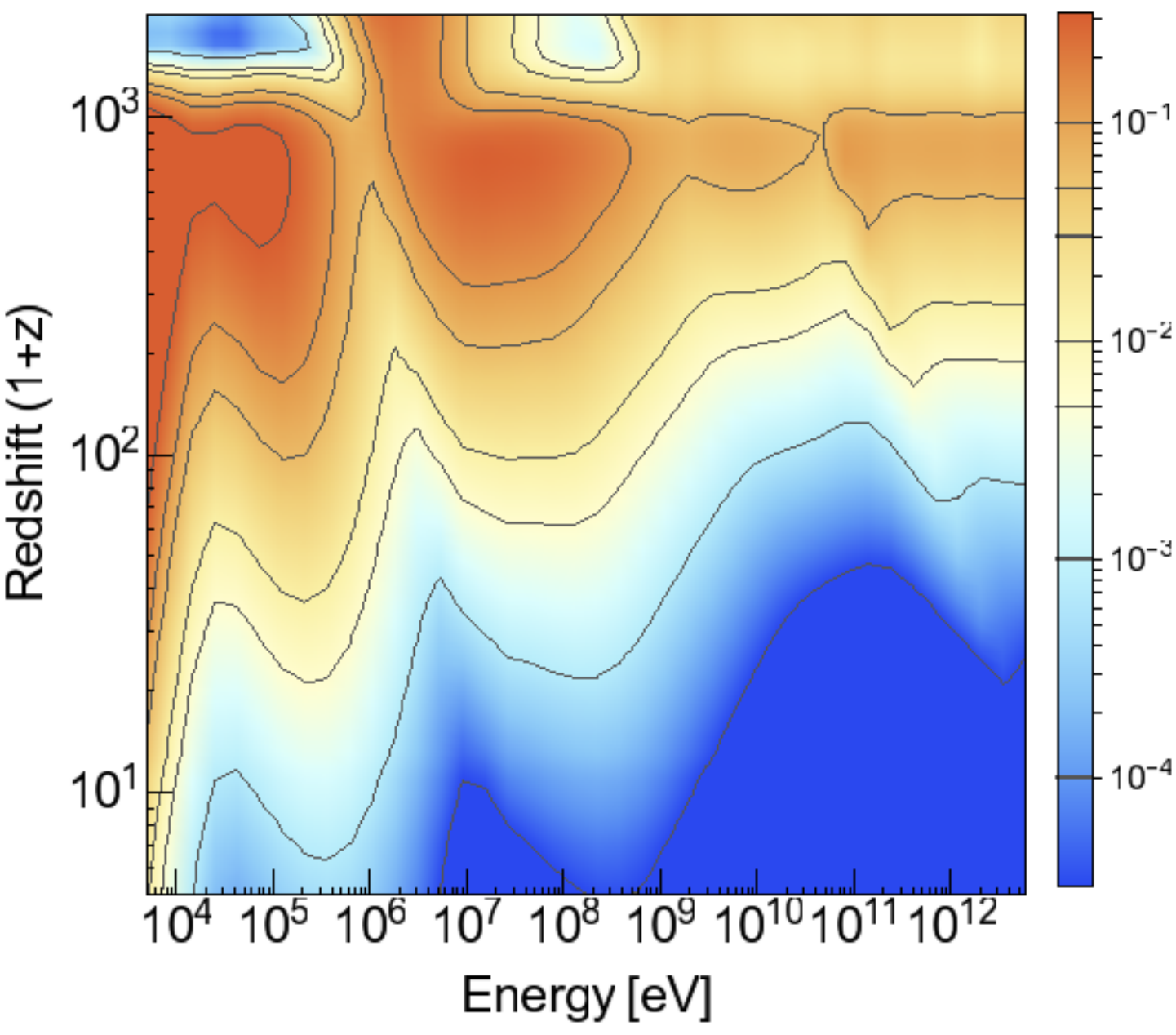}
	} \\
	\subfigure{
		\includegraphics[scale=0.32]{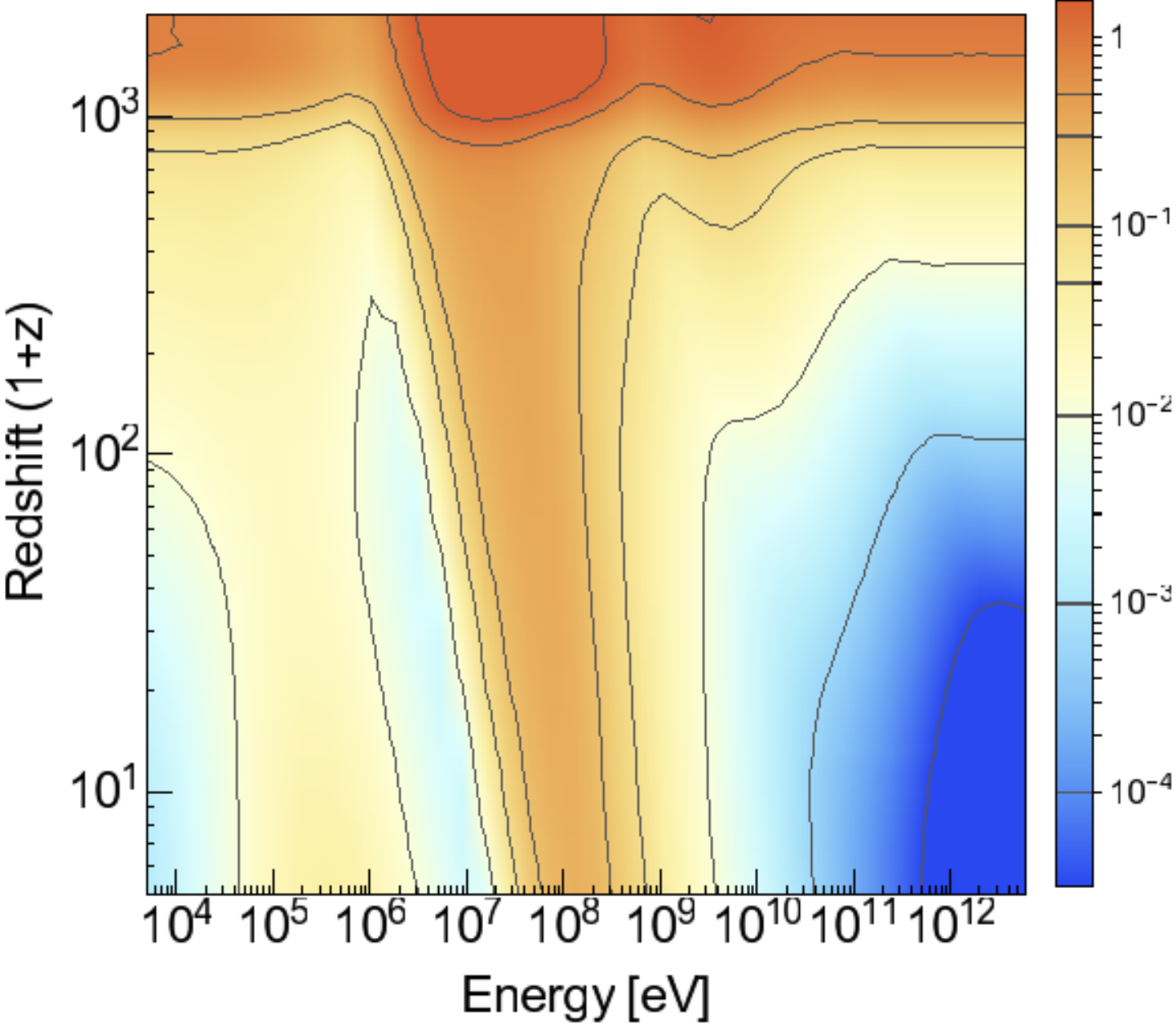}
	} & 
	\subfigure{
		\includegraphics[scale=0.32]{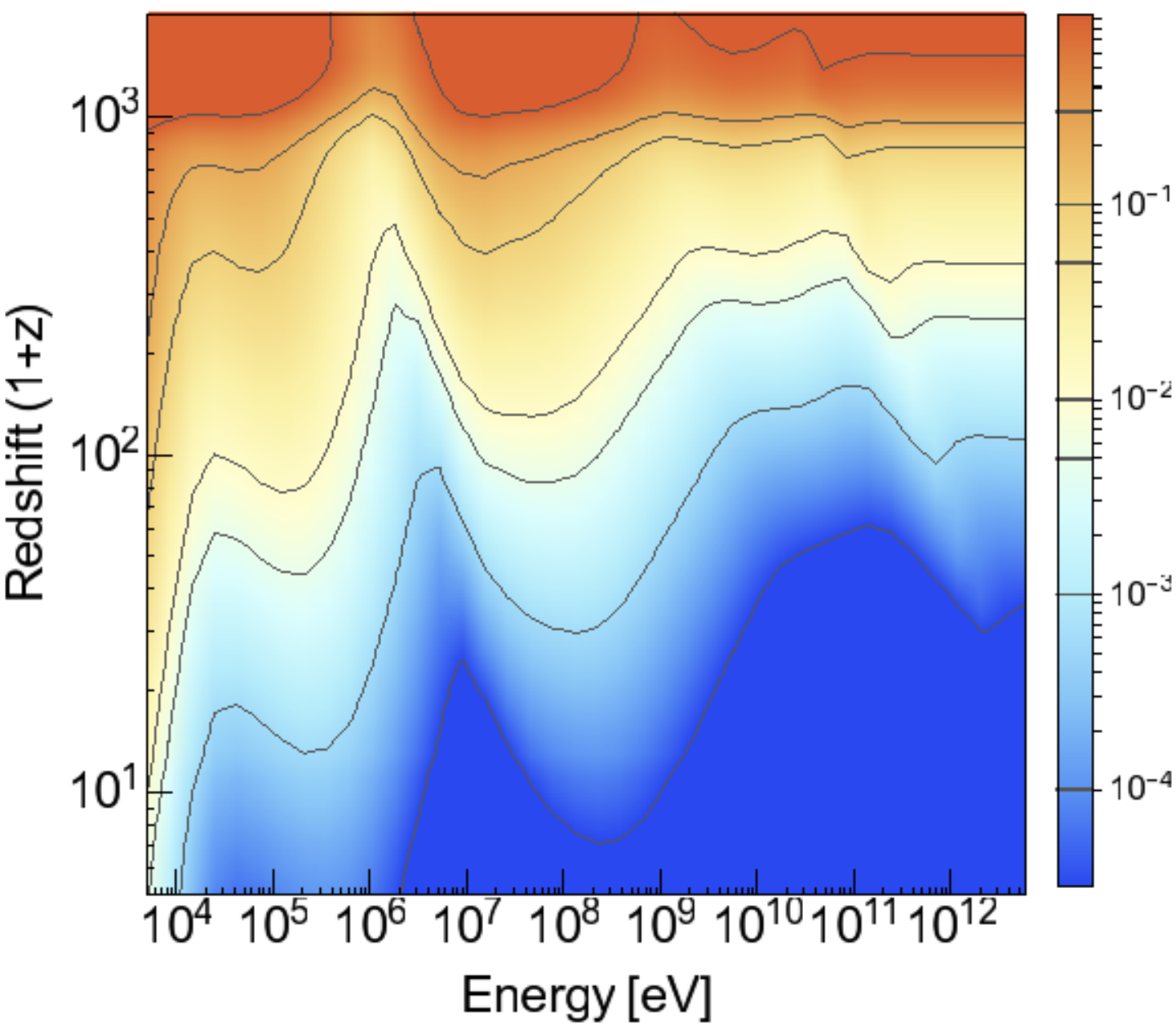}
	} \\
	\subfigure{
		\includegraphics[scale=0.32]{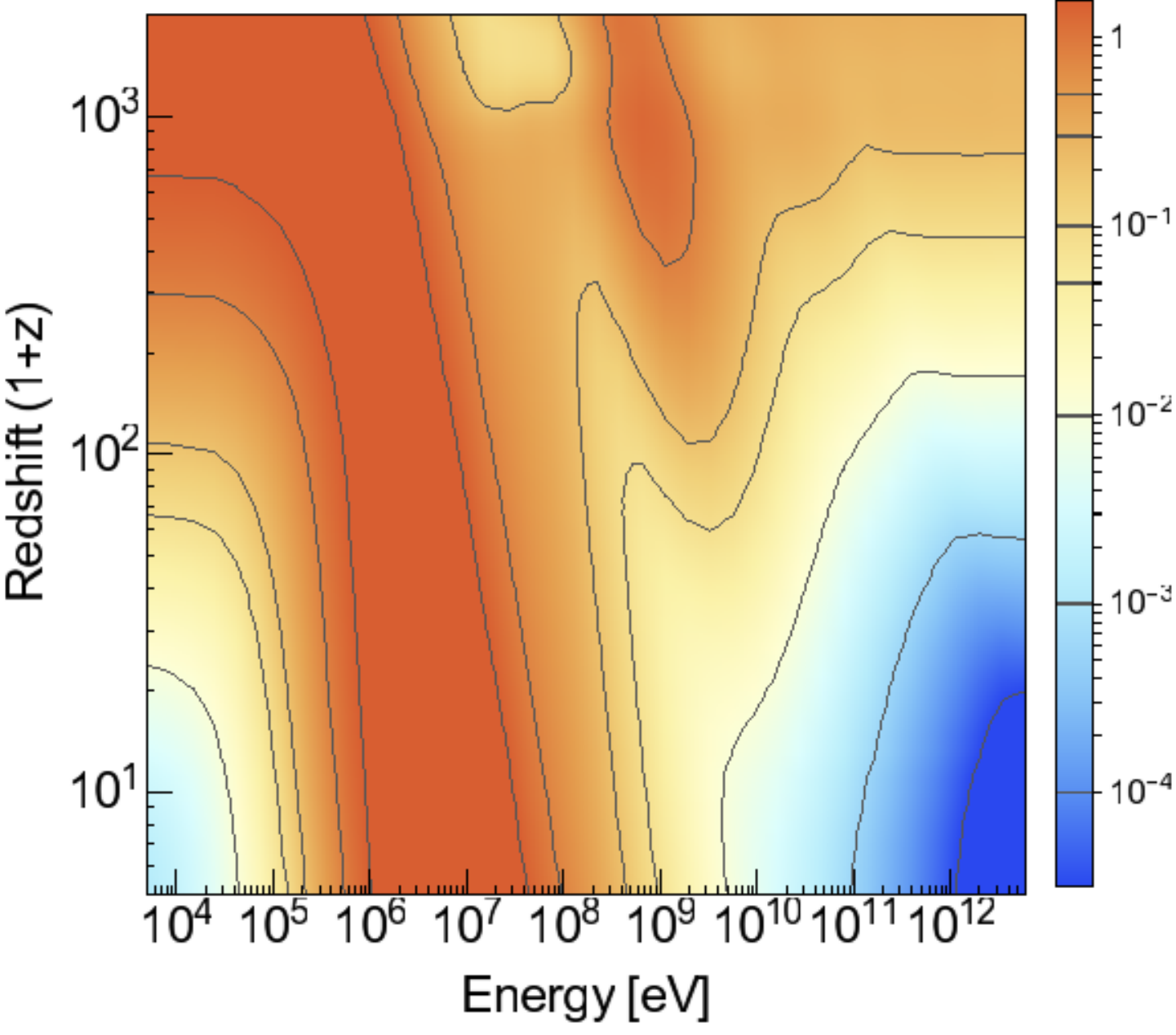}
	} & 
	\subfigure{
		\includegraphics[scale=0.32]{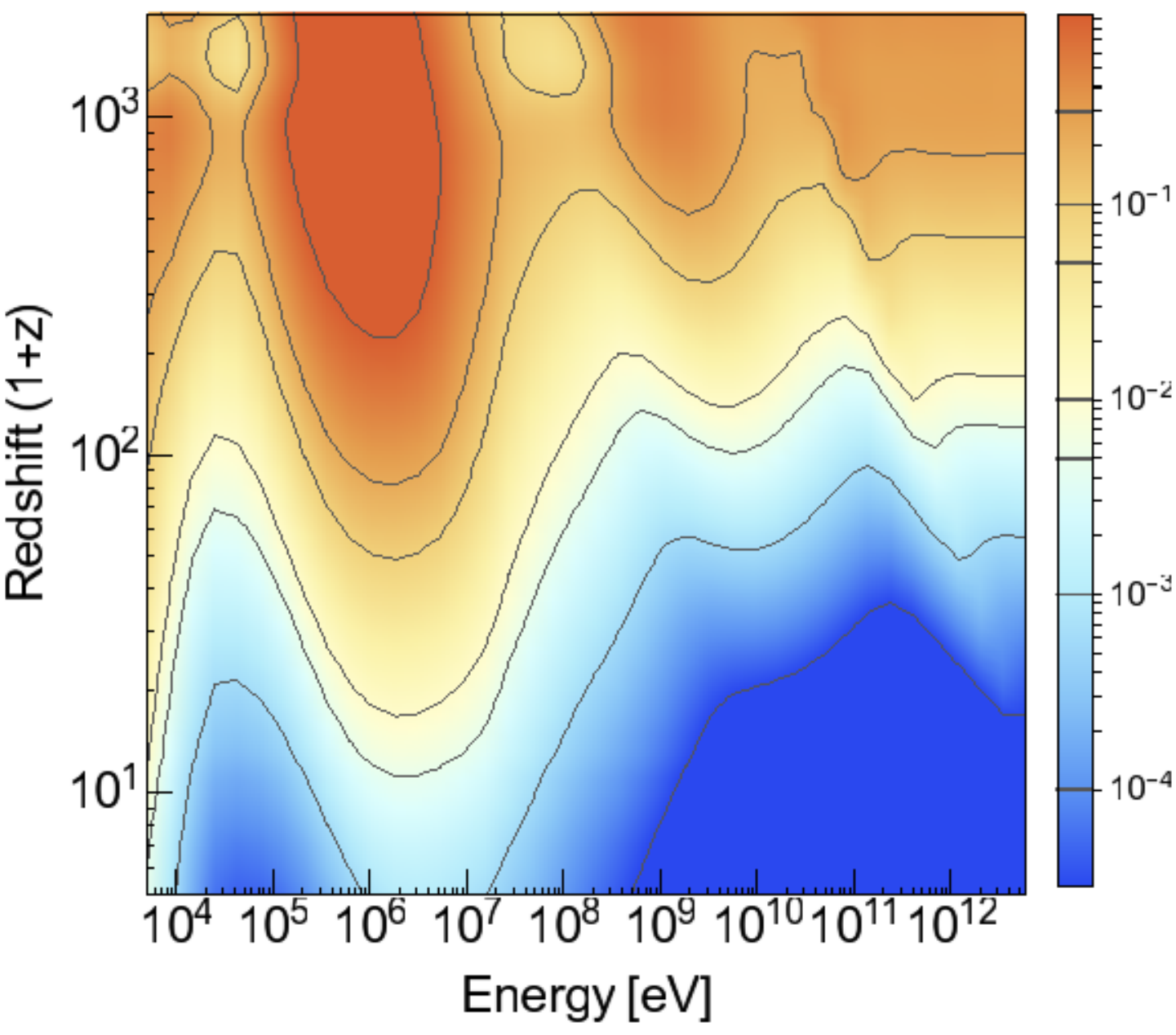}
	} \\
	\end{tabular}
	\caption{\footnotesize{Contour plots of $f_c(z)$ for $\chi \to e^+ e^-$ (left) and $\chi \to \gamma \gamma$ (right) decays into (from top to bottom) H ionization; He ionization; Lyman-$\alpha$; heating; and sub-10.2 eV continuum photons as a function of injection energy and redshift. Lines on the bar legend indicate the value of $f_c(z)$ at which contours are drawn. No reionization is assumed.  }}
	\label{fig:fz_decay}
\end{figure*}

Figures \ref{fig:fz_sWave}, \ref{fig:fz_pWave} and \ref{fig:fz_decay} shows contour plots of $f_c(z)$ for annihilations or decays into $e^+e^-$ and $\gamma \gamma$ as a function of redshift and injection energy, based on equation (\ref{eqn:fz}). No reionization is assumed in all of these plots, and for scenarios where structure formation is important, the prescription with the largest boost is used in the calculation.

\bibliography{ionization_jzf_hl}

\end{document}